\documentclass[prd,preprintnumbers,nofootinbib,aps]{revtex4-1}
\usepackage{graphicx}
\usepackage{dcolumn}
\usepackage{bm}
\usepackage{epsfig,amsmath}
\usepackage{amssymb}
\usepackage{subfigure}
\usepackage{float}
\usepackage[usenames,dvipsnames]{color}
\usepackage[pagebackref=false, colorlinks=true]{hyperref}
\definecolor{redish}{rgb}{0.7,0.2,0.0}  
\definecolor{bluish}{rgb}{0.2,0.5,0.8}

\hypersetup{linkcolor=redish,          
                  citecolor=blue,        
                  filecolor=magenta,      
                  urlcolor=bluish}          

\DeclareFontFamily{U}{rsfs}{}         
\DeclareFontShape{U}{rsfs}{m}{n}{<5> rsfs5 <6><7> rsfs7          %
  <8><9><10><10.95><12><14.4><17.28><20.74><24.88> rsfs10}{}     %
\DeclareMathAlphabet{\mathfs}{U}{rsfs}{m}{n}

\def \O{\Omega}

\def \f{\frac}
\def \w{\wedge}
\def \o{\omega}

\def \a{\alpha}
\def \b{\beta}
\def \t{\tilde}
\def \g{\gamma}
\def \s{\sigma}

\def \p{\partial}
\def \d{\Delta}

\def \l{\lambda}
\def \th{\theta}
\def \r{\rho}
\def \v{\varepsilon}
\def \k{\kappa}

\begin{document}
\title{Distinguishing Kerr naked singularities and black holes
using the spin precession of a test gyro in strong gravitational fields}
\author{Chandrachur Chakraborty$^1$}
\email{chandrachur.chakraborty@tifr.res.in}
\author{Prashant Kocherlakota$^1$}
\email{k.prashant@tifr.res.in}
\author{Mandar Patil$^2$}
\email{mpatil@impan.pl}
\author{Sudip Bhattacharyya$^1$}
\email{sudip@tifr.res.in}
\author{Pankaj S. Joshi$^1$}
\email{psj@tifr.res.in}
\author{Andrzej Kr\'olak$^2$}
\email{krolak@impan.pl}

\affiliation{$^1$Tata Institute of Fundamental Research, Mumbai 400005, India}
\affiliation{$^2$Institute of Mathematics of Polish Academy of Sciences, 
Sniadeckich 8, 00-956 Warsaw, Poland}

\begin{abstract}
We study here the precession of the spin of a test gyroscope 
attached to a stationary observer in the Kerr spacetime, specifically, to
distinguish naked singularity (NS) from black hole (BH). It 
was shown recently that for gyros attached to static observers,
their precession frequency became arbitrarily large in the limit of approach 
to the ergosurface. For gyros attached to stationary observers that 
move with non-zero angular velocity $\Omega$, 
this divergence at the ergosurface can be avoided. Specifically, for such gyros, 
the precession frequencies diverge on the event
horizon of a BH, but are finite and regular for NS everywhere except at the singularity itself. 
Therefore a genuine detection of the event 
horizon becomes possible in this case. We also show that for a near-extremal NS ($1<a_* < 1.1$), 
characteristic features appear in the radial profiles of the 
precession frequency, using which we can further distinguish a near-extremal
NS from a BH, or even from NS with larger angular momentum. We then investigate
the Lense-Thirring (LT) precession or nodal plane precession frequency
of the accretion disk around a BH and NS 
to show that clear distinctions exist for these configurations in terms 
of radial variation features. The LT precession in equatorial circular orbits 
increases with approach to a BH, whereas for NS it increases, attains a peak
and then decreases. Interestingly, for $a_*=1.089$, it decreases until it vanishes
at a certain radius, and acquires negative values for $a_* > 1.089$ for a certain  
range of $r$. For $1<a_*<1.089$, a peak appears, but the LT frequency remains
positive definite. There are important differences in accretion disk LT 
frequencies for BH and NS and since LT frequencies are intimately related
to observed QPOs, these features
might allow us to determine whether a given rotating compact astrophysical
object is BH or NS. 
\end{abstract}

\maketitle

\section{Introduction}
An important issue in relativistic astrophysics and gravitation theory today
has been to rule out the existence of naked singularities in the gravitational collapse of massive 
matter clouds. Alternatively, if NSs do exist as endstates of massive stars collapsing 
under self-gravity towards the end of their life-cycles,
the important physical question would then be, how can one decide whether 
a particular astrophysical compact object is a BH or NS.
This is a key issue, really at the heart 
of making physical predictions about
very strong gravity regions in the universe, which major astrophysical 
missions are probing currently \cite{eht}.

In this connection, it is important to examine in some detail
the Kerr geometry configurations, and find physical quantities that can be
used to differentiate the BH from NS. The Kerr spacetime describes either a 
rotating BH or NS, depending on the Kerr spin parameter $a$, which 
is the specific angular momentum ($J/M$).  The Kerr BH possesses 
two event horizons and two ergoregions which are the outer and inner event horizons
and outer and inner ergoregions. The region between 
the outer event horizon and outer ergoradius is called the 
outer ergoregion or simply the ergoregion.  
Here, we will be primarily concerned with this region.
Similarly, the region between the 
inner event horizon and the inner ergoradius is called the 
inner ergoregion. The inner ergoregion is often dismissed as being unphysical 
since we do not really know reliably what happens behind the outer event
horizon of a rotating black hole. The rotational energy of the BH
can be extracted by a particle
from the outer ergoregion (henceforth, just ergoregion) and this is
called the Penrose process.

The ergoregion is responsible for several interesting phenomena, of which 
one has been recently discussed in Ref.\cite{cm} where
the exact Lense-Thirring (LT) precession frequency of a test gyroscope
in Kerr spacetime has been derived and it was shown that the 
LT precession frequency diverges on the boundary of the 
ergoregion (henceforth, \lq ergosurface\rq) of a BH. For BHs, the dimensionless Kerr parameter satisfies $a_*=J/M^2 \leq 1$. In another work \cite{ckj}, the spin precession frequency was
discussed in detail in the case of NS ($a_*>1$). For NS, the entire region 
between the inner and outer ergo radii is defined to be the ergoregion. 
It has been discussed there that the drastic change in geometry of the 
ergoregion for $a_*>1$, as opposed to $a_*\leq 1$, could allow for a 
differentiation of the two space-times using a physical quantity, namely the precession 
frequency of gyroscopes attached to static observers placed around these
compact objects. Specifically, for the NS configuration, 
along the pole a finite angle opens up 
where the ergoregion is absent and $r=0$ can be accessible 
via this region without passing through the ergoregion.
This region broadens on increasing the value of $a_*$ beyond $1$
and the ergoregion shrinks toward the equator. It was discussed that the precession 
frequency always diverges on the ergosurface. Since the ergoregion completely bounds a BH, on 
\lq approaching\rq\ it in any direction, one would find a divergence. Whereas, in the case of  
NS, the frequency remains finite and regular in the \lq opening
angle\rq\ due to the absence of the ergoregion. We find this to be a possible
experiment to distinguish the two qualitatively distinct Kerr configurations.

It was discussed in Ref.\cite{jmn} how 
a non-rotating black hole can be distinguished from a naked singularity.
In this paper, we discuss how one can distinguish a rotating BH from 
NS using the behavior of the precession frequency of 
the spin of a test gyro
which moves, in general, along a non-geodetic orbit around such a Kerr compact object, thus generalizing the earlier work on distinguishing black holes 
from naked singularities.
In this regard, we find it useful to mention that recently Bini et al. have 
analyzed 
the precession of a test gyroscope along bound \cite{binib} and unbound
\cite{biniu} equatorial plane 
geodesic orbits around a BH with respect to a
static reference frame whose axes point towards \lq fixed stars.\rq\
It is well known that the paths followed by spinning test particles are not,
in general, geodesics \cite{abk, hoj}. 
 
In the present article, we discuss the precession of
a gyroscope both outside and inside the ergoregion of a BH and NS. We find
that the divergence of the spin precession frequency on the ergosurface 
reported in Ref.\cite{ckj} can be avoided if the test gyro moves with
a non-zero angular velocity $\O$. In Ref.\cite{ckj}, the motivation 
was to study the precession of spins of gyroscopes attached to stationary 
observers, that is, to answer physical questions like \lq how the 
gyroscope of an astronaut holding his spaceship at a constant distance 
from a Kerr compact object will behave.\rq\
The four-velocity $(u)$ of \lq static gyros\rq\ on the ergosurface 
satisfies $u.u=0$, that is, it becomes null. However, if one introduces 
an azimuthal component to the four-velocity, the norm can be made non-zero
and time-like. We can therefore extend the study of the behavior of a
gyro into the ergoregion. We find that the precession frequency of the 
gyro behaves differently inside the ergoregion of a BH versus that
of a NS, due to the presence of the event horizon,
therefore rendering it a viable physical quantity 
that can be used to distinguish a BH from NS.

We note that the Killing vectors of the Kerr spacetime provide invariant characterizations
of the ergosurfaces and horizons. The time translation Killing vector $\p_0$ 
is the unique Killing vector that approaches a unit timelike vector at 
spatial infinity and its vanishing norm gives the invariant location of
the ergosurface. Further, one can construct a Killing vector from a
linear combination, with constant coefficients, of $\p_0$ and the azimuthal
Killing vector $\p_\phi$. The vanishing norm of this Killing vector, for
some fixed $\O$ (say $\O=\O_H$), can be used to invariantly determine the location
of the horizon. With this motivation, we consider here the precession of
the spin of gyroscopes attached to stationary
observers, whose velocity vectors are proportional to the Killing vectors
$K=\p_0 + \O\p_\phi$. These gyroscopes move along circular orbits around
the central compact object with a constant angular velocity $\O$ which at
any given $(r,\theta)$ can be chosen to be in a particular range, so that
$K$ is timelike. Indeed, when the spacetime describes a BH,
this frequency becomes arbitrarily large for observers located
infinitesimally close to the horizon, in all coordinates. We interpret 
this as a coordinate invariant method to locate the horizon. For NS, 
a divergence in the precession frequency occurs at the singularity itself, which is present in the equitorial plane.
In \cite{ckj}, the precession frequency of gyroscopes attached 
to static observers was considered, namely observers that do not change their spatial 
position with time. The velocity vectors of these observers are proportional
to $\p_0$ and we found that the precession frequency for static observers located
infinitesimally close to the ergosurface became arbitrarily large, in all
coordinates, thus invariantly indicating the location of the ergosurface.

In a later section, we examine the Lense-Thirring precession 
frequency and the fundamental frequencies of an accretion disk around a Kerr
compact object as the other distinguishing criterion of a NS from BH. Stable
circular orbits in the equatorial plane for 
both the BH and NS cases are investigated, and we show 
that there are important characteristic differences which can be used
to distinguish the BH and NS configurations from
each other. Specifically, we find the features of stable
circular orbits in the equatorial plane for both the BH and
NS cases. We also show that the radial 
and epicyclic frequencies show rather distinct features in the BH and NS cases.
Further, in observed QPOs from accretion disks, if one finds a clear peak at 
some radius, then the existence of NSs could possibly 
be established. 

The scheme of the paper is as follows. In Sec.\ref{s2}, we discuss
stationary observers. We outline the derivation for
the spin precession frequency of a test gyro
which rotates with a non-zero angular velocity in a general stationary
and axisymmetric spacetime in Sec.\ref{formalism}. We use this expression for the 
precession frequency specifically for the Kerr spacetime 
and discuss its features in various useful regimes of the spacetime in
Sec.\ref{spe}. Characteristic features of the precession frequency of a gyroscope
orbiting NSs and BHs that can be used to distinguish them, in principle,
are described in Sec.\ref{5}. Further, in Sec.\ref{nens}, features which can be 
used to distinguish near-extremal NSs from NSs with higher angular momentum are discussed.  
In Sec.\ref{Lense-Thirring}, we use the Lense-Thirring precession (orbital plane
precession) of an 
accretion disk around a Kerr spacetime as another distinguishing
criterion. We also discuss observational aspects related to QPOs.
Finally Sec.\ref{con} outlines our results and conclusions.

\section{Stationary observers in Kerr spacetime \label{s2}}
In a rotating spacetime, observers can remain still without 
changing their location with respect to infinity only outside the ergoregion. 
Such observers are called static observers \cite{mtw, str} and their 
four-velocity is given as
\begin{eqnarray}
 u^{\s}_{\rm static}=u^t_{\rm static}~(1,0,0,0).
\end{eqnarray}
Once inside the ergoregion, it is impossible to stay fixed to a point 
without changing all three spatial coordinates 
(say, $r, \th, \phi$) of their world lines with time. The ergoregion 
is a characteristic feature of non-static stationary spacetimes. 
Specifically, in the case of the BH, its event horizon 
lies inside the ergoregion. This means that, in general, static observers 
cannot exist arbitrarily close to the horizon of black hole. 
In contrast, an observer can hover very close to the horizon of 
a Schwarzschild black hole, remaining fixed. However, it is 
possible for observers to fix ($r, \th$) and rotate (prograde only) 
around a BH or NS inside their ergoregion, with respect to infinity. 
Such observers are called stationary observers and their four-velocity is given as
\begin{eqnarray}
 u^{\s}_{\rm statio}=u^t_{\rm statio}~(1,0,0,\O)
 \label{ui}
\end{eqnarray}
where $t$ is the time coordinate and $\O$ is the angular velocity 
of the observer. Since we are interested in studying timelike observers, 
the values that $\O$ takes are restricted. This is true outside the ergoregion as well and 
retrograde rotating observers are then allowed. Let us henceforth call
gyroscopes attached to static and stationary observers static gyroscopes and 
stationary gyroscopes respectively, for brevity.

\section{Spin precession of a test gyroscope: Formalism \label{formalism}}~
Let us consider a test gyroscope attached to a stationary 
observer, which moves along a Killing trajectory in a stationary spacetime. 
The spin of such a gyroscope undergoes Fermi-Walker transport along, 
\begin{equation}
u=(-K^2)^{-\f{1}{2}} K,
\end{equation}
where $K$ is the timelike Killing vector field.  In this special situation, 
it is known that the gyroscope precession
frequency coincides with the vorticity field associated with the Killing
congruence. That is, this gyroscope
rotates relative to a corotating frame with an angular 
velocity and this effect is generally known as the \lq\lq gravitomagnetic
precession,\rq\rq\ since the vorticity vector plays the role of a magnetic field in the $3+1$
splitting of spacetime \cite{jan}. Thus, the general spin precession 
frequency of a test gyro, $\O_s$, is the rescaled vorticity field of the
observer congruence and can be expressed as \cite{str}
\begin{eqnarray}
\t{\O}_s &=& \f{1}{2K^2}*(\t{K} \w d\t{K})\  \nonumber\\
\mbox{or,}\,\,\,\, (\O_s)_\mu &=& \frac{1}{2K^2}\eta_\mu^{\ \l\beta\delta}K_\l\p_\beta K_\delta,
 \label{b}
\end{eqnarray}
where $\O_s$ is the spin precession frequency in coordinate basis, $*$ represents the Hodge star 
operator or Hodge dual, $\eta^{\mu\l\beta\delta}$ represent the components of the volume-form in spacetime and $\t{K}$, $\t{\O}_s$ are the one-forms of $K$ and $\O_s$ respectively. In any stationary spacetime,
$K$ can be chosen to be $K=\p_0$ 
for which, from Eq.(\ref{b}), $\O_s$ becomes $\O_{\rm LT}$, the Lense-Thirring (LT) precession frequency. This can be expressed as \cite{str, cm},
\begin{equation}
\O_{\rm LT} = \frac{1}{2} \frac{\varepsilon_{ijl}}{\sqrt {-g}} \left[g_{0i,j}\left(\p_l - 
\frac{g_{0l}}{g_{00}}\p_0\right) -\frac{g_{0i}}{g_{00}} g_{00,j}\p_l\right] .
\label{bs}
\end{equation}
In a static spacetime, LT precession vanishes since $g_{0i}=0$. On the other hand, 
it does not vanish in a stationary spacetime. Moreover, 
due to the presence of $K^2=g_{00}$ in the denominator, Eq.(\ref{b}) 
and Eq.(\ref{bs})
diverge if $g_{00}$ vanishes. In a stationary and axisymmetric spacetime,
this happens only on the ergosurface, which makes $K$ a null vector there. 
Inside the ergoregion, $K$ is no longer timelike but becomes spacelike.
Thus, Eq.(\ref{bs}) is invalid inside
the ergoregion as well as on its boundary.

Since the focal point of this paper is to study spin precession in the Kerr
spacetime, we point out here that it has two Killing vectors: the time translation Killing vector $\p_0$ 
and the azimuthal Killing vector $\p_{\phi}$. Any constant coefficient linear combination $K=\p_0+\O \p_\phi$ is also a Killing vector and this exhausts the set of Killing vectors in the Kerr spacetime. 
As stated earlier, the Killing vector $\p_0$ is the unique Killing vector that approaches 
a unit timelike vector at spatial infinity. We now consider the full timelike Killing
vector of the Kerr space-time and study spin precession for observers with $u$, that is
stationary observers. This is in contrast to \cite{ckj}, where the velocity was chosen
to be proportional $\p_0$ and these described static observers. Therefore, for a general
stationary spacetime which also possess a spacelike Killing vector we can write the most general timelike Killing vector as,
\begin{eqnarray}
 K=\p_0+\O\ \p_c.
 \label{K}
\end{eqnarray}
where $\p_c$ is a spacelike Killing vector of that stationary spacetime and $\O$, for an observer moving along integral curves of $K$, represents the angular velocity. 
The metric of this particular spacetime is independent of $x^0$ and $x^c$ coordinates. 
The corresponding co-vector of $K$ is,
\begin{eqnarray}
 \t{K}=g_{0\nu}dx^{\nu}+\O g_{\g c}dx^{\g},
\end{eqnarray}
where $\g , \nu=0,c,2,3$ in 4-dimensional spacetime. 
Separating space and time components we can write $\t{K}$ as
\begin{eqnarray}
  \t{K}=(g_{00}dx^0+g_{0c}dx^c+g_{0i}dx^i)+\O (g_{0c}dx^0+g_{cc}dx^c+g_{ic}dx^i)
  \label{kt}
\end{eqnarray}
where $i=2,3$. Since we are mainly interested in the ergoregion of a stationary and
axisymmetric spacetime,
we can abolish $g_{0i}$ and $g_{ic}$ terms. Finally, we obtain
\begin{eqnarray}
  \t{K}=(g_{00}dx^0+g_{0c}dx^c)+\O (g_{0c}dx^0+g_{cc}dx^c)
  \label{kt}
\end{eqnarray}
and 
\begin{eqnarray}
  d\t{K}=(g_{00,k}dx^k \w dx^0+g_{0c,k}dx^k \w dx^c)
  +\O (g_{0c,k}dx^k \w dx^0+g_{cc,k}dx^k \w dx^c) .
  \label{kto}
\end{eqnarray}
Now, Eq.(\ref{b}) can be modified as 
\begin{eqnarray}
 \t{\O}_p=\f{1}{2K^2}*(\t{K} \w d\t{K}).
 \label{be}
\end{eqnarray}
Substituting the expressions of $\t{K}$ and $d\t{K}$ in Eq.(\ref{be}),
we obtain the one-form of the precession frequency {\footnote{$\O_p$ is not the LT 
precession frequency of the gyro. Since the gyro has a non-zero angular velocity
$\O$, the precession frequency $\O_p$ is modified. If we set $\O=0$, we 
have $\O_p=\O_{\rm LT}$. In this work, the expression of $\O_p$ describes the 
overall precession which includes the LT effect as well as some other effects (for e.g., geodetic precession)
which we will discuss as we proceed.}} as:
\begin{eqnarray}
 \t{\O}_p&=&\f{\v_{ckl}g_{l\mu}dx^{\mu}}{2\sqrt {-g}\left(1+2\O\f{g_{0c}}{g_{00}}
 +\O^2\f{g_{cc}}{g_{00}}\right)}. \nonumber
 \\
 &&\left[\left(g_{0c,k}
-\f{g_{0c}}{g_{00}} g_{00,k}\right)+\O\left(g_{cc,k}
-\f{g_{cc}}{g_{00}} g_{00,k}\right)+ \O^2 \left(\f{g_{0c}}{g_{00}}g_{cc,k}
-\f{g_{cc}}{g_{00}} g_{0c,k}\right) \right]
\end{eqnarray}
where we have used $*(dx^0 \w dx^k \w dx^c)=\eta^{0kcl}g_{l\mu}dx^{\mu}
=-\f{1}{\sqrt{-g}}\v_{kcl}g_{l\mu}dx^{\mu}$ and 
$K^2=g_{00}+2\O g_{0c}+\O^2 g_{cc}$. Corresponding vector ($\O_p$) of the 
co-vector $\t{\O}_p$ is
\begin{eqnarray}
\O_p&=&\f{\v_{ckl}}{2\sqrt {-g}\left(1+2\O\f{g_{0c}}{g_{00}}
 +\O^2\f{g_{cc}}{g_{00}}\right)}. \nonumber
 \\
 &&\left[\left(g_{0c,k}
-\f{g_{0c}}{g_{00}} g_{00,k}\right)+\O\left(g_{cc,k}
-\f{g_{cc}}{g_{00}} g_{00,k}\right)+ \O^2 \left(\f{g_{0c}}{g_{00}}g_{cc,k}
-\f{g_{cc}}{g_{00}} g_{0c,k}\right) \right]\p_l .
\label{elt}
\end{eqnarray}
In a stationary and axisymmetric spacetime with coordinates
$0, r, \th, \phi$, Eq.(\ref{elt}) reduces to
\begin{eqnarray}
 \vec{\O}_p&=&\f{1}{2\sqrt {-g}\left(1+2\O\f{g_{0\phi}}{g_{00}}+\O^2\f{g_{\phi\phi}}{g_{00}}\right)}. \nonumber
 \\
 &&\left[-\sqrt{g_{rr}}\left[\left(g_{0\phi,\th}
-\f{g_{0\phi}}{g_{00}} g_{00,\th}\right)+\O\left(g_{\phi\phi,\th}
-\f{g_{\phi\phi}}{g_{00}} g_{00,\th}\right)+ \O^2 \left(\f{g_{0\phi}}{g_{00}}g_{\phi\phi,\th}
-\f{g_{\phi\phi}}{g_{00}} g_{0\phi,\th}\right) \right]\hat{r} \right. \nonumber
\\
&& \left. +\sqrt{g_{\th\th}}\left[\left(g_{0\phi,r}
-\f{g_{0\phi}}{g_{00}} g_{00,r}\right)+\O\left(g_{\phi\phi,r}
-\f{g_{\phi\phi}}{g_{00}} g_{00,r}\right)+ \O^2 \left(\f{g_{0\phi}}{g_{00}}g_{\phi\phi,r}
-\f{g_{\phi\phi}}{g_{00}} g_{0\phi,r}\right) \right]\hat{\th}\right] .
\label{ltp}
\end{eqnarray}
For $\O=0$, Eq.(\ref{ltp}) reduces to 
\begin{equation}
 \vec{\O}_p|_{\O=0}=\f{1}{2\sqrt {-g}}\left[-\sqrt{g_{rr}}\left(g_{0\phi,\th}
-\f{g_{0\phi}}{g_{00}} g_{00,\th}\right)\hat{r}
+\sqrt{g_{\th\th}}\left(g_{0\phi,r}-\f{g_{0\phi}}{g_{00}}
g_{00,r}\right)\hat{\th}\right],
\label{ltp0}
\end{equation}
which is only applicable outside the ergoregion. This is the LT 
precession frequency of a test gyro due to the rotation of 
any stationary and axisymmetric spacetime \cite{cmb, ccb}.

\subsection{Application to Kerr Spacetime}
We now apply the above formalism to the Kerr spacetime to describe
the behavior of a test gyro both inside and outside ergoregion. The Kerr metric in Boyer-Lindquist coordinates can be written as
\begin{equation} 
ds^2=-\left(1-\f{2Mr}{\rho^2}\right)dt^2-\f{4Mar \sin^2\theta}{\rho^2}d\phi dt 
+\f{\rho^2}{\Delta}dr^2 +\rho^2 d\theta^2+\left( r^2+a^2+\f{2Mra^2 \sin^2\theta}
{\rho^2}\right) \sin^2\theta d\phi^2 
\label{kerr}
\end{equation}
where $a$ is the specific angular momentum, defined as $a=J/M$ and,
\begin{equation}
 \rho^2=r^2+a^2 \cos^2\theta,        \,\,\,\,\,\,\,\,          \Delta=r^2-2Mr+a^2 .
\end{equation}
For convenience, we also define the dimensionless Kerr parameter $a_{*}=a/M=J/M^{2}$, 
which we shall use almost exclusively. The various metric components can be read
off from Eq.(\ref{kerr}) and we have,
\begin{equation}
\sqrt{-g}=\rho^2 \sin\theta.
\end{equation}
Substituting the metric components of Kerr spacetime into Eq.(\ref{ltp}), 
we obtain the spin precession frequency of a gyroscope to be,
\begin{eqnarray}
 \vec{\O}_p=\f{A~\sqrt{\d}\cos\th~ \hat{r}+B~\sin\th~\hat{\th}}
 {\r^3\left[(\r^2-2Mr)+4\O Mar\sin^2\th-\O^2\sin^2\th
 [\r^2(r^2+a^2)+2Ma^2r\sin^2\th]\right]},
 \label{genex}
\end{eqnarray}
where,
\begin{eqnarray}
 A&=&2aMr-\f{\O}{8}\left[8r^4+8a^2r^2+16a^2Mr+3a^4+4a^2(2\d-a^2)\cos2\th
 +a^4\cos4\th \right]+ 2\O^2a^3Mr\sin^4\th , \nonumber
 \\
 B&=&aM(r^2-a^2\cos^2\th)+\O \left[a^4r\cos^4\th
 +r^2(r^3-3Mr^2-a^2M(1+\sin^2\th)) \right. \nonumber
 \\
&& \left. +a^2\cos^2\th (2r^3-Mr^2+a^2M(1+\sin^2\th))\right]
 +\O^2 aM\sin^2\th[r^2(3r^2+a^2)+a^2\cos^2\th(r^2-a^2)].
 \label{AB}
\end{eqnarray}

\subsection{Range of $\O$}
Eq.(\ref{genex}) is valid both inside and outside the ergoregion.
From the requirement that $u$ be timelike,
\begin{eqnarray}
 K^2 = g_{\phi\phi}\O^2+2g_{t\phi}\O+g_{tt} & < & 0,
\end{eqnarray} 
we can calculate the restricted range of $\O$. Therefore, the allowed values of $\O$ 
at any fixed $(r,\theta)$ are,
\begin{eqnarray}
 \O_-(r,\theta) < \O(r,\theta) < \O_+(r,\theta)
 \label{oeg}
\end{eqnarray}
with,
\begin{eqnarray}
 \O_{\pm} =\f{-g_{t\phi}\pm \sqrt{g_{t\phi}^2-g_{\phi\phi}g_{tt}}}{g_{\phi\phi}}.
 \label{oek}
\end{eqnarray}
\\
Specifically, in the Kerr spacetime,
\begin{eqnarray}
 \O_{\pm}=\f{2Mar\sin\th \pm \r^2\sqrt{\d}}{\sin\th[\r^2(r^2+a^2)+2Ma^2r\sin^2\th]},
 \label{oe}
\end{eqnarray}
which shows that the range of allowed values for $\O$ becomes increasingly limited as 
the observer is located close to the horizon, that is $r \sim r_+$, and is eventually limited to the single value at the horizon of the BH,
\begin{eqnarray}
 \O_H=\f{a}{2Mr_+}.
\end{eqnarray}
Further, in the equatorial plane ($\th=\pi/2$), Eq.(\ref{oe}) becomes
\begin{eqnarray}
 \O_{\pm}|_{\th=\pi/2}=\f{2Ma\pm r\sqrt{\d}}{r(r^2+a^2)+2Ma^2}.
 \label{oeeq}
\end{eqnarray}
Since the ergoregion for NSs always extends all the way up to the ring singularity in
the equitorial plane, we can evaluate $\O_{\pm}|_{\th=\pi/2}$ at $r=0$ and, from
Eq. (\ref{oeeq}), it is evident that these two frequencies match at the singularity.
Therefore, we must point out that both at the horizon ($r_+$) and at the ring singularity
($r=0,\theta=\pi/2$), there exist no valid values for $\O$ (because of the strict 
inequality) implying that no time-like stationary observer can exist at these points.
Our precession frequency expression is not valid at these points but it is still meaningful and illuminating to study and plot its limiting values at these points.
\\

\begin{figure}[h!]
\begin{center}
\subfigure[$a_*=0.9$]{
\includegraphics[width=3in,angle=0]{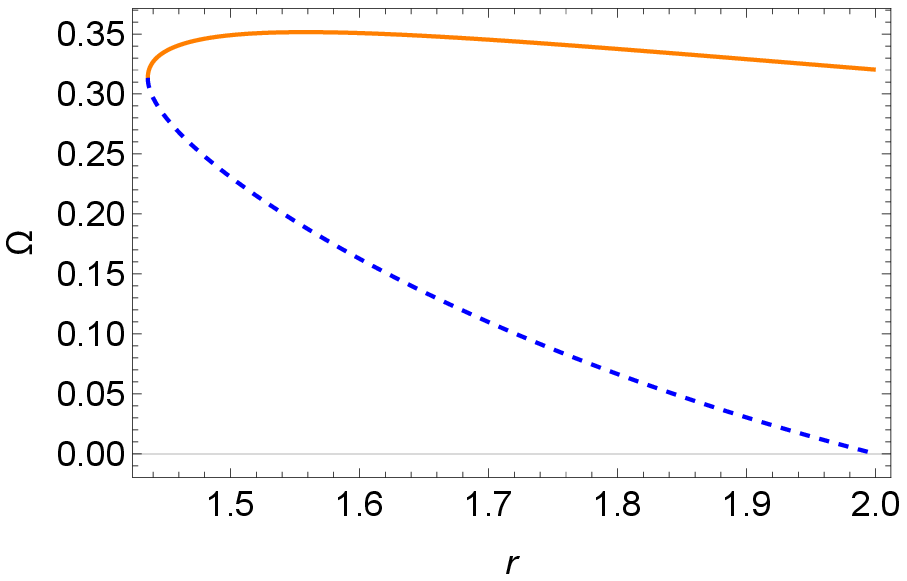}}
\subfigure[$a_*=2$]{
\includegraphics[width=3in,angle=0]{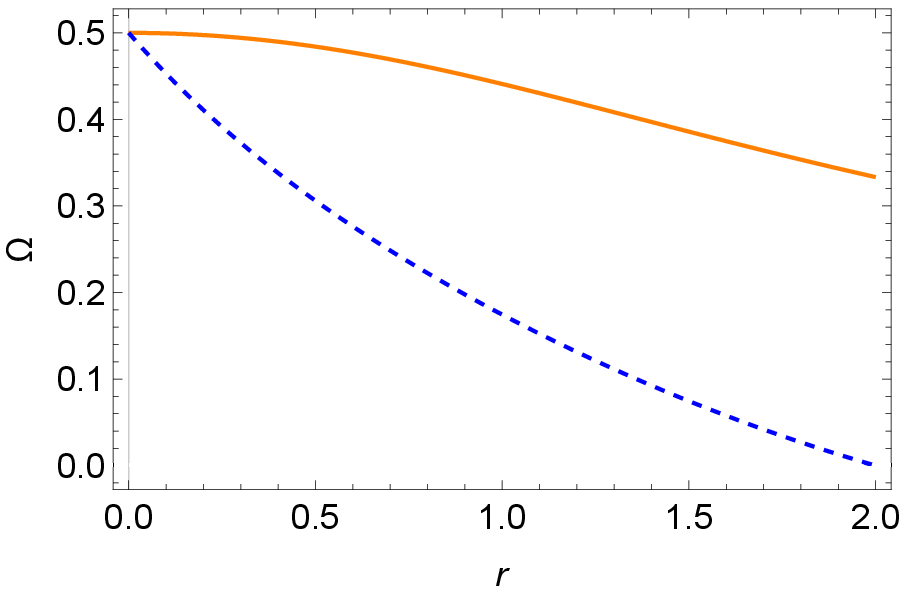}}
\subfigure[$a_*=1.001$]{
\includegraphics[width=3in,angle=0]{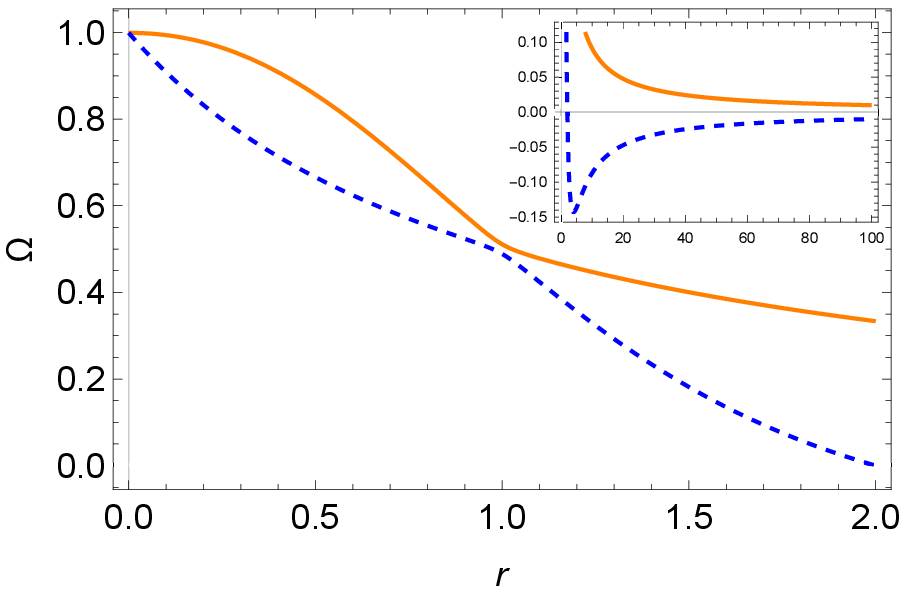}}
\caption{\label{Omega_pm} The frequency of a stationary gyroscope $\O$ can only
take values in the range: $(\O_-,\O_+)$ at any $(r,\theta)$. $\O_-$ and $\O_+$ (in $M^{-1}$) are in dashed blue and orange respectively and have been plotted specifically inside the ergoregion, in the equatorial plane ($\theta = \pi/2$), as a function of $r$ (in $M$) for different values of Kerr parameter $a_*$. The inset in panel $(c)$ shows $\O_{\pm}$ outside the ergoregion, for reference. This is the boundary of the region in which the most general Killing vector in Kerr spacetime, $K=\p_0 + \O\p_\phi$, is time-like. For black holes, we can see from panel $(a)$ that $\O_{\pm}$ meet at the horizon.
For NSs, we can see from panels $(b),(c)$ that $\O_{\pm}$ meet at the singularity. Further, it can be seen from $(c)$, that for 
near-extremal NSs ($a_*\sim 1$), $\O_{\pm}$ take close values near $r=1$, which was
the location of the horizon of the extremal BH ($a_*=1$).}
\end{center}
\end{figure}

We can see from panel (a) of Fig.\ref{Omega_pm} that $\O_+|_{\th=\pi/2}$ and 
$\O_-|_{\th=\pi/2}$ match at $r \rightarrow r_+$, for a BH with $a_*=0.9$, and the 
value of $\O$ becomes $\O_H \approx 0.31$. Panel (c) of the same figure shows that a 
small gap appears between the same two quantities at $r \approx M$ in case of the 
near-extremal NS with $a_*=1.001$ and the two curves match with one another at
$r \rightarrow 0$ with the value of $\O=1/a_*$. In the case of NS with higher angular
momentum, say, for $a_*=2$ (see panel (b)), the small gap disappears at $r=M$ and the two curves
match at $r \rightarrow 0$, as usual for NSs.\\

To compare the behavior of the gyro inside the ergoregion of a BH and  NS, one can now 
plot $\O_p=|\vec{\O}_p|$ with $r$ at different $\theta$ for observers with varying $\O$.
For this purpose, we introduce the parameter $q$ to scan the range of allowed values for
$\O$ as follows,
\begin{eqnarray}
 \O = q~\O_+ + (1-q)~\O_- &=&\o -(1-2q)\sqrt{\o^2-\f{g_{tt}}{g_{\phi\phi}}} 
=\f{2Mar\sin\th - (1-2q)~\r^2\sqrt{\d}}{\sin\th[\r^2(r^2+a^2)+2Ma^2r\sin^2\th]}
\label{q}
\end{eqnarray}
where $0 < q < 1$ and $\o=-g_{t\phi}/g_{\phi\phi}$. Clearly, the range of
$q$ covers the entire range of $\O$ from $\O_+$ to $\O_-$. Now, using the expression for
$\O$, from Eq.(\ref{q}), we can simplify the denominator of Eq.(\ref{genex}) and obtain
a nice and compact expression for the spin precession frequency,
\begin{eqnarray}
\vec{\O}_p=\f{(r^2+a^2)^2-a^2\d \sin^2\th}
 {4q(1-q)~\r^7 \d}\left[A~\sqrt{\d}\cos\th~ \hat{r}+B~\sin\th~\hat{\th}\right],
 \label{gnx}
\end{eqnarray}
where $0 < q <1$ and $A$ and $B$ have been given in Eq.(\ref{AB}).\\
\\
 If we use the expression for $\O$ from Eq.(\ref{q}) in Eq.(\ref{K}), we obtain
\begin{eqnarray}
 K=\f{q K_+ + (1-q) K_-}{|| q K_+ + (1-q) K_- ||}
 =\f{q K_+ + (1-q) K_-}{\sqrt{2q (1-q) K_+ . K_-}}
\end{eqnarray}
where $K_{\pm}$, the two null vectors associated with $\O_{\pm}$, are given as (see Eq.\ref{K})
\begin{eqnarray}
 K_{\pm}=\p_t + \O_{\pm} \p_{\phi}.
\end{eqnarray}
Further, in a general stationary and axisymmetric spacetime, we can write 
$K_+ . K_- = 2 (g_{tt}-\o^2 g_{\phi\phi})$.

\subsection{Acceleration of the test gyroscope}~
The gyroscopes we consider here have a specific velocity $u$ and, in general, experience 
non-zero acceleration. They follow helical paths tangent to the Killing trajectory and
the acceleration experienced by these gyroscopes is not arbitrary. This acceleration 
might be provided by using large amounts of rocket power or some other source of thrust
and we can calculate the necessary acceleration scalar, for our gyros, using the 
following expression of the 4-acceleration,
\begin{eqnarray}
 \a_{\b}=\f{1}{2}\nabla_{\b}{\rm ln}~|K^2|.
 \label{acc}
\end{eqnarray}
We find the acceleration scalar for $u$ to be
\begin{eqnarray}
 \a=\sqrt{g^{\b\g}\a_{\b}\a_{\g}}
 &=&-\f{1}{\r^5 K^2} \left[\d\left\{M(2r^2-\rho^2)\left[1 - 
a\O\sin^2{\theta}\right]^2 - r\rho^4 \O^2\sin^2{\theta} \right\}^2 \right.
\nonumber \\
&& \left.  + \sin^2\th \cos^2\th \left\{2Mr\left[a -
\O(r^2+a^2)\right]^2 + \O^2\rho^4\Delta\right\}^2 \right]^{\f{1}{2}},
\label{acck}
\end{eqnarray}
where $K^2$ is
\begin{eqnarray}
 K^2= g_{\phi\phi}\O^2+2g_{t\phi}\O+g_{tt}
 =\f{-4q(1-q)\r^2\d}{(r^2+a^2)^2-a^2\d \sin^2\th}.
\end{eqnarray}
Therefore, Eq.(\ref{acck}) reduces to 
\begin{eqnarray}
 \a
 &=&\f{(r^2+a^2)^2-a^2\d \sin^2\th}{4q(1-q)\r^7 \d} \left[\d\left\{M(2r^2-\rho^2)\left[1 - 
a\O\sin^2{\theta}\right]^2 - r\rho^4 \O^2\sin^2{\theta} \right\}^2 \right.
\nonumber \\
&& \left.  + \sin^2\th \cos^2\th \left\{2Mr\left[a -
\O(r^2+a^2)\right]^2 + \O^2\rho^4\Delta\right\}^2 \right]^{\f{1}{2}}.
\label{acce}
\end{eqnarray}
Eq.(\ref{acce}) represents the amount of acceleration that is needed to move the 
test gyro in the Kerr spacetime. It is also evident from the  above expression that the acceleration becomes arbitrarily high, blowing 
up as the event horizon is approached from any direction. Since NSs do not possess horizons, the acceleration of the gyro remains finite all along including at $r=0$ ($\theta \neq \pi/2$), which we can see from,
\begin{eqnarray}
 \a|_{r=0} =\f{1}{4q(1-q)a^2 \cos^3\th} \left[M^2(1 - a\O \sin^2 \th)^4 + 
 a^6 \O^4 \sin^2\th \cos^2\th \right]^{\f{1}{2}}.
\label{accr0}
\end{eqnarray}
However, the above expression diverges close to the ring singularity $(r=0, \th=\pi/2)$. The acceleration of the gyro (Eq.\ref{acck}) vanishes if it rotates in a geodesic with the Kepler frequency 
\begin{eqnarray}
\O=\O_{\phi} =\pm \f{M^{\f{1}{2}}}{r^{\f{3}{2}} \pm a M^{\f{1}{2}}}.
\end{eqnarray}

\subsection{Zero Angular Momentum Observer \label{zamo}}
We note that for $q=0.5$, $\O$ becomes the characteristic ZAMO frequency $\o$,
\begin{eqnarray}
\o =\f{2Mar}{(r^2+a^2)^2-a^2\d \sin^2\th}=-\f{g_{t\phi}}{g_{\phi\phi}}.
\label{ch}
\end{eqnarray}
In this case, test gyros attached to stationary observers regard both $+\phi$ and $-\phi$ directions equivalently, in terms of the local geometry, and see photons symmetrically \cite{mtw}. These gyros are non-rotating relative to the local spacetime geometry. The angular momentum of such a \lq\lq locally non-rotating\rq\rq\ observer is zero and is therefore called a zero angular momentum observer (ZAMO), first introduced
by Bardeen \cite{bd,mtw}. Bardeen et al.\cite{bpt} showed that the ZAMO frame 
is a powerful tool in the analysis of physical processes near astrophysical objects.
Here, we should note that Eq.(\ref{acc}) reduces to Eq.(33.23) of (see Exercise 33.4) 
\cite{mtw} in the case of a ZAMO in a general stationary and axisymmetric spacetime.

\section{Useful limits \label{spe}} 
In this section, we will explore various useful limits of the spin precession frequency 
(Eq.\ref{gnx}). We discuss what happens to the frequency as $r\rightarrow 0$ 
(specifically, the ring singularity is at $r=0$, $\theta = \pi/2$), how it looks like 
in the equatorial plane $\theta=\pi/2$ and whether this frequency vanishes in the 
Schwarzschild spacetime.

\subsection{Behavior of $\vec{\O}_p$ at $r= 0$.} 
As discussed earlier, our expression for $\vec{\O}_p$ is not valid at the ring singularity ($r=0,  \theta=\pi/2$). However, we can still study its behavior in its vicinity, that is, in the region $r=0$, $0 \leq \th < 90^0$. We note here that this region is strictly completely outside the ergoregion since the ergosurface meets the ring singularity
(see Figs.1 and 2 of Ref.\cite{ckj}). At $r=0$,
Eq.(\ref{genex}) becomes
\begin{eqnarray}
 \vec{\O}_p|_{r= 0}=\f{-a^2\O[3+4\cos2\th+\cos4\th]\ \hat{r}
 -4M\sin2\th[1-a\O(1+\sin^2\th)+a^2\O^2\sin^2\th]\ \hat{\th}}{8a^2\cos^4\th[1-a^2\O^2\sin^2\th]},
 \label{genexr0}
\end{eqnarray}
where we have used, from Eq.(\ref{AB}),
\begin{eqnarray}
 A|_{r= 0}&=&-\f{a^4\O}{8}\left[3+4\cos2\th+\cos4\th \right], \nonumber
 \\
 B|_{r= 0}&=&-Ma^3\cos^2\th\left[1-a\O(1+\sin^2\th)+a^2\O^2\sin^2\th \right].
\end{eqnarray}
In the above expressions, the allowed range of $\O$ is 
\begin{eqnarray}
 -\f{1}{a\sin\th} < \O < \f{1}{a\sin\th}.
\end{eqnarray}
Since we are outside the ergoregion, we can consider static observers, that is we set
$\O=0$ and Eq.(\ref{genexr0}) reduces to
\begin{eqnarray}
 |\vec{\O}_p|=\f{M}{a^2}\tan\th\sec^2\th,
\end{eqnarray}
and matches with our earlier calculations of Ref.\cite{ckj} and can be seen from Eq.(6) and (7) therewith. Therefore, $\O_p$ varies from $0 \leq \O_p < \infty$ for $0 \leq \th < 90^0$ at $r=0$. That is, it diverges only on the ring singularity (which is at $x^2+y^2=a^2,\ z=0$ in Cartesian Kerr-Schild coordinates) but is finite \textit{inside} it ($x^2+y^2<a^2,\ z=0$).

It is useful to mention here that one can smoothly, in principle, go over to the region with
\lq negative $r$\rq\ (i.e., $r < 0)$ in Kerr spacetime, which is tantamount
to passing through the ring singularity. 
but we stop at $r=0$ and avoid probing negative values of $r$. The reason behind stopping at $r=0$ is that it is fairly widely accepted that quantum gravity will  resolve the singularity resulting in a compact overspinning object with boundary at a positive value of $r$, which is referred to as \lq superspinar\rq\ \cite{horava}. Thus the region with 
negative values of $r$ will be excised and pathological features such as closed timelike curves which occur in $r < 0$ region will not arise. Thus we restrict our probe of the Kerr spacetime to $r\geq 0$.

\subsection{Behavior of $\vec{\O}_p$ in the equatorial plane, $\theta=\pi/2$}
The precession frequency in the equatorial plane is,
\begin{eqnarray} 
 \vec{\O}_p|_{\th=\pi/2}&=&\f{aM+\O(r^3-3Mr^2-2Ma^2)+aM\O^2(3r^2+a^2)}
 {r^2\left[(r-2M)+4\O Ma-\O^2[r(r^2+a^2)+2Ma^2]\right]},
 \label{eq}
 \end{eqnarray}
with the range $\O$ determined from Eq.(\ref{oeeq}). Specifically, at the ergosurface $(r=2M)$ the precession frequency becomes
\begin{eqnarray}
\vec{\O}_p|_{\th=\pi/2, r=2M}=\f{a-2\O(a^2+2M^2)+a\O^2(a^2+12M^2)}
  {16\O M^2[a-\O(a^2+2M^2)]}
\end{eqnarray}\label{eqr2m}
with $\O$ being restricted to (see FIG.\ref{Omega_pm} also)
\begin{eqnarray}
 0 < \O < \f{a}{a^2+2M^2}.
\end{eqnarray}
That is, if the mass or angular momentum of the central object increases,
the allowed range of $\O$ at the ergosurface, in the equitorial plane, decreases.\\

For extremal BHs $(a_*=1)$, the precession frequencies at the outer ergoregion and at the outer event horizon respectively can be obtained as 
\begin{eqnarray}
\vec{\O}_p|_{\th=\pi/2, r=2M, a_*=1}=\f{1-6\O M+13\O^2M^2}{16\O M^2[1-3\O M)]}\ ;\ \vec{\O}_p|_{\th=\pi/2, r=M, a_*=1}=-\f{1}{M}.
\end{eqnarray}

\subsection{Non-zero $\vec{\O}_p$ in the Schwarzschild spacetime}
Now, if we set $a=0$, the Kerr spacetime reduces to the Schwarzschild spacetime, which is {\it non-rotating}. From Eq.(\ref{genex}), we obtain 
\begin{eqnarray}
 \vec{\O}_p|_{a=0}=\O ~ \f{(r-3M)\sin \th ~\hat{\th}
 -(r^2-2Mr)^{\f{1}{2}}\cos\th ~\hat{r}}{r-2M-r^3\O^2\sin^2\th}
 \label{sc}
\end{eqnarray}
where $\O$ can take any value such that $u$ is timelike. Since the Schwarzschild spacetime is spherically symmetric, we can write Eq.(\ref{sc}) for $\th=\pi/2$ as 
\begin{eqnarray}
\O_p|_{a=0}=\O ~ \f{r-3M}{r-2M-r^3\O^2}.
 \label{sce}
\end{eqnarray}
This means that a gyroscope moving in the Schwarzschild spacetime, which is a static
spacetime, will precess. Now, if the gyro moves along a circular geodesic $\O$ should be
the Kepler frequency, i.e., $\O_{Kep}=(M/r^3)^{1/2}$ and Eq.(\ref{sce})
reduces to 
\begin{eqnarray}
 \O_p \Bigr|_{a=0,\ \O=\O_{Kep}}=\O= \left(\f{M}{r^{3}}\right)^{\f{1}{2}} .
\end{eqnarray}
The above expression gives the precession frequency in the Copernican 
frame, computed with respect to the proper time $\tau$. The proper time $\tau$, 
measured in the Copernican frame, is related to the coordinate time $t$ via 
$d\tau=\sqrt{1-\frac{3M}{r}}dt$ and we can obtain the precession frequency in the 
coordinate basis $\O^{'}$ as,
\begin{equation}
 \O^{'}=\left(\f{M}{r^3}\right)^{\f{1}{2}}\sqrt{1-\frac{3M}{r}}.
\end{equation}
We can now find the frequency associated with the change in the angle of the spin vector
over, say, one complete revolution around the central object. This is just the difference
of $\O^{\prime}$ and $\O$ \cite{jh}, and we get
\begin{equation}
 \O_{\rm geodetic}=\left(\f{M}{r^3}\right)^{\f{1}{2}} \left(1- \sqrt{1-\frac{3M}{r}}\right),
\end{equation}
where we have identified above that this precession is just due to geodetic 
precession ($\O_{\rm geodetic}$). This agrees with standard results \cite{chiba}.

\section{Distinguishing Kerr naked singularities from Kerr black holes using the precession of a test gyro \label{5}}

In this section, we point out the characteristic differences in the behavior of the modulus of the spin precession frequency ($\O_p$) of stationary gyroscopes for BHs and NSs. We show that the value of $\O_p$ becomes arbitrarily large for such gyroscopes, located arbitrarily close to the horizon of a BH, that is $r \sim r_+$, for all values of $q$ except $q=0.5$, the ZAMO frequency. However, for a NS, $\O_p$ always remains finite upto $r=0$, except for $r=0$ and $\theta \sim \pi/2$, i.e., near the singularity. 

We obtain distinguishing characteristic features specifically in the radial profile of
$\O_p$ for both BH and NS cases, which we will discuss as we proceed. Further, we also 
obtain features in the radial profile of $\O_p$ that could help
distinguish near-extremal NSs from those with higher spin.
We explore the details of such features and provide a criterion to separate near-extremal NS 
($1 < a_* < 1.1$) from those with higher spins ($a_* \geq 1.1$) in Sec.\ref{nens}.

\begin{figure}[!h]
\begin{center}
\subfigure[~BH with $a_*=0.9$, $q=0.1~(\O < \o)$]{
\includegraphics[width=3in,angle=0]{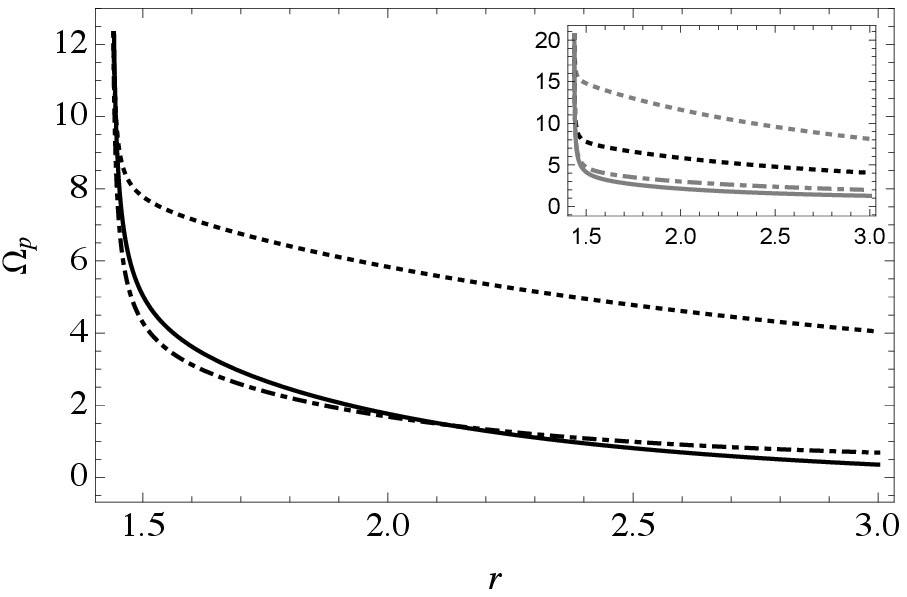}}
\subfigure[~NS with $a_*=1.1$, $q=0.1~(\O < \o)$]{
\includegraphics[width=3in,angle=0]{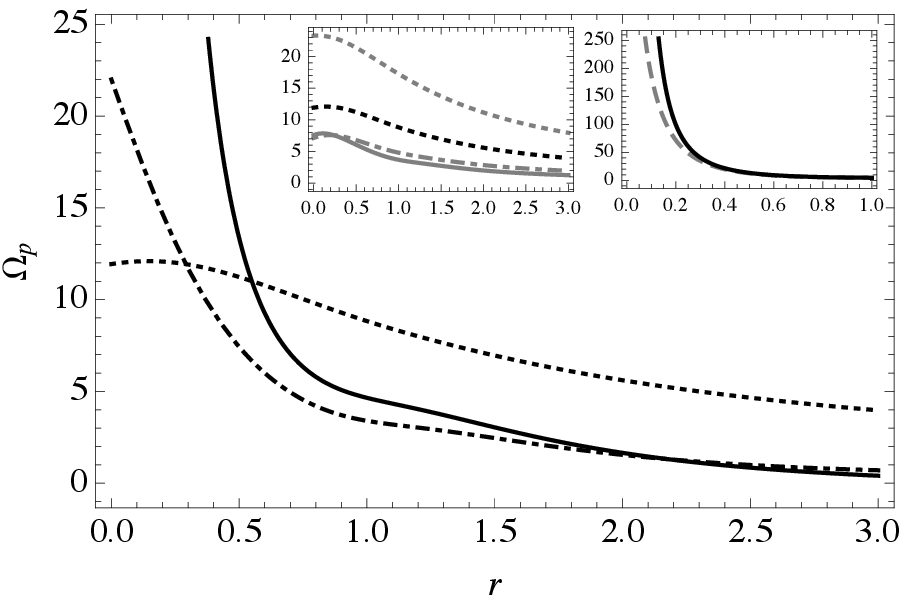}} 
\subfigure[~BH with $a_*=0.9$, $q=0.5~(\O = \o)$]{
\includegraphics[width=3in,angle=0]{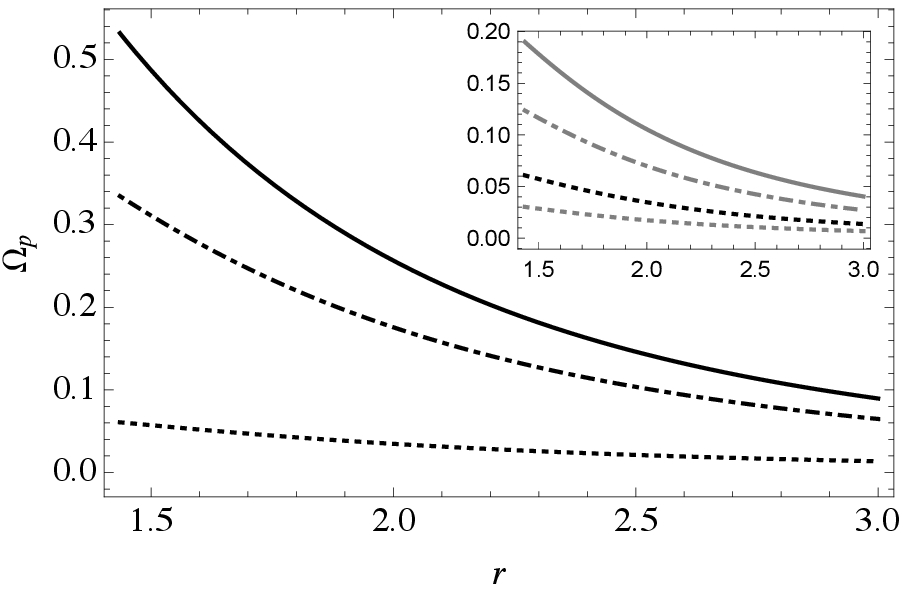}} 
\subfigure[~NS with $a_*=1.1$, $q=0.5~(\O = \o)$]{
\includegraphics[width=3in,angle=0]{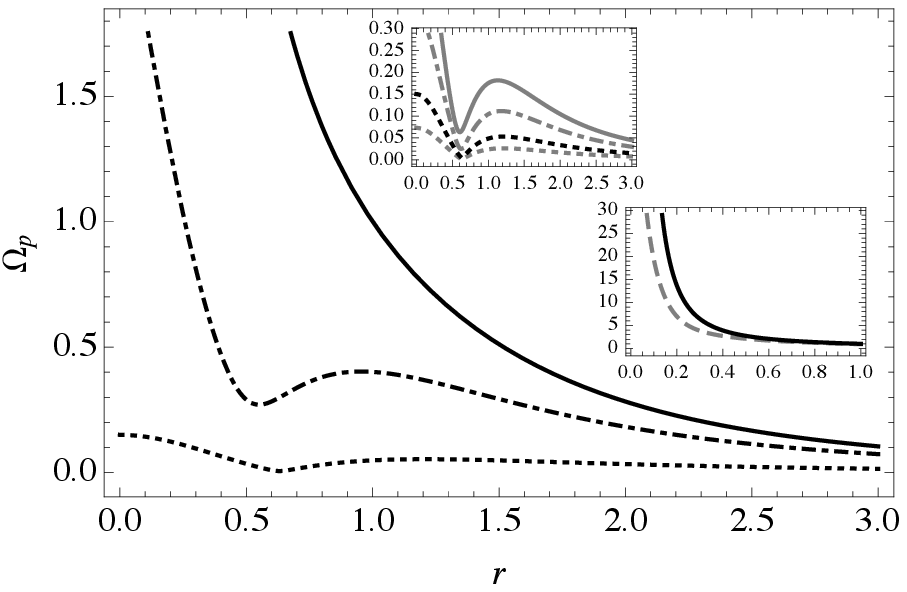}} 
\subfigure[~BH with $a_*=0.9$, $q=0.9~(\O > \o)$]{
\includegraphics[width=3in,angle=0]{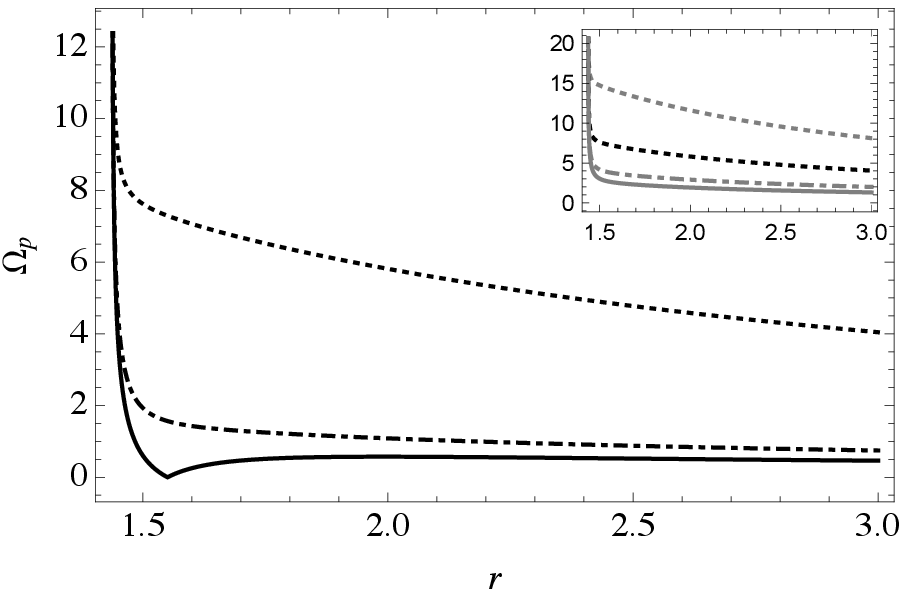}} 
\subfigure[~NS with $a_*=1.1$, $q=0.9~(\O > \o)$]{
\includegraphics[width=3in,angle=0]{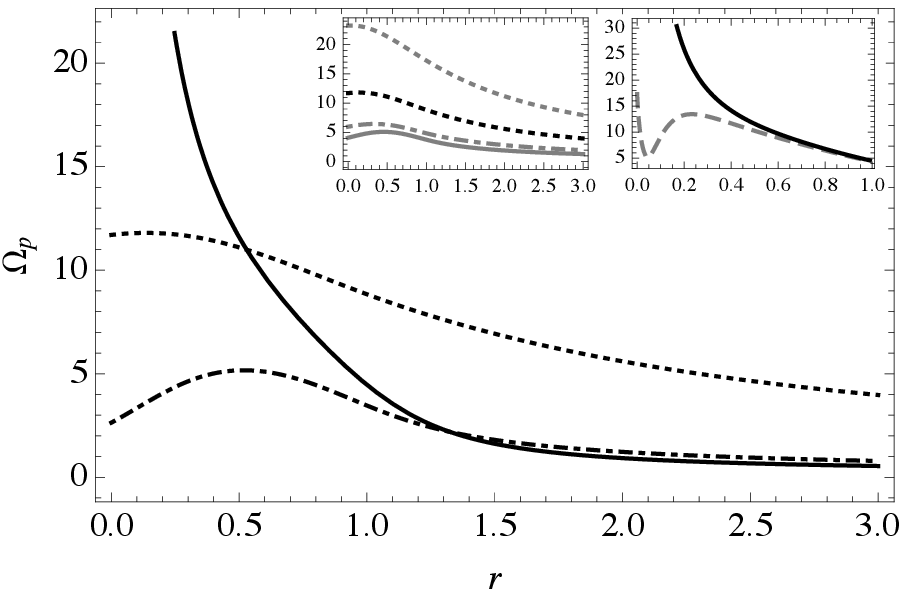}}
\caption{\label{BHvNS} We have plotted in each panel the modulus of the precession frequency of stationary gyroscopes $\Omega_p$ (in $M^{-1}$) versus $r$ (in $M$) around a black hole (left panels) with $a_*=.9$ and a naked singularity (right panels) with $a_* = 1.1$ for different $q$ and $\theta$. Each of the rows has a different value of $q$,  which measures its angular velocity, and in each panel, $\theta$ takes values $10^0, 50^0, 90^0$ represented in dotted black, dot-dashed black and black respectively. For the BH, $r$ ranges from the horizon radius (which is at $\sim 1.44$, in this case) to $3$. For the NS, the plots begin from $r=0$ (specifically, the singularity is at $r=0$ \textit{and} $\theta=90^0$) to $3$ and the ergosurface is at $2$ for $\th = 90^0$. 
It can be seen that there is a much bigger drop in $\O_p$ from $10^0$ to $50^0$ than from $50^0$ to $90^0$. We have therefore inset plots (left inset for NS panels) for additional $\theta$ values (close to the pole) of $5^0, 20^0$ and $30^0$ in dotted gray, dot-dashed gray and gray along with $10^0$ in dotted black, same as the main panel. Further, for the NS case, since the singularity is at $\theta=90^0$ in these coordinates, as $r\rightarrow 0, \theta \rightarrow 90^0$, the frequency blows up. We have zoomed in on the range of $r$ between $0$ and $1$ and inset (on the right in the NS panels) the plots for $\theta = 80^0, 90^0$ in dashed gray and black 
to demonstrate how quickly $\O_p$ increases relative to angles much smaller than $90^0$.}
\end{center}
\end{figure}

\begin{figure}[!h]
\begin{center}
\subfigure[~$q=0.1, \theta=10^0$]{
\includegraphics[width=2in,angle=0]{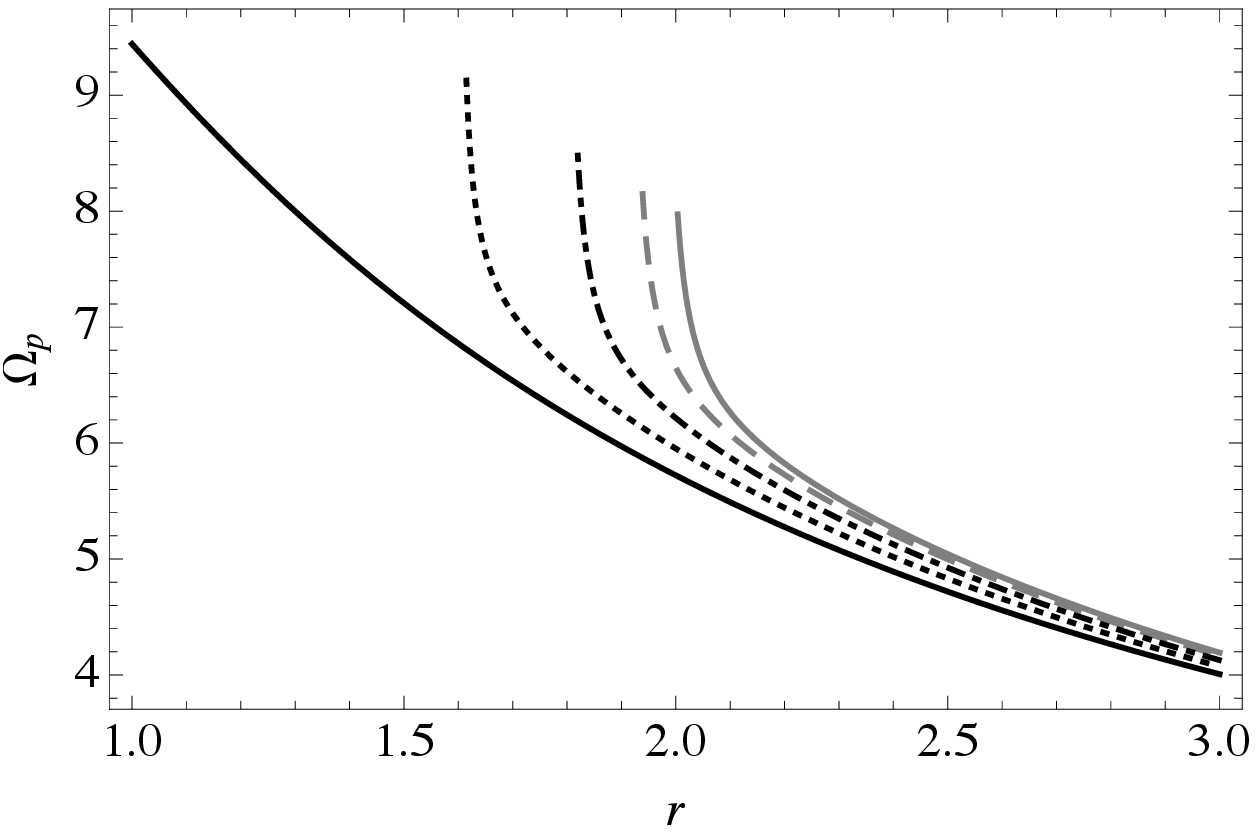}}
\subfigure[~$q=0.1, \theta=50^0$]{
\includegraphics[width=2in,angle=0]{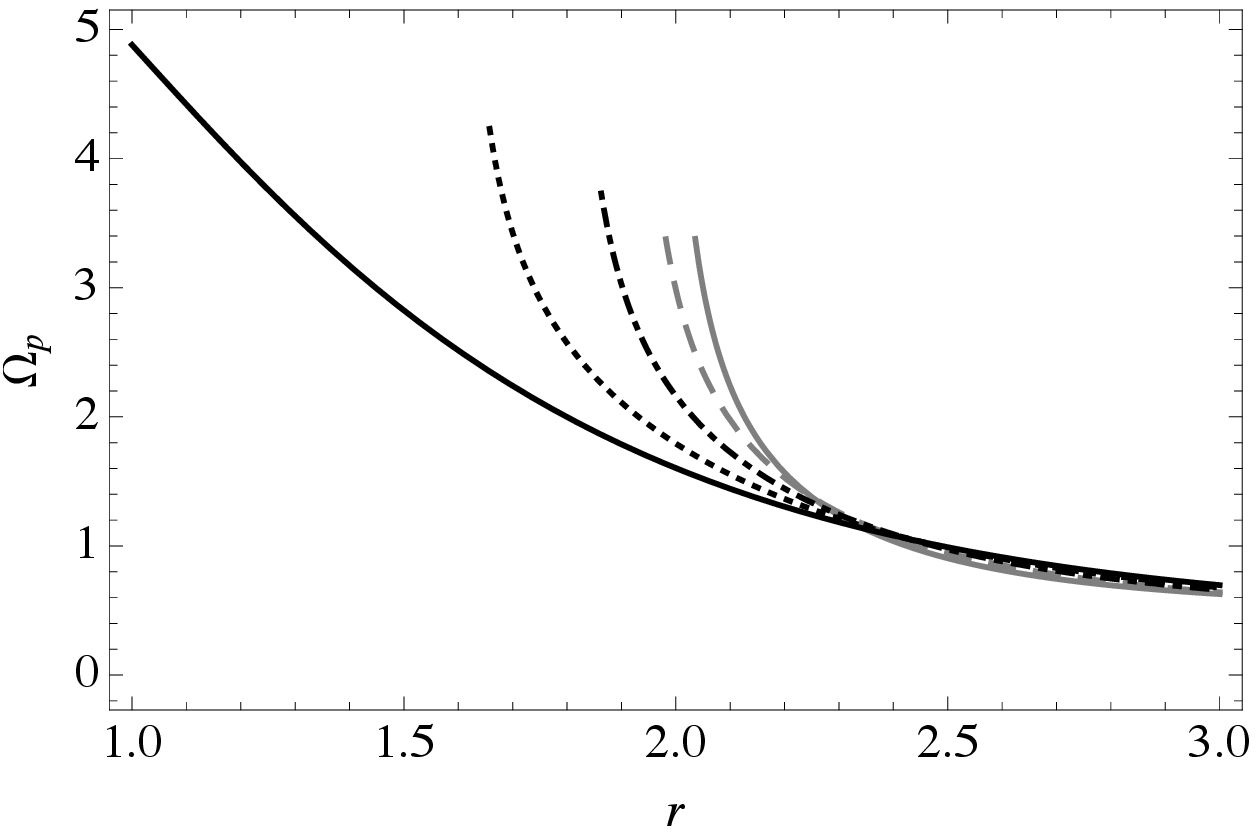}}
\subfigure[~$q=0.1, \theta=90^0$]{
\includegraphics[width=2in,angle=0]{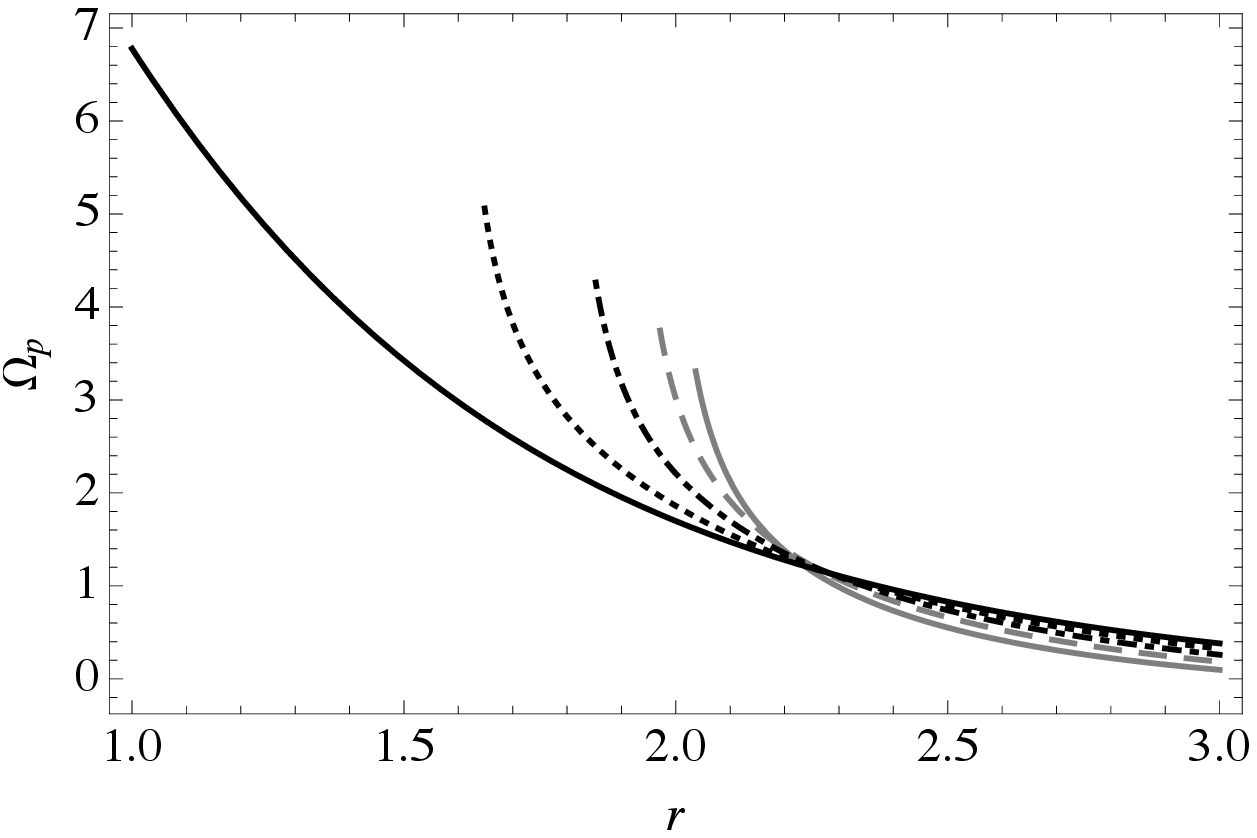}}
\subfigure[~$q=0.5, \theta=10^0$]{
\includegraphics[width=2in,angle=0]{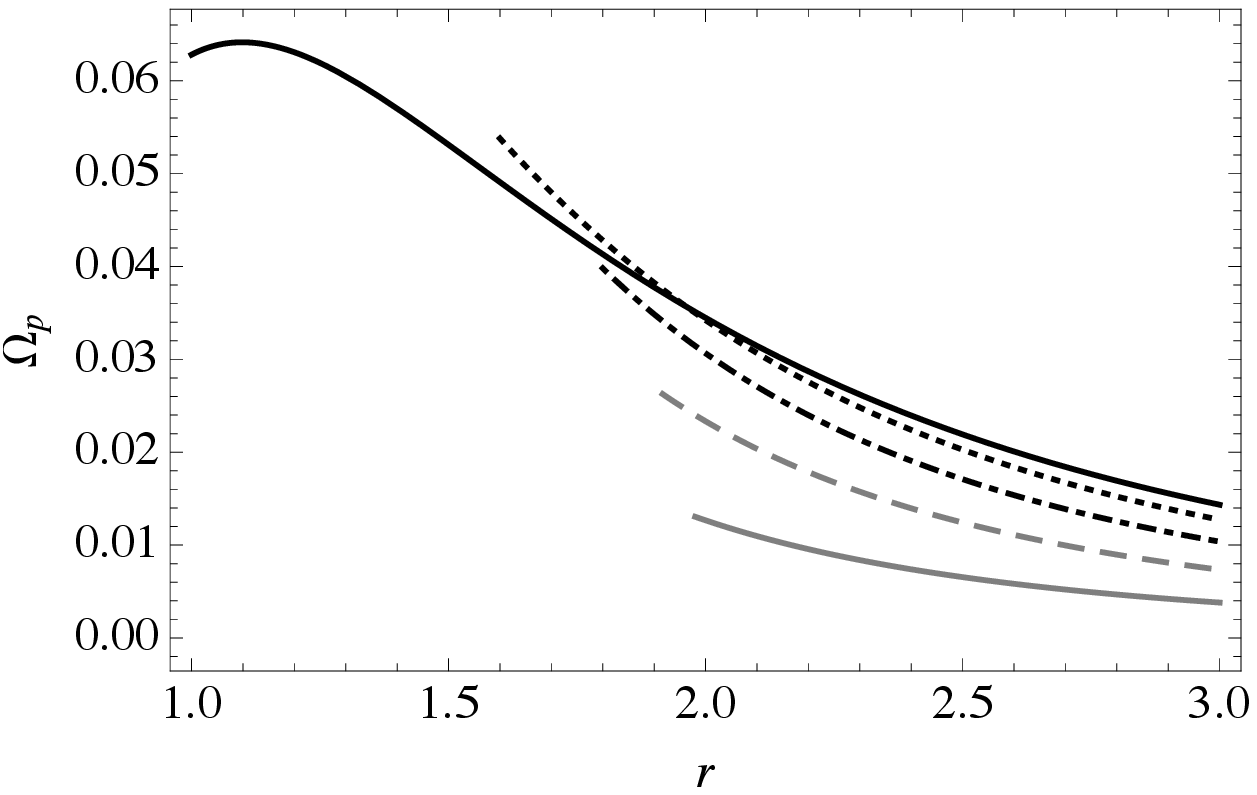}}
\subfigure[~$q=0.5, \theta=50^0$]{
\includegraphics[width=2in,angle=0]{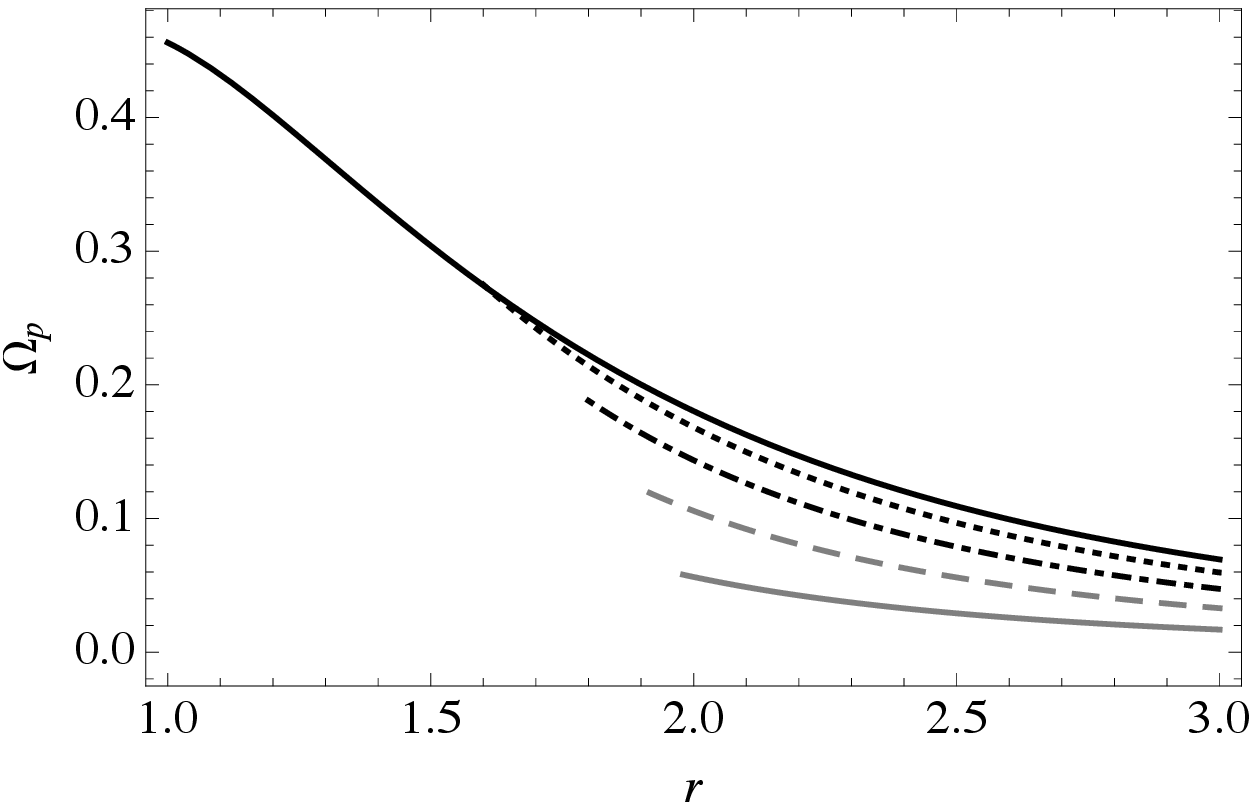}}
\subfigure[~$q=0.5, \theta=90^0$]{
\includegraphics[width=2in,angle=0]{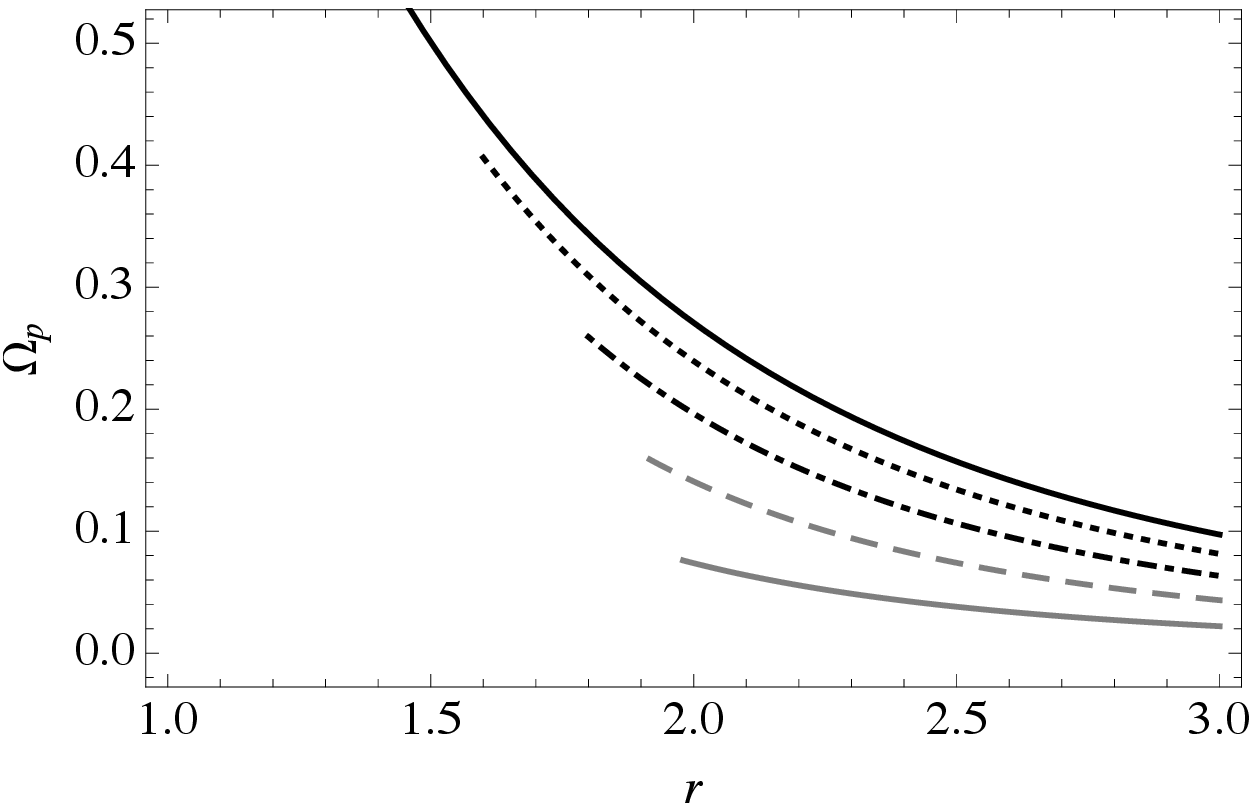}}
\subfigure[~$q=0.9, \theta=10^0$]{
\includegraphics[width=2in,angle=0]{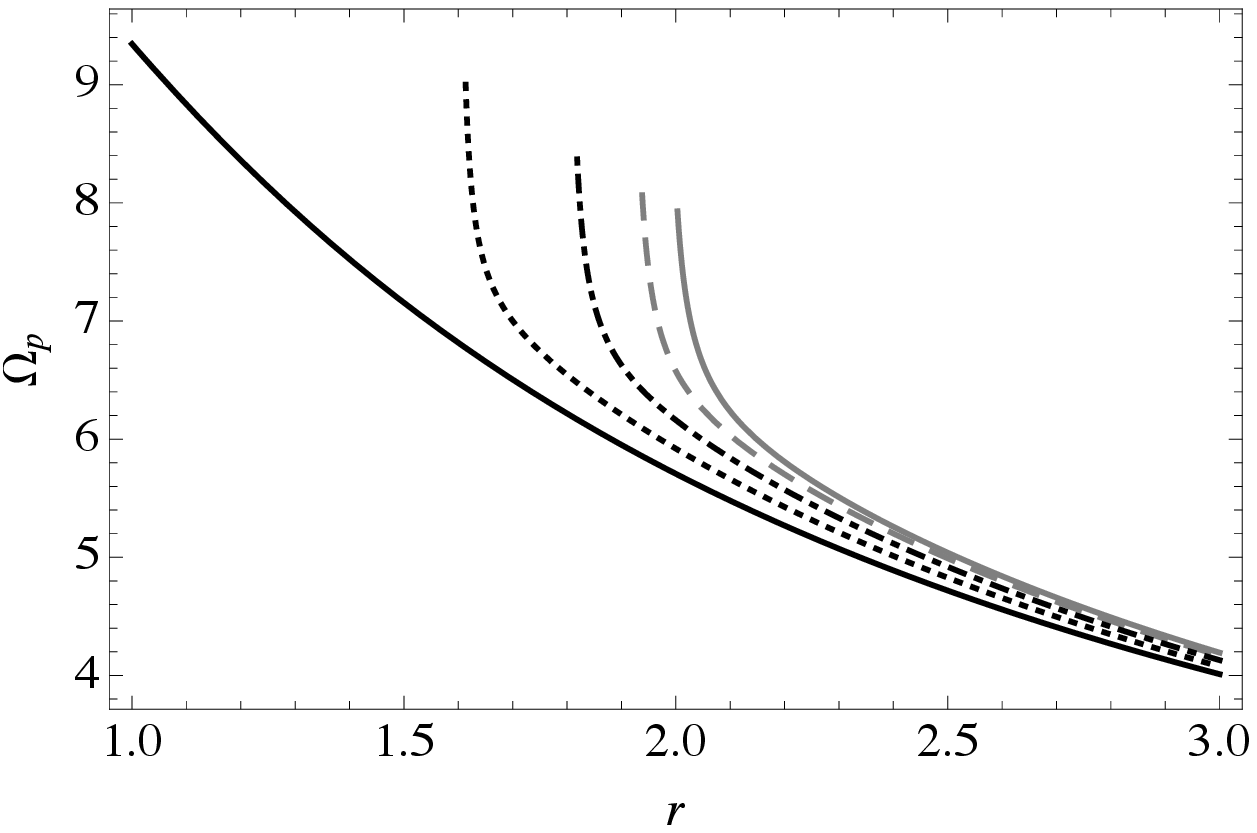}}
\subfigure[~$q=0.9, \theta=50^0$]{
\includegraphics[width=2in,angle=0]{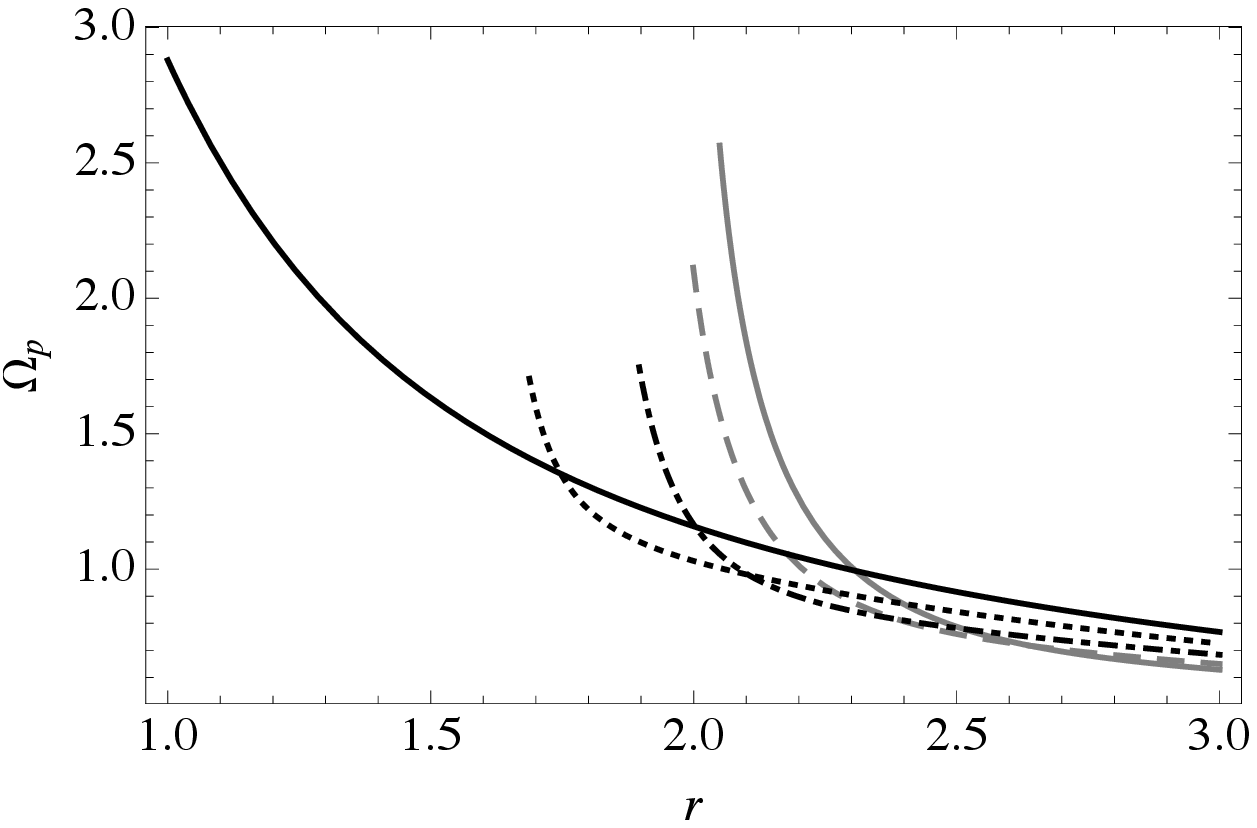}}
\subfigure[~$q=0.9, \theta=90^0$]{
\includegraphics[width=2in,angle=0]{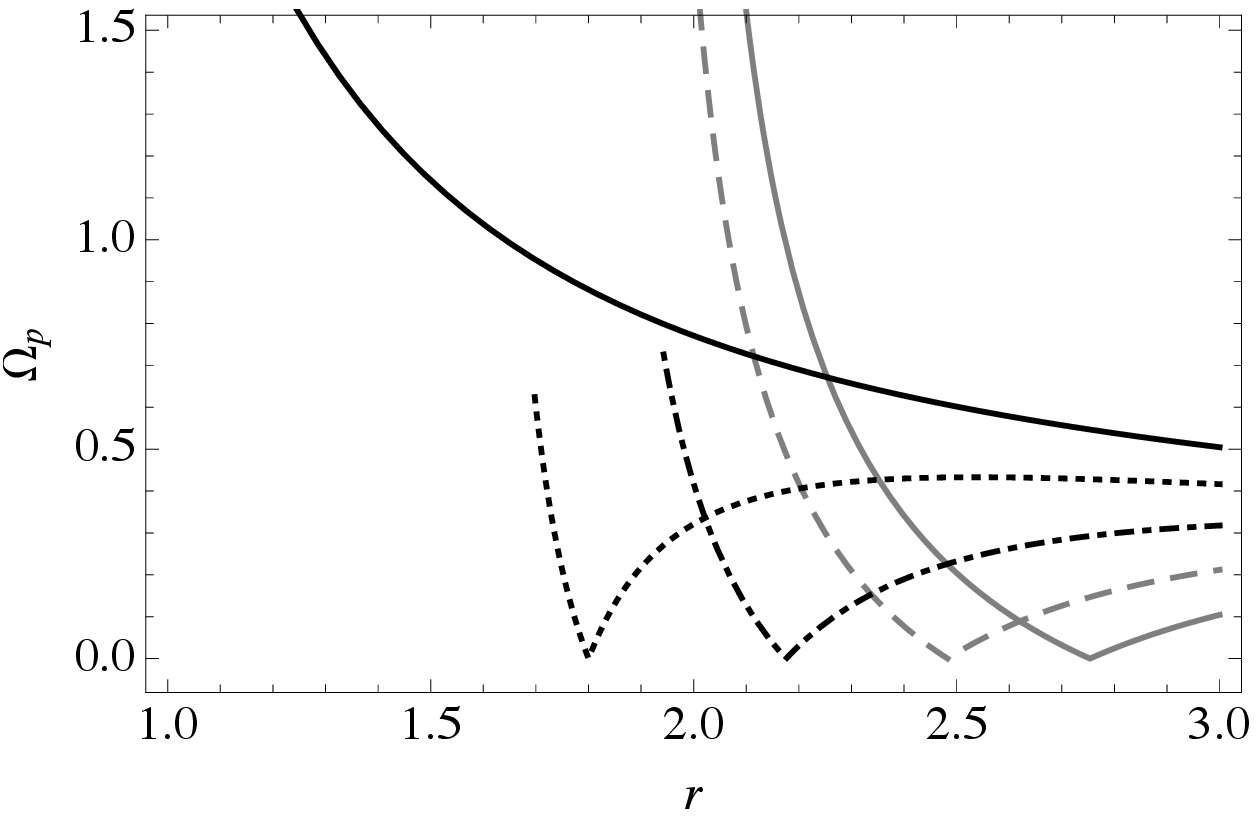}}
\caption{\label{BHGen} We have plotted here, for black holes, the modulus of the 
precession frequency $\O_p$ (in $M^{-1}$) vs $r$ (in $M$) for different $a_*, q$ and
$\theta$. We have $q=0.1, 0.5, 0.9$ in the top, middle and bottom rows respectively
and $\theta = 10^0, 50^0, 90^0$ in the left, centre and right columns respectively. 
In each panel, the line style is gray, dashed gray, dot-dashed black, dotted black 
and black for $a_* = .2, .4, .6, .8, 1$ respectively. We have plotted $\O_p$ for each
BH (with different $a_*$) between its horizon radius ($r_+$) and $r=3$. The ergoregion
is at $r=2$ for $\theta = 90^0$ (bottom row), for reference. This figure clearly 
demonstrates that for all values of $a_*,  q, \theta$, the precession frequency 
$\O_p$ becomes arbitrarily large near the event horizon, in general. As can be seen 
from the bottom row, for $q > 0.5$, minimas appear. Specifically, from panel (i), it 
can be seen that the sharpness of the minimas increases with $a_*$, with extremal black
holes as exceptions.}
\end{center}
\end{figure}

\begin{figure}[!h]
\begin{center}
\subfigure[~$q=0.1, \theta=10^0$]{
\includegraphics[width=2in,angle=0]{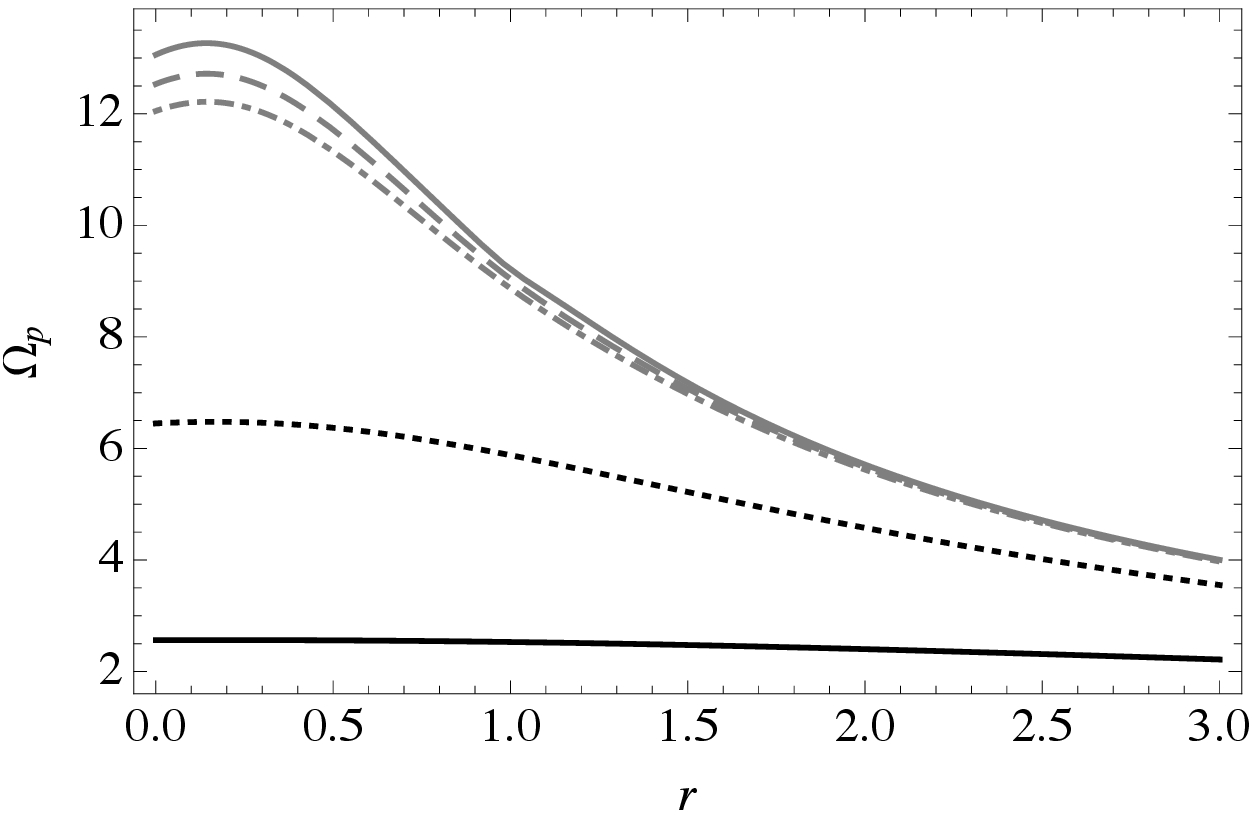}}
\subfigure[~$q=0.1, \theta=50^0$]{
\includegraphics[width=2in,angle=0]{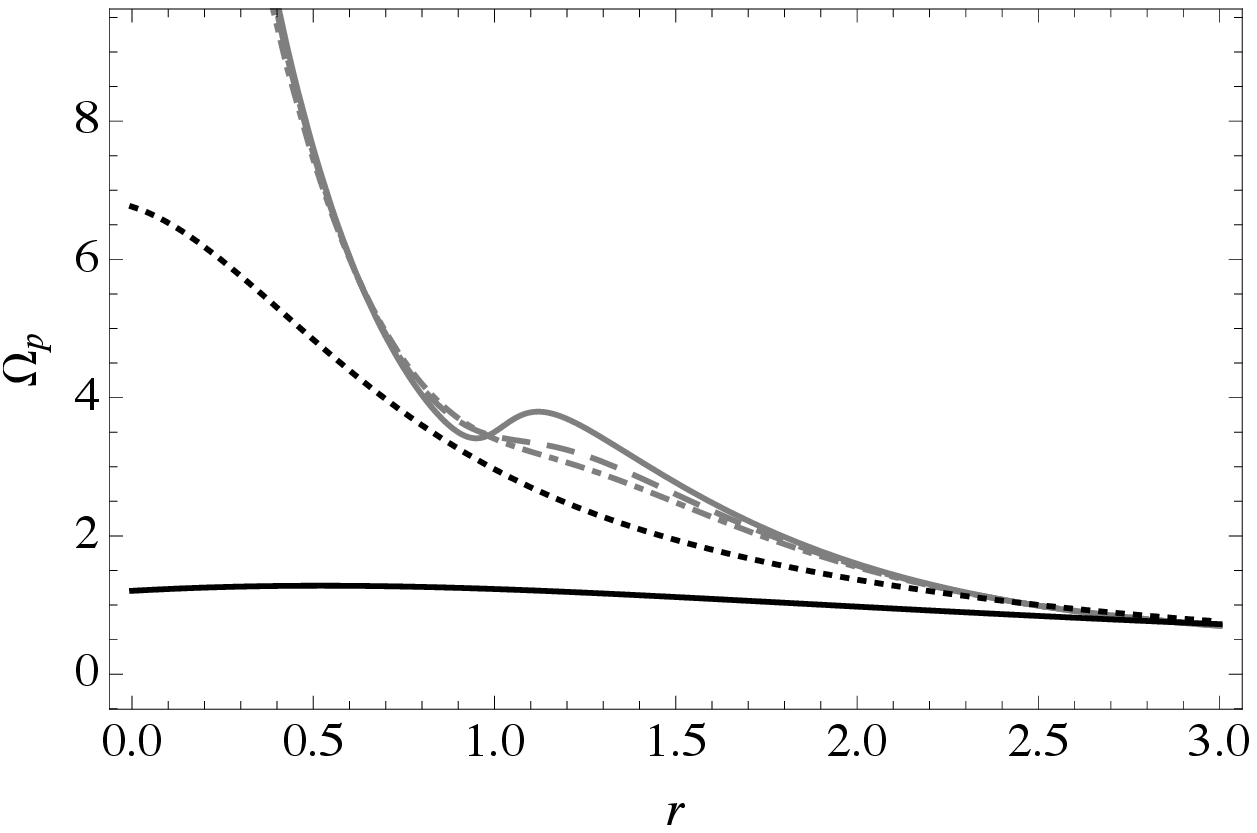}}
\subfigure[~$q=0.1, \theta=90^0$]{
\includegraphics[width=2in,angle=0]{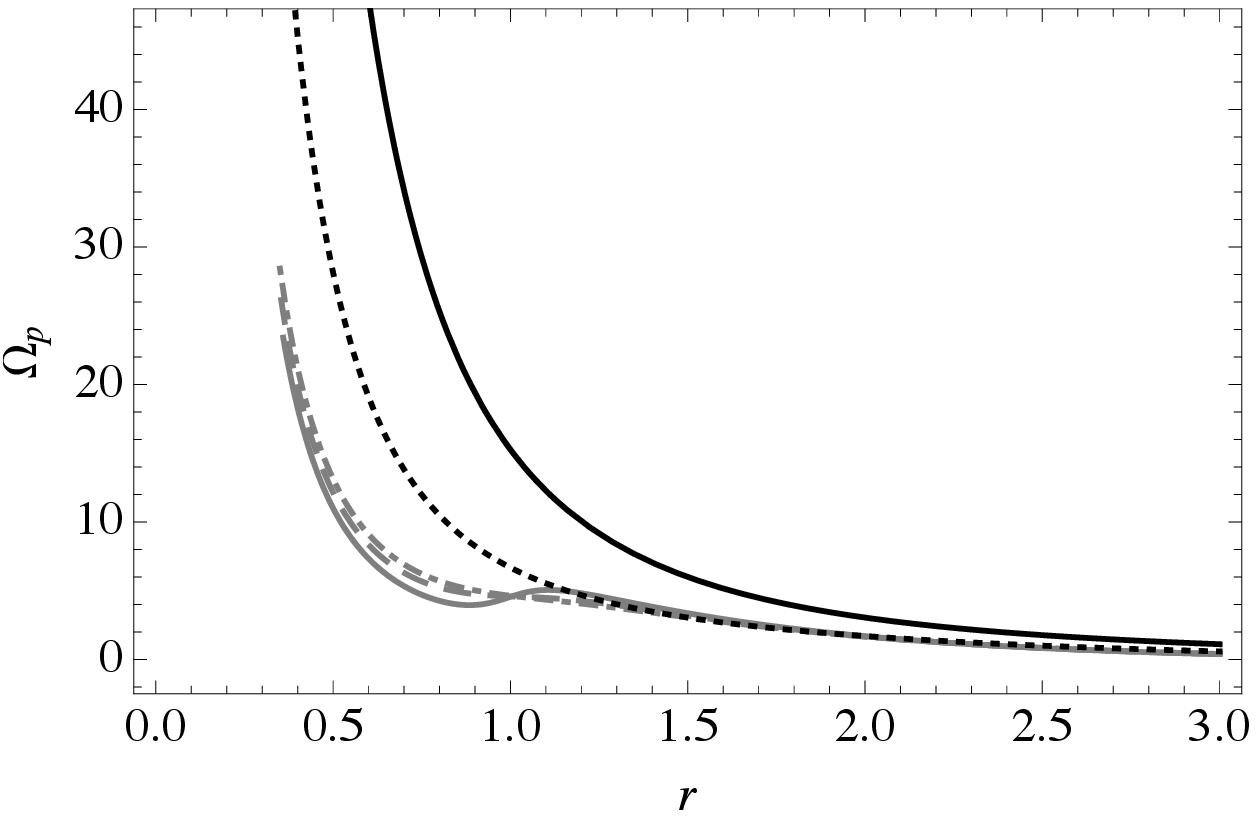}}
\subfigure[~$q=0.5, \theta=10^0$]{
\includegraphics[width=2in,angle=0]{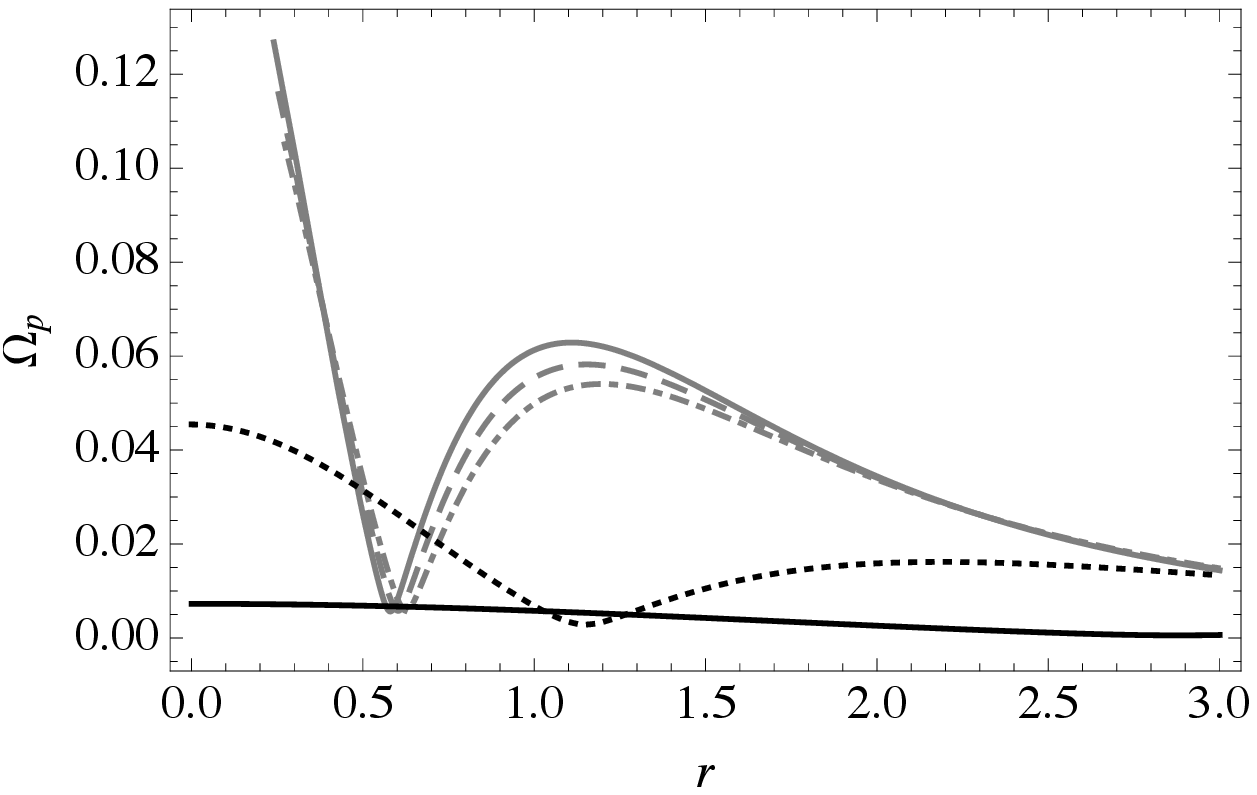}}
\subfigure[~$q=0.5, \theta=50^0$]{
\includegraphics[width=2in,angle=0]{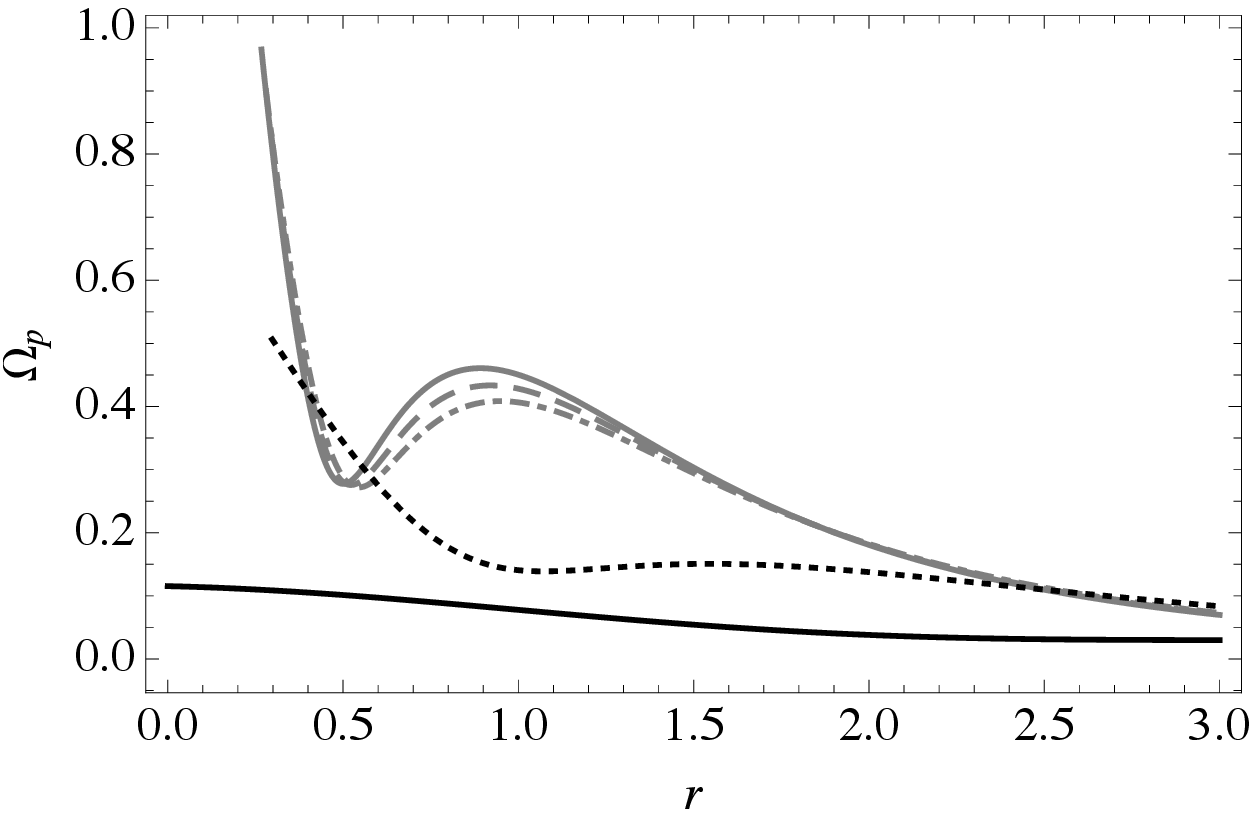}}
\subfigure[~$q=0.5, \theta=90^0$]{
\includegraphics[width=2in,angle=0]{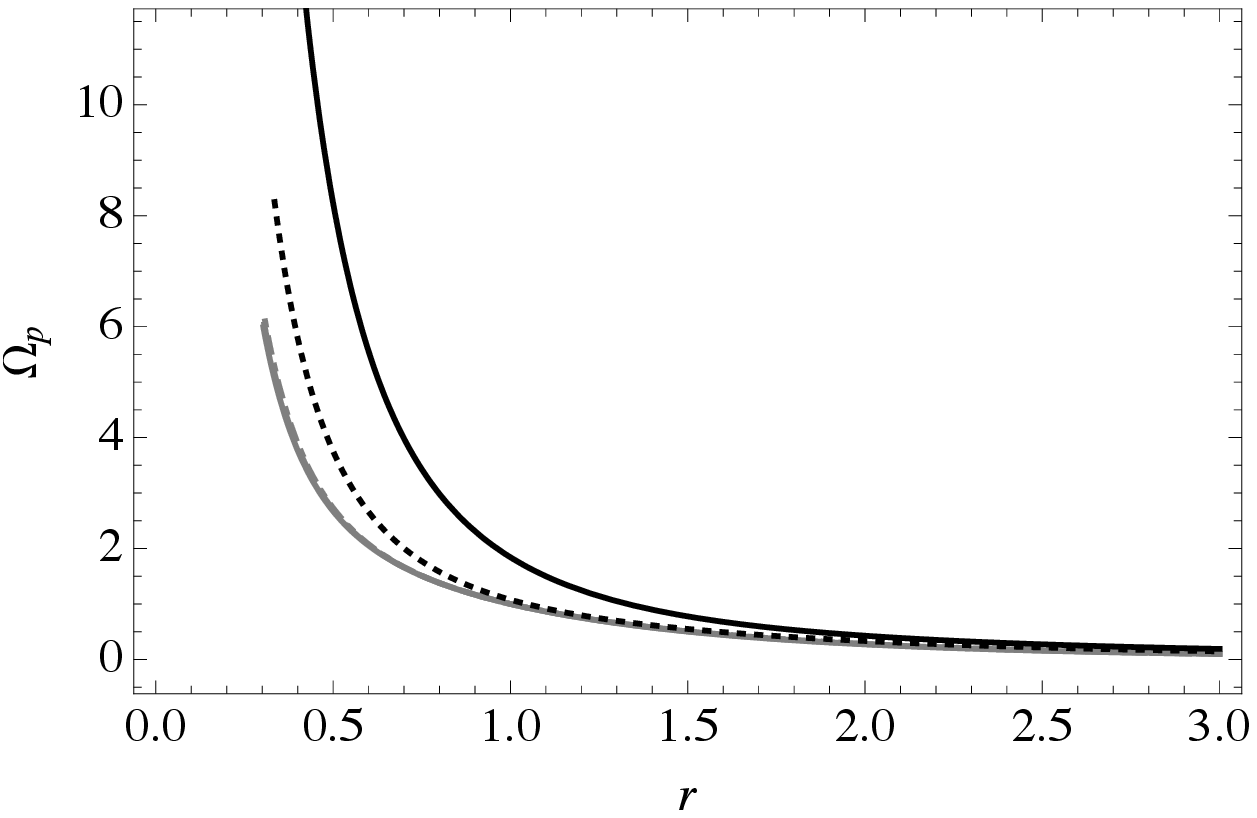}}
\subfigure[~$q=0.9, \theta=10^0$]{
\includegraphics[width=2in,angle=0]{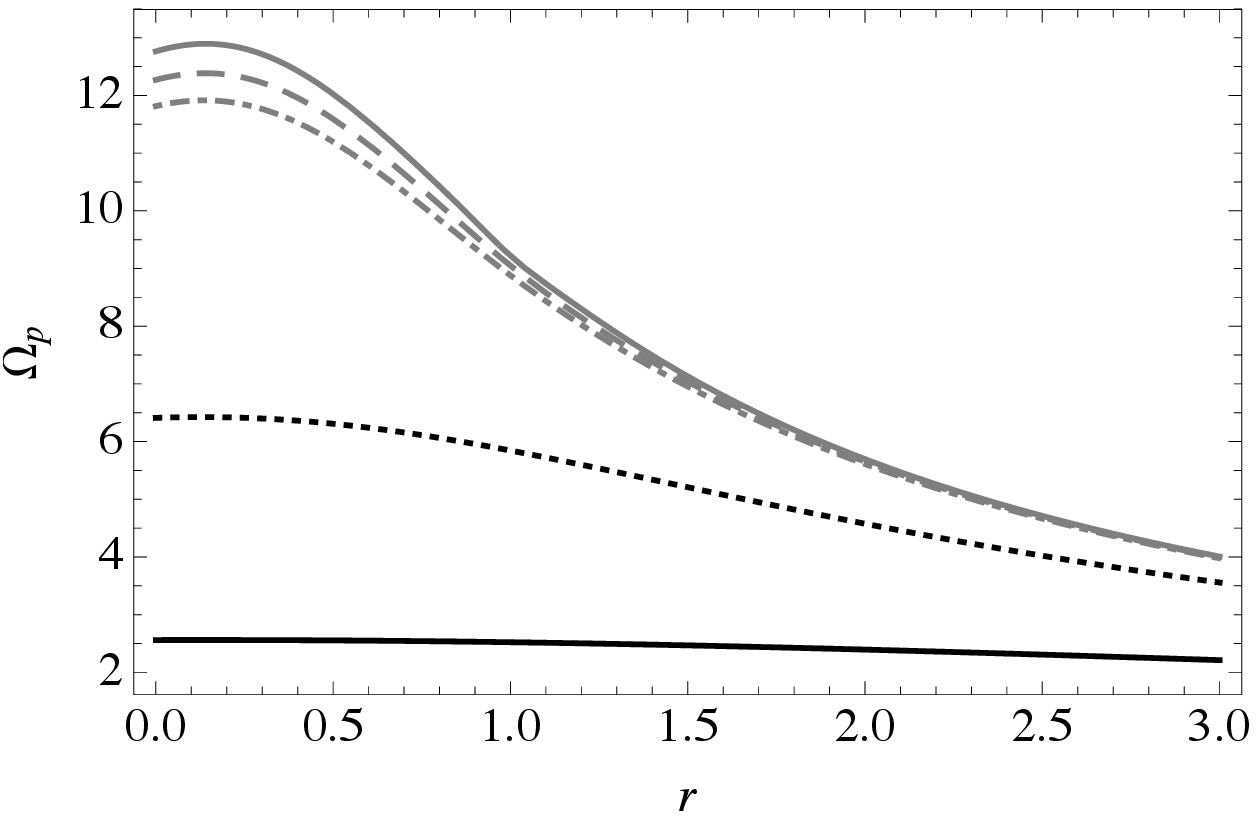}}
\subfigure[~$q=0.9, \theta=50^0$]{
\includegraphics[width=2in,angle=0]{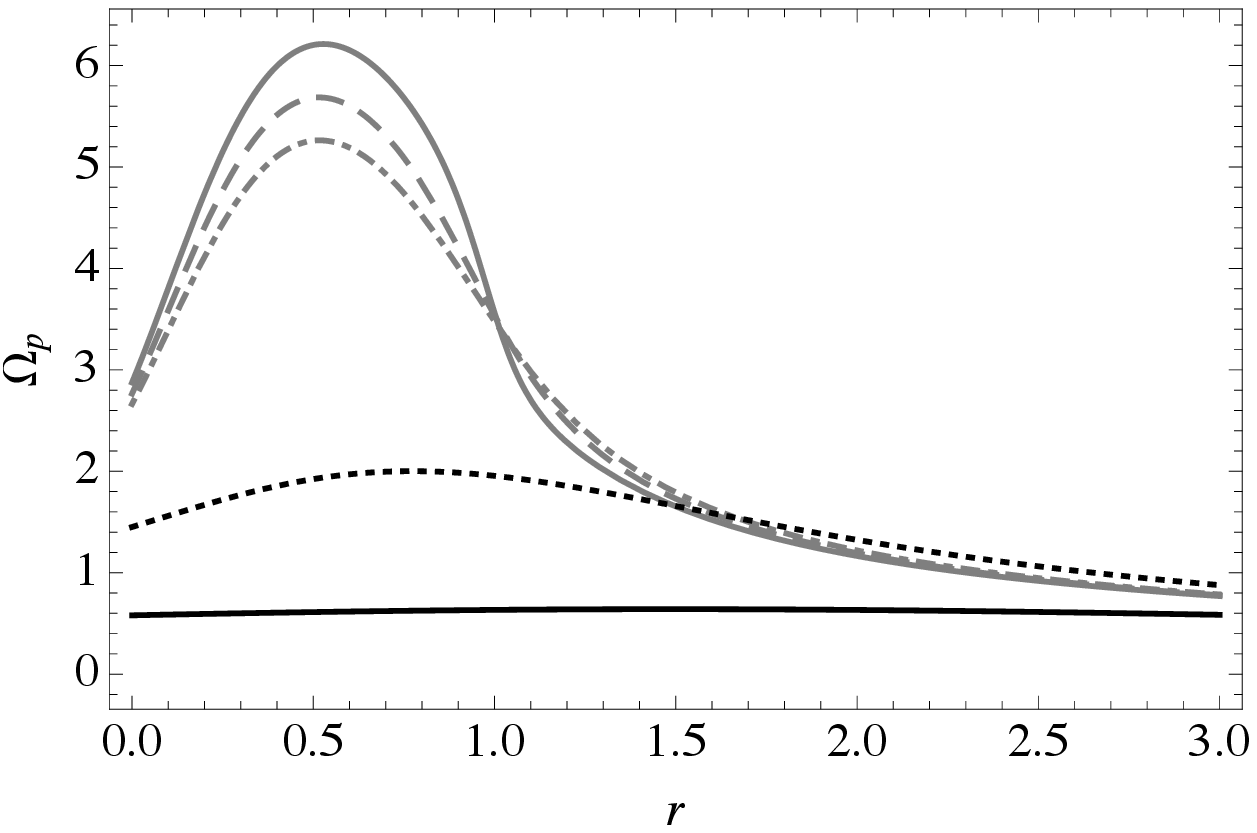}}
\subfigure[~$q=0.9, \theta=90^0$]{
\includegraphics[width=2in,angle=0]{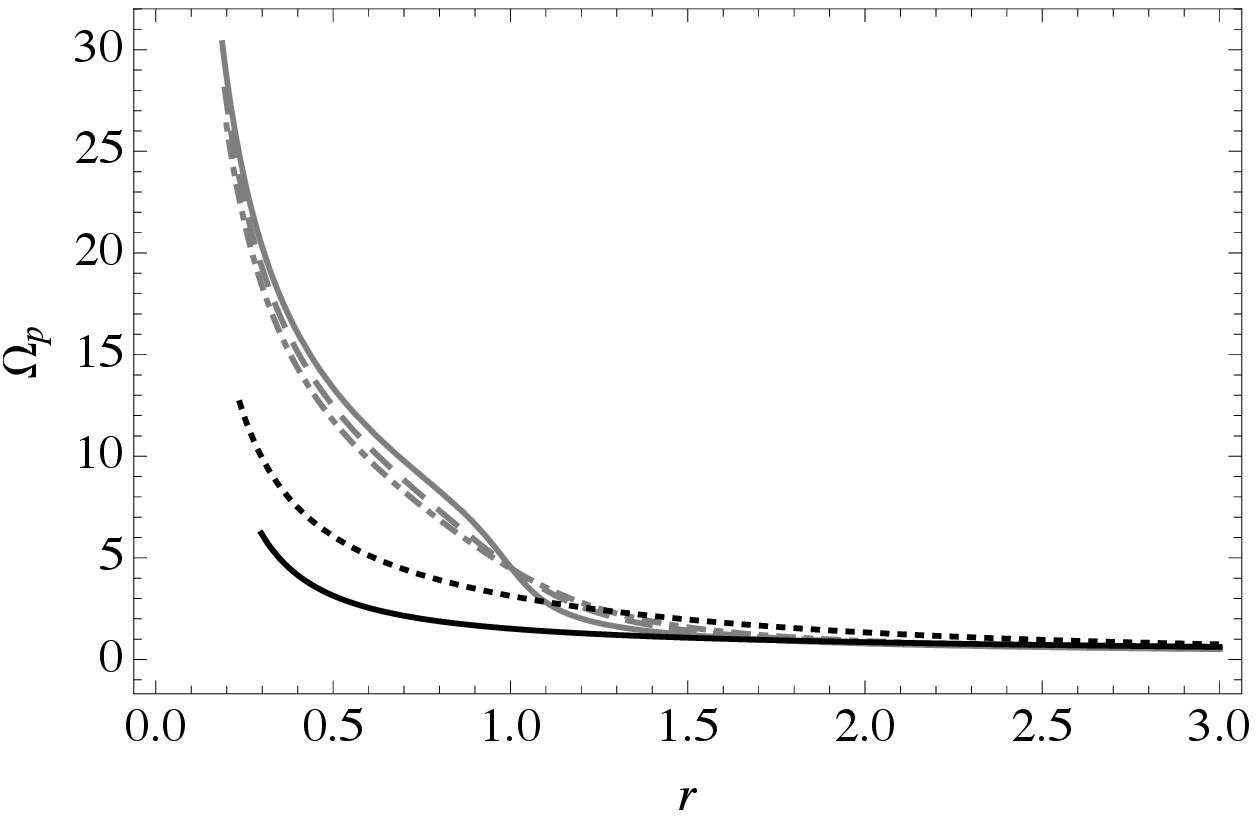}}
\caption{\label{NSGen} We have plotted here, for naked singularities, the modulus of the
precession frequency $\O_p$ (in $M^{-1}$) vs $r$ (in $M$) for different $a_*, q$ and 
$\theta$. We have $q=0.1, 0.5, 0.9$ in the top, middle and bottom rows respectively and
$\theta = 10^0, 50^0, 90^0$ in the left, centre and right columns respectively. In each
panel, the line style is gray, dashed gray, dot-dashed gray, dotted black and black for
$a_* = 1.01, 1.05, 1.09, 2, 5$ respectively. We have plotted $\O_p$ for each NS (with 
different $a_*$) between $r=0$ and $r=3$. The ergoregion is at $r=2$ for $\theta = 90^0$
(bottom row), for reference. This figure clearly demonstrates that for all values of 
$a_*,  q, \theta$, the precession frequency $\O_p$ becomes arbitrarily large near the 
ring singularity. Further, the gray lines are all near-extremal NSs and this figure 
demonstrates how near-extremal NSs appear to have additional characteristic features
that can be used to distinguish them from NSs with higher angular momentum. Motivated
thus, we explore this distinction in greater detail in Section.\ref{nens} since 
near-extremal naked singularities are of great interest from an observational standpoint,
if they exist.}
\end{center}
\end{figure}

We plot the modulus of the precession frequency of stationary gyros $\O_p=|\vec{\O}_p|$,
gotten from Eq.(\ref{gnx}),
\begin{equation} \label{eq:modOp}
\O_p = \f{(r^2+a^2)^2-a^2\d \sin^2\th}{4q(1-q)\r^7 \d}\sqrt{A^2\d\cos^2\th + B^2\sin^2\th},
\end{equation}
and study its variation with $a_*, q, r$ and $\theta$. Briefly, from the above 
expression, one can see that the denominator of $\O_p$ vanishes for $\rho=0, \Delta=0$ 
or $q=0,1$. Since it can be seen from Eq.(\ref{AB}) that $A$ and $B$ are finite valued, 
$\O_p$ becomes arbitrarily large when its denominator vanishes. $\rho = 0$ is the ring
singularity, $\Delta = 0$ is the event horizon and $q=0, 1$ are the (excluded) 
boundaries of the allowed values of $\O$.

For a BH with $a_* = 0.9$, we can see from the left panels of Fig.\ref{BHvNS} that the 
precession frequency indeed becomes arbitrarily large close to the horizon for all 
values of $a_*, q$ and $\theta$, except $q=0.5$. From panel (a), we can see that for 
$q < 0.5$ the radial variation of $\O_p$ 
is monotonic, with no maxima or minima. However, for $q > 0.5$, a minima starts
appearing, which can be seen from panels (e) and this minima is sharp for 
$\theta = \pi/2$. For observers at the ZAMO frequency, $q=0.5$, the precession frequency
remains smooth and finite, as can be seen from panel (c), even for 
gyros orbiting close to the horizon. We note that the ZAMO frequency exhibits consistently peculiar behaviour which might lead to interesting insights on exploring further. On the other hand, for a NS with $a_* = 1.1$, as
can be seen from the plots on the right in Fig.\ref{BHvNS}, the spin precession frequency
does not diverge. It remains finite and regular even as one approaches $r = 0$ for all
angles $0 < \th \lnsim 90^0$. Near $r=0, \th=90^0$, the precession frequency becomes arbitrarily large because of the presence of the ring singularity. This is also in stark contrast to the BH case 
in the present paper, for which we obtain a divergence in the precession frequency close
to the event horizon, \lq far away\rq\ from $r=0$. One also finds that a local minima 
and a local maxima appear for $q \geq 0.5$ for some angles, i.e. there are additional features that might help to ascertain the angular velocity of a stationary observer w.r.t. the ZAMO frequency. We also note here that it can be seen from the $y$-axis scales in the panels (a), (b) and (e), (f) of Fig.\ref{BHvNS} relative to the 
scales in the other panels that $\O_p$ rises sharply as the angular momentum of the 
stationary observer $\O$ nears its allowed bounding values $\O_\pm$. These panels represent $q=0.1, 0.9$ 
respectively for BH and NS.

In Fig.\ref{BHGen}, we demonstrate that the features obtained for
$a_* = 0.9$ are characteristic to all BHs by plotting $\O_p$ for other values of 
$a_* = 0.2, 0.4, 0.6, 0.8, 1$. We show that the spin precession frequency is 
finite and smooth both inside and outside the ergoregion but it diverges near
the horizon for all $a_*, q, r$ and $\theta$, except for $q=0.5$. 
Finally, in Fig.\ref{NSGen}, we demonstrate that the features obtained for $a_* = 1.1$ are characteristic of 
NSs, in general, by considering other values of $a_* = 1.01, 1.05, 1.09, 2, 5$. We have picked these values at non-uniform intervals anticipating additional features in the plots for near-extremal NSs. 
We show that the spin precession frequency is finite and smooth both inside and 
outside of the ergoregion, same as the BH case, but it diverges near the ring 
singularity for all $a_*, q, r$ and $\theta$. This is different from the BH case,
as we have already mentioned above. 
Indeed, we also note here that near-extremal NSs appear to have 
additional characteristic features which could be used to distinguish them 
from other generic higher spin NSs, as can be seen clearly from the panels of 
this figure, and we explore this observation in the following section.

We now describe our experiment to distinguish a Kerr black hole from a Kerr naked 
singularity. Consider gyroscopes attached to stationary observers with a non-zero 
azimuthal component ($\O$) to their four-velocities $u$. These are observers moving
along circles at constant $r$ and $\theta$, with a constant angular velocity $\O$. For every orbit or trajectory at fixed $(r,\theta)$, we can find the range of allowed $\O(r,\theta)$ by finding the lower and upper bounds, $\O_{-}(r,\theta)$ and $\O_{+}(r,\theta)$. We can represent this angular velocity equivalently by the parameter $q$, which gives the absolute relation of that observer with respect to the ZAMO ($q=0.5$). Consider observers along two directions, say $\theta = \theta_1, \theta_2$ ($\theta_1 \nsim \theta_2$). From the modulus of the precession frequencies $\O_p$ of gyroscopes attached to timelike stationary observers orbiting a Kerr compact object at different $r$ along these two directions, we can make the following statements: (i) if $\O_p$ becomes arbitrarily large in the limit of approach to the central object for both $\theta_1, \theta_2$, then the spacetime contains a black hole, whereas (ii) if $\O_p$ becomes arbitrarily large in the limit of approach to the central object for at most one of the two directions $\theta_1, \theta_2$, then the spacetime contains a naked singularity. The reason for this distinction is as follows. For a black hole, $\O_p$ becomes arbitrarily large in the limit of approach to the event horizon, which exists in all directions (i.e., for all $\theta$) and therefore, observers approaching the black hole in all directions will display a divergence. However, for a naked singularity, since the divergence occurs only close to the ring singularity, which exists in the equitorial plane ($\theta=\pi/2$), only those observers that approach the compact object approximately along this direction will see a divergence. Therefore, if one of $\theta_1, \theta_2 = \pi/2$, then we will see a divergence only along that direction. Or, if neither $\theta_1, \theta_2 \neq \pi/2$, we will not see any divergence. Therefore, a divergence along at most one direction will indicate that the spacetime contains a naked singularity. We also note here that our statements are qualitatively independent of the mass of the compact object.

Finally, we expect that these results can possibly be extended to any black hole or 
naked singularity with symmetries. We have used 
in \cite{ckj} and in this work the result that there exist invariant characterizations 
of ergoregions (for rotating spacetimes) and horizons respectively in terms of the 
Killing vectors. We set up observers equipped with gyroscopes along integral curves
of these Killing vectors and then study the precession behaviour of these gyros and 
interpret any arbitrarily large growth in the modulus of the precession frequency 
that is obtained as indicators of the presence of an ergosurface, as in \cite{ckj},
or a horizon, as we have discussed here.

\section{Distinguishing near-extremal Kerr naked singularities from ones with 
higher angular momentum \label{nens}}~

\begin{figure}[!h]
\begin{center}
\subfigure[~$a_*=1.0001, q=0.3$ ($\O<\o$)]{
\includegraphics[width=2.85in,angle=0]{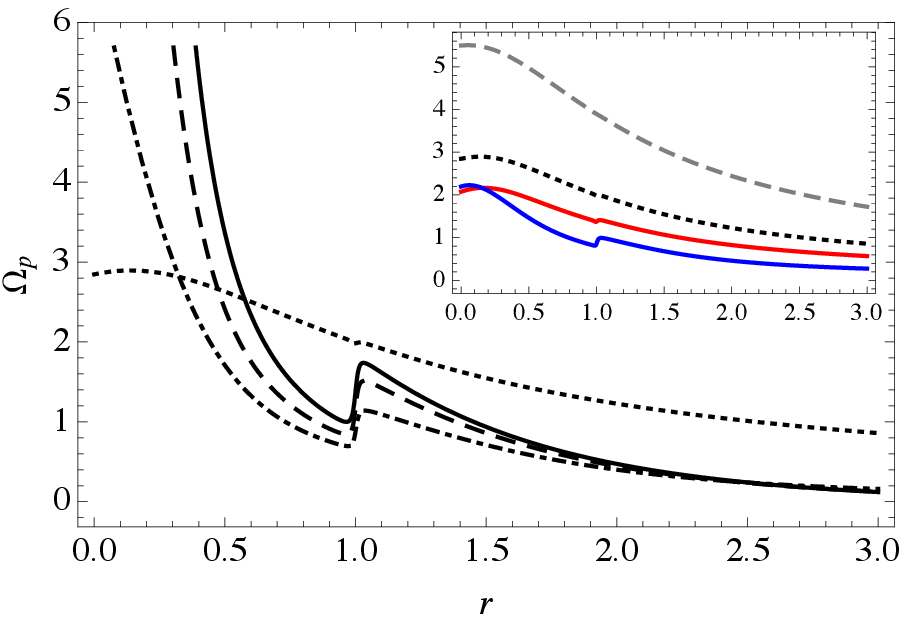}} 
\subfigure[~$a_*=1.01, q=0.3$ ($\O<\o$)]{
\includegraphics[width=2.85in,angle=0]{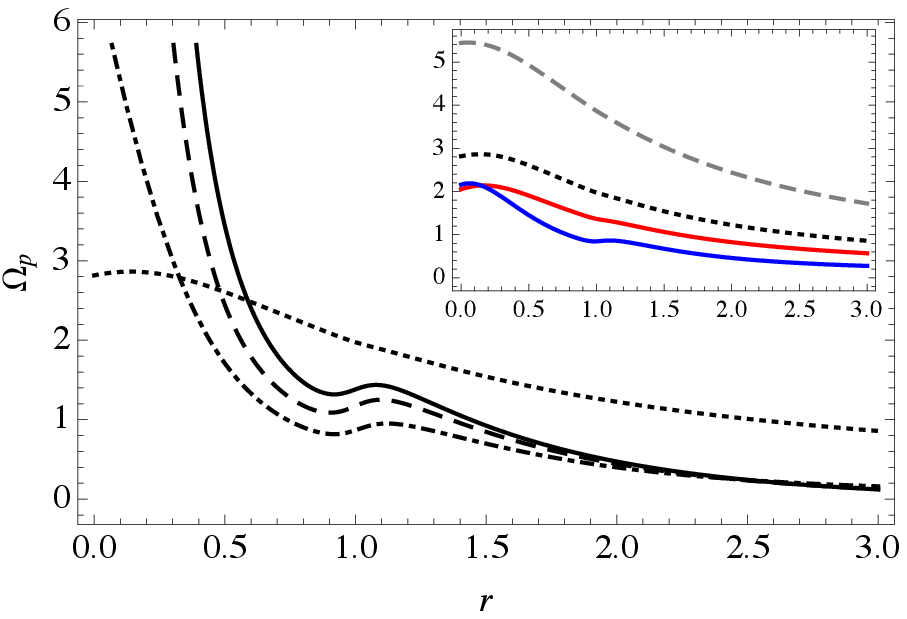}} 
\subfigure[~$a_*=1.0001, q=0.5$ ($\O=\o$)]{
\includegraphics[width=2.85in,angle=0]{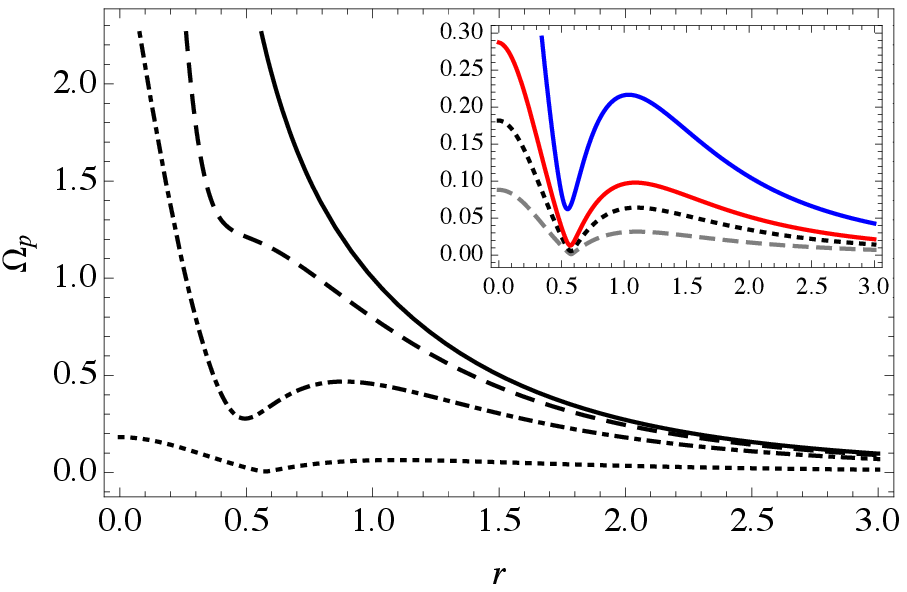}}
\subfigure[~$a_*=1.01, q=0.5$ ($\O=\o$)]{
\includegraphics[width=2.85in,angle=0]{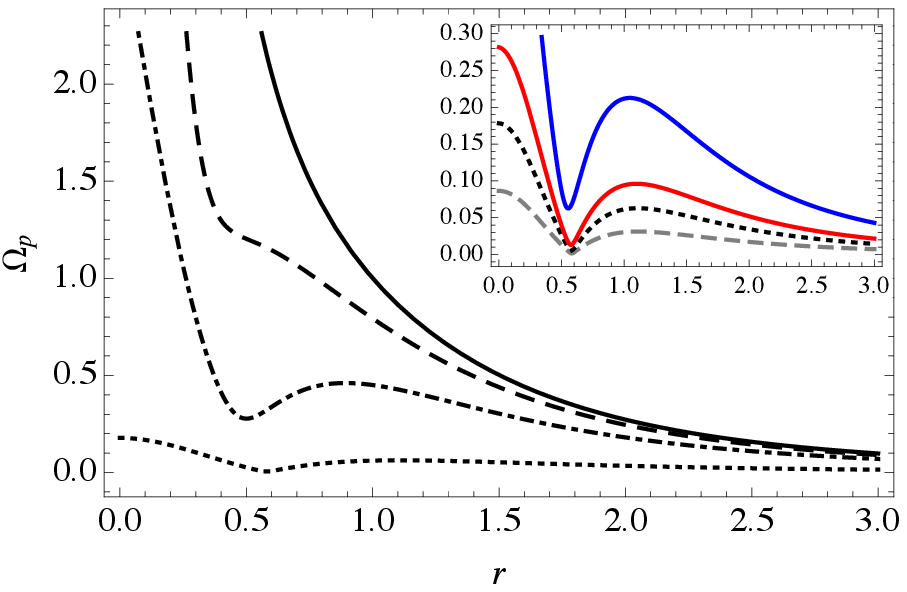}}
\subfigure[~$a_*=1.0001, q=0.7$ ($\O>\o$)]{
\includegraphics[width=2.85in,angle=0]{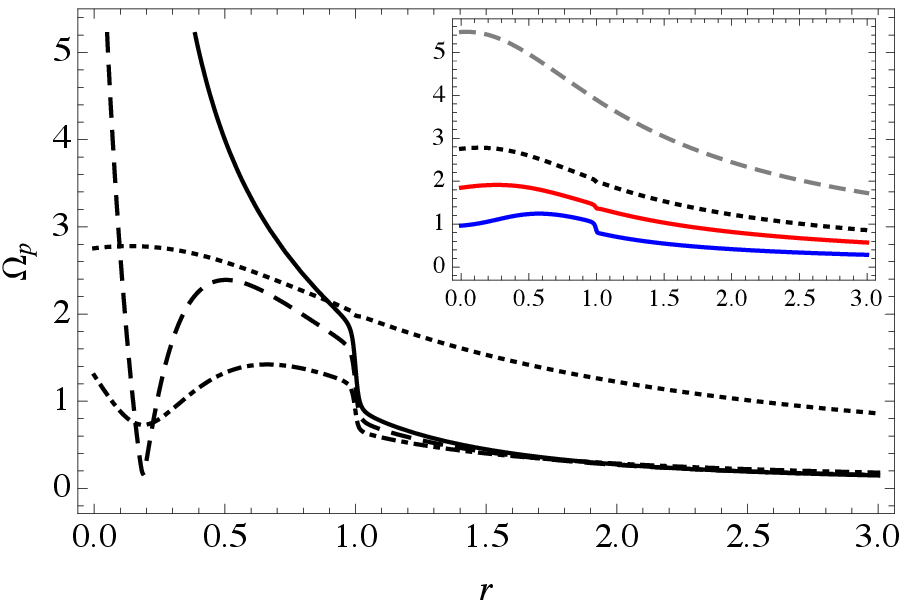}}
\subfigure[~$a_*=1.01, q=0.7$ ($\O>\o$)]{
\includegraphics[width=2.85in,angle=0]{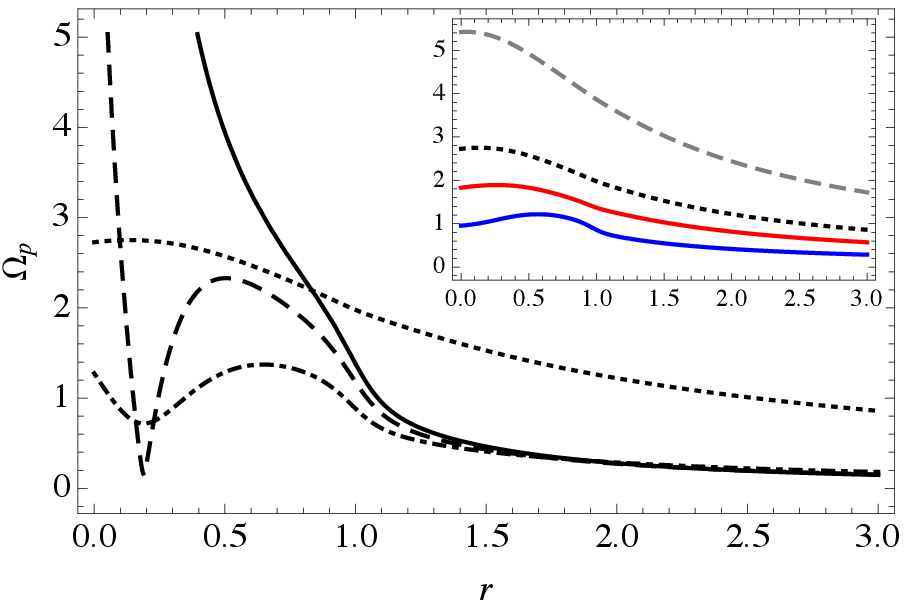}}
\caption{\label{NENS} The modulus of the precession frequency $\O_p$ 
(in $M^{-1}$) versus $r$ (in $M$) has been plotted for near-extremal naked singularities
with two different $a_*$ for different  $q, \theta$. We have used $a_*=1.0001, 1.01$ in
the plots in the left and right columns and  $q=0.3, 0.5, 0.7$ in the top, middle and 
bottom rows, which are representative of $\O<\o, \O=\o$ and $\O>\o$ respectively. In
each panel, the black dotted, dot-dashed, dashed and regular lines represent
$\theta=10^0, 50^0, 70^0, 90^0$ respectively. This plot shows that for all $q$, at
$\theta \sim 0^0$, the radial variation of $\O_p$ is smooth. From panels (a),(e), 
for $\O \nsim \o$, a clear maxima-minima pair appears around $r=1$ resulting in a 
sharp drop/rise in $\O_p$ at that radius. The event horizon of an extremal black hole
$a_*=1$ is located at $r=1$ and we link this sharp feature to this observation. At
$\O=\o$ itself $\O_p$ is smooth, devoid of this particular feature. We discuss 
$\O \sim \o$ in the next figure since these $q$ values have richer features. As can be
seen from panels (b),(f), this sharp rise/drop in $\O_p$ gets smoother with increasing
$a_*$. By $a_* \sim 1.1$, these features completely vanish and we interpret this feature
as providing an important criterion based on which one can distinguish a near-extremal NS 
($1<a_*<1.1$) from one with a higher spin. In the inset, we display approximately at 
what angle $\theta$ this sharp $r=1$ feature starts to appear from and we have used
$\theta=5^0, 10^0, 15^0, 30^0$ for the gray dashed, black dotted (same as the main panel), red and blue lines respectively.}
\end{center}
\end{figure}

\begin{figure}[!h]
\begin{center}
\subfigure[~$a_*=1.0001, q=0.49$ ($\O \lesssim \o$)]{
 \includegraphics[width=3in,angle=0]{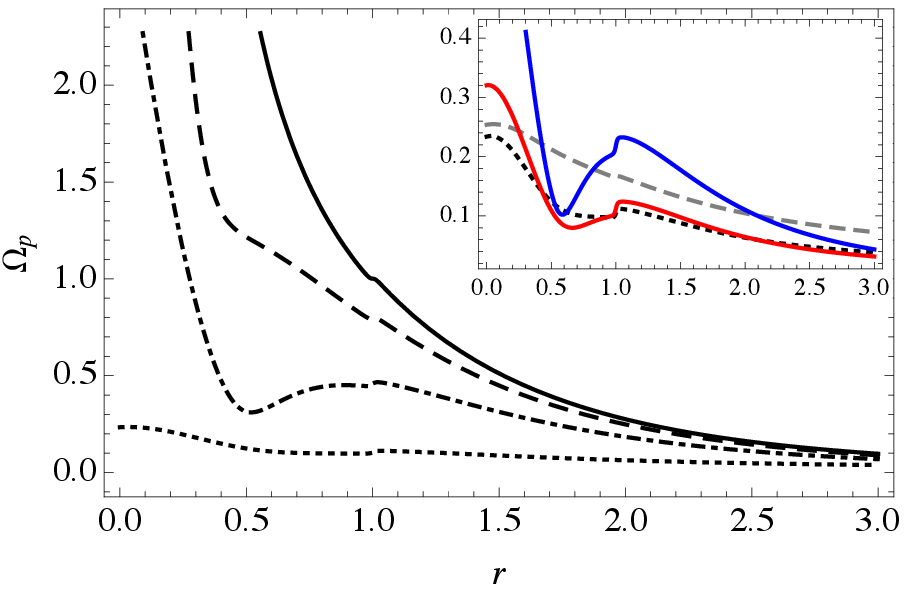}}
\subfigure[~$a_*=1.01, q=0.49$ ($\O \lesssim \o$)]{
\includegraphics[width=3in,angle=0]{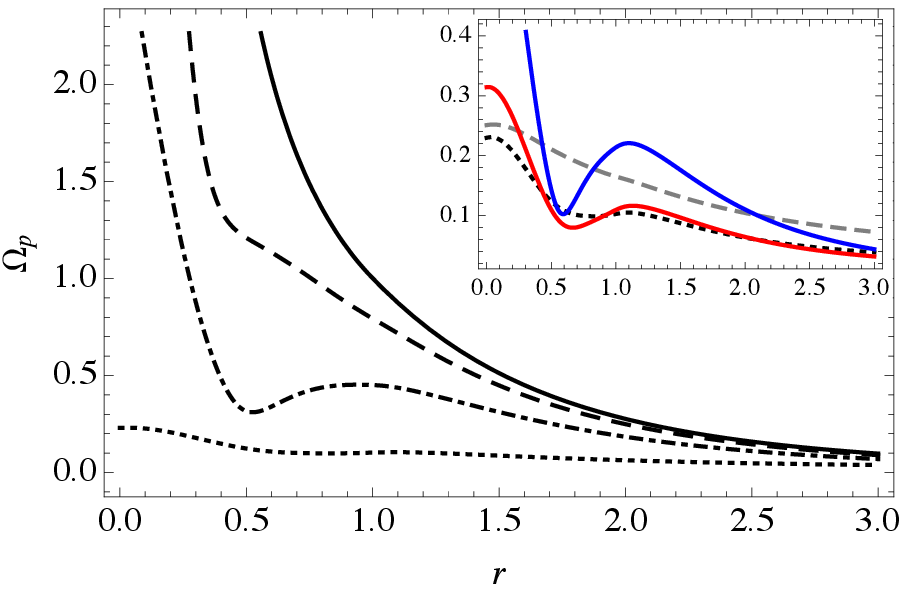}}
\subfigure[~$a_*=1.0001, q=0.51$ ($\O \gtrsim \o$)]{
\includegraphics[width=3in,angle=0]{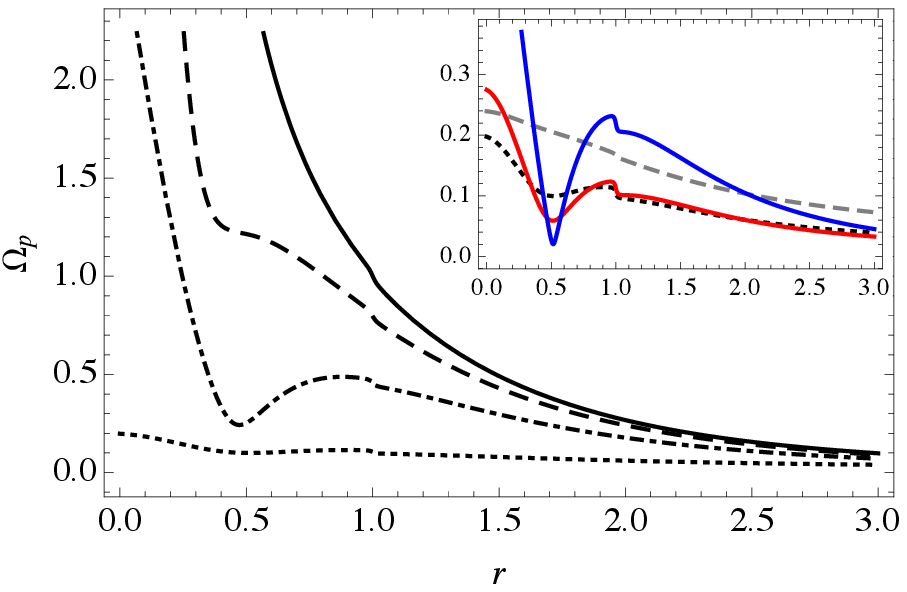}}
\subfigure[~$a_*=1.01, q=0.51$ ($\O \gtrsim \o$)]{
\includegraphics[width=3in,angle=0]{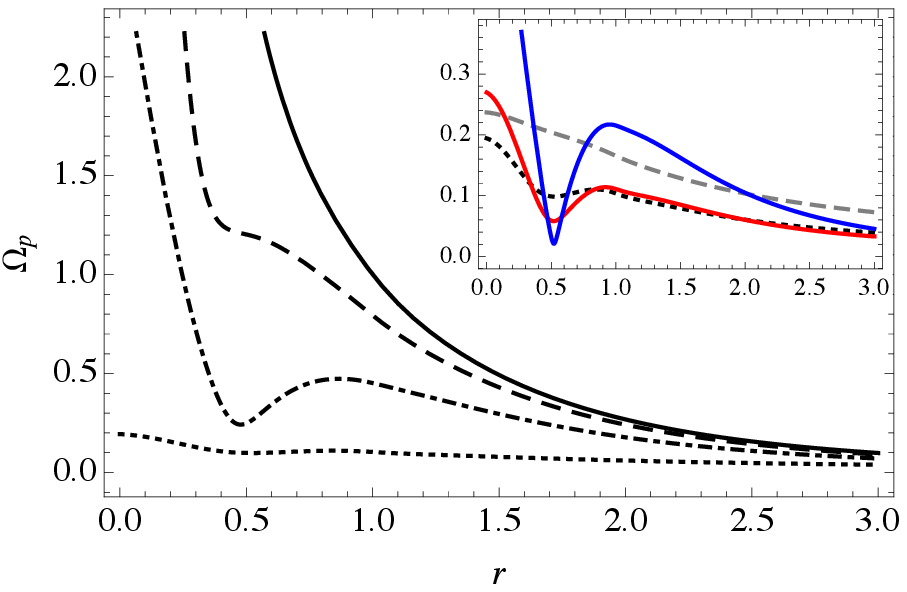}}
\caption{\label{NENS_NearZAMO} We plot now $\O_p$ (in $M^{-1}$) vs r (in $M$) for near-extremal naked singularities with $a_*=1.0001$ and $1.01$ at different $\theta$, for
$q \sim .5$ ($q\neq .5$). The line styles are the same as Fig.\ref{NENS} and the insets
in this figure demonstrate that for these values of $q$, that is $q$ close to $.5$ but
not equal to, there are richer features at smaller angles $\theta$. It can be seen from
this figure, as was from Fig.\ref{NENS}, that with increase in $a_*$, these features all
get smoothed out.}
\end{center}
\end{figure}

In the previous section, we discussed how the spin precession frequency, in the case of
a BH, becomes arbitrarily large at the horizon for all values of $q$ barring $q\sim 0.5$.
For a NS, it diverges only near the ring singularity $r=0, \theta = \pi/2$ and remains 
finite even for $r=0$ for $0 < \th \lnsim \pi/2$. We pointed out that this can be used to 
distinguish a BH from a NS. Further, from Fig.\ref{NSGen}, a general feature that
emerges is that with increase in $a_*$, the radial profile of $\O_p$ becomes 
increasingly \lq smoother.\rq\ This motivates us to use the \lq sharp\rq\ features 
that appear for naked singularities with $a_* \gtrsim 1$ to separate them from those
with $a_* \gg 1$. Indeed, we find that one can use the features that appear for
$1 < a_* < 1.1$ to characterize them and potentially distinguish them from 
$a_*>1$. In this section we highlight these features in $\O_p(r)$ with change in
$a_*, q, \theta$. 
\\

We lay emphasis on this study because of the importance of near extremal naked 
singularities in general relativity. For example, in a black hole binary collision,
like the one studied by LIGO recently, the angular momentum of the compact object during
the collision could temporarily exceed the Kerr bound and result in a temporary
near-extremal naked singularity, a scenario that is of interest to 
\cite{JoshiMalafarina}. Further, if a thick accretion disk could spin up a near 
extremal black hole, it would likely form a near-extremal naked singularity due
to its proximity to the black hole geometry in the $a_*$ parameter space. We mention in relation to this point that using the Polish doughnut model (not for Kerr spacetime), Li and Bambi \cite{lb} 
showed that the overspinning compact objects can be generated by thick accretion disks.
\\

The radial variation of the modulus of the precession frequency for all $q$, at 
$\theta \sim 0^0$ is smooth, as can be seen from Fig.\ref{NENS} and 
Fig.\ref{NENS_NearZAMO}. For an observer moving with an angular velocity that is not
close to the ZAMO frequency ($q\nsim 0.5$), a clear maxima-minima pair appear around 
$r=1$ resulting in a sharp drop/rise in $\O_p$ at that radius, as can be seen from 
panels (a) and (e) of Fig.\ref{NENS}. The event horizon of an extremal black hole $a_*=1$ 
is located at $r=1$ and we link this sharp feature to this observation. We will discuss this in some more detail in the following subsection. Roughly, however, the reason for these sharp features is as follows. From 
Eq.(\ref{eq:modOp}), we see that $\Delta$ appears in the denominator. We know that
$\Delta = 0$ marks the location of the horizon for $a_*\leq 1$ and specifically, 
because of this, for $a_*=1$, $\O_p$ exhibits a divergence at $r=1$. This divergence 
is avoided for $a_*\gtrsim 1$ since $\Delta \neq 0$ but $\Delta$ changes only slightly
from $0$ and hence we see a sharp change at $r \sim 1$. The dependence on $\theta$ is 
due to the other factors in Eq.(\ref{eq:modOp}). That is, $a_*\gtrsim 1$ naked 
singularities feel the \lq phantom effects\rq\ of the extremal event horizon. Further,
we can ascertain whether an observer is rotating with an angular frequency $\O$ above 
or below the ZAMO frequency $\o$, that is we can distinguish whether $\O>\o$ or $\O<\o$,
by looking at the additional maxima-minima structure in the region $r<1$. We note here
that at $\O=\o$ itself $\O_p$ is smooth, as can be seen from panels (c) and (d) of Fig.\ref{NENS} and it is devoid of the sharp features that are obtained at $r=1$ for $q\nsim .5$. Features for $q\sim .5$ are highlighted in Fig.\ref{NENS_NearZAMO}.

On moving closer to the compact object, that is on decreasing $r$, in any direction 
$\theta$, $\O_p$ always increases. Further, for $r\rightarrow 0$, on increasing $\theta$,
observers get closer to the ring singularity and therefore see a rapidly rising $\O_p$.
Specifically, as $\theta\rightarrow\pi/2$, $\O_p$ becomes unbounded. With increase in 
$a_*$, as is demonstrated in both Figs.\ref{NENS} and \ref{NENS_NearZAMO}, we see that
this $r=1$ feature becomes smoother. By $a_* \sim 1.1$, these features completely vanish
and we interpret this result as providing an important criterion based on which one can
distinguish a near-extremal NS ($1<a_* < 1.1$) from one with a higher spin. In the insets
of both figures, we explore approximately at what angle $\theta$ this sharp $r=1$ feature
starts to appear from and this value of $\theta$ depends on $a_*$, in general.

\subsection{Behavior of gyro frequency for near-extremal naked singularities}
In this subsection, we will discuss the reasons for the features that are exhibited by 
near-extremal naked singularities that are different from those with higher spins.
\\
Define $\chi$ from Eq.(\ref{gnx}) for convenience as,
\begin{equation}
\chi = \frac{(r^2+a^2)^2 - a^2 \Delta \sin^2\theta}{4q(1-q)~\rho^7 \Delta},
\end{equation}
so that the precession frequency $\vec{\Omega}_p$ becomes,
\begin{eqnarray} \label{eq:Op_Chi}
\vec{\Omega}_p = \chi\left[A\sqrt{\Delta}\cos\theta~\hat{r} + B\sin\theta~\hat{\theta}\right].
\end{eqnarray}
First, we de-dimensionalise the above expression by replacing $a_* = a/M$ and 
introducing the dimensionless radial variable $y=r/M$. We can then write $\chi$ as,
\begin{equation}
\chi = \frac{(y^2+a_*^2)^2-a_*^2\Delta_* \sin^2\theta}{4q(1-q)\rho_*^7 \Delta_*}~M^{-5}.
\end{equation}
Therefore, $\chi$ has mass dimension $-5$. Similarly, $A$ and $B$ have mass dimensions $3$ and $4$ respectively, $\sqrt{\Delta}$ has mass dimension $1$ and the term in the square 
braces of Eq.(\ref{eq:Op_Chi}) has a total mass dimension of $4$. Therefore, $\vec{\Omega}_p$ has an overall mass dimension of $-1$. In this section, henceforth, we work exclusively with dimensionless quantities and simply drop all factors of $M$. We write down now the dimensionless expressions for a near-extremal NS by replacing $a_* = 1+\epsilon$ ($\epsilon > 0$),
\begin{eqnarray} \label{eq:NENS_Dimensionless} 
\chi_{\rm ne} &=& \frac{(y^2 + 1 + 2\epsilon)^2 - (1 + 2\epsilon)\Delta_{\rm ne} \sin^2\theta}{4q(1-q)\rho_{\rm ne}^7 \Delta_{\rm ne}},\\ 
A_{\rm ne} &=& 2(1 + \epsilon)y  - \frac{\Omega_{\rm ne}}{8}\left\{8y^4 + 8(1 + 2\epsilon)y^2 + 16(1 + 2\epsilon)y + 3(1+4\epsilon)  \right.\nonumber\\
&& \left. \ \ \ \ \ \ \ \ +~4(1 + 2\epsilon)(2\Delta_{\rm ne} -1 - 2\epsilon)\cos2\theta + (1 + 4\epsilon)\cos4\theta \right\} + 2\Omega_{\rm ne}^2 (1 + 3\epsilon) y\sin^4\theta,\nonumber\\
B_{\rm ne} &=& (1 + \epsilon)(y^2- (1 + 2\epsilon)\cos^2\theta) + \Omega_{\rm ne}\left\{(1 + 4\epsilon)y\cos^4\theta + y^2(y^3 - 3y^2 - (1+2\epsilon)(1+\sin^2\theta)) \right.\nonumber\\
&& \left. +~(1 + 2\epsilon)\cos^2\theta (2y^3 - y^2 + (1 + 2\epsilon)(1+\sin^2\theta))\right\} \nonumber\\
&& +~\Omega_{\rm ne}^2 (1 + \epsilon)\sin^2\theta[y^2(3y^2 + 1 + 2\epsilon) + (1 + 2\epsilon)\cos^2\theta(y^2 - 1 - 2\epsilon)],\nonumber\\
\Omega_{\rm ne} &=& \frac{2(1 + \epsilon)y\sin\theta - (1-2q)~\rho_{\rm ne}^2\sqrt{\Delta_{\rm ne}}}{\sin\theta[\rho_{\rm ne}^2(y^2 + 1 + 2\epsilon) + 2(1 + 2\epsilon) y\sin^2\theta]},\nonumber\\
\Delta_{\rm ne} &=&y^2 - 2y + 1 + 2\epsilon,\nonumber\\
\rho_{\rm ne}^2 &=& y^2 + \cos^2\theta + 2\epsilon\cos^2\theta \nonumber
\end{eqnarray}
where subscript `ne' stands for `near-extremal.' For some constant $\k$, we can write,
\begin{eqnarray}\nonumber
(\Delta_{\rm ne})^\k \approx &~(y-1)^{2\k}\left[1 + \epsilon~\frac{2\k}{(y-1)^2}\right]&,~~
\mbox{if}\ |y-1|\gg \epsilon\\
\approx &~\epsilon^\k &,~~\mbox{otherwise},\nonumber \\
(\rho_{\rm ne}^2)^\k  \approx &~(y^2 + \cos^2\theta)^\k\left[1 + \epsilon~\frac{2\cos^2\theta}{y^2 + \cos^2\theta}\right]^\k &,~~\mbox{if}\ |y^2 + \cos^2\theta| \gg \epsilon\nonumber\\
\approx &~(\epsilon \cos^2\th)^\k &,~~\mbox{otherwise}.
\end{eqnarray}
$y=1$ was the location of the event horizon for an extremal black hole ($a_* = 1$), 
which vanished as $a_*$ was changed slightly from $1$. As can be seen from the above 
expressions, this is a special point for $\Delta_{\rm ne}$. For $\rho_{\rm ne}$, the two cases
correspond to being far from and near the ring singularity at $r=0, \theta=\pi/2$ 
respectively.

As can be seen from Eq.(\ref{eq:NENS_Dimensionless}), $\Omega_{\rm ne}$ is
finite and smooth always (remember that at the pole i.e., for $\theta = 0$, the only
allowed value of $\O$ is $\Omega = 0$ and hence, $\Omega_{\rm ne, \theta=0}=0$). Therefore, $A$
and $B$ are also finite and smooth and we can restrict ourselves to studying 
$\chi_{\rm ne}$ to find any interesting \lq sharp\rq\ features in the radial profile of
the modulus of the precession frequency $\Omega_p$, for a near-extremal NS. Indeed, 
we can see from $\chi_{\rm ne}$ given in Eq. (\ref{eq:NENS_Dimensionless}) that the factor 
of $\Delta_{\rm ne}$ in the denominator will drive $\Omega_p$ to rise sharply near $y=1$ 
for near-extremal naked singularities. Specific maxima/minima structure in the radial 
profile of $\Omega_p$ can also be ascertained from Eq.(\ref{eq:NENS_Dimensionless}).

\subsection{Near-extremal overspinning Kerr geometry and ultra-high energy collisions}
Many interesting physical processes occur in near-extremal Kerr geometry at $r=M$. 
These processes include ultra-high energy 
particle collisions  and collisional Penrose process with extremely large efficiency
of energy extraction.

In \cite{Patil1,Patil2}, we considered two particles which follow geodesic motion 
on the equatorial plane of overspinning Kerr geometry starting from rest at infinity and 
undergo a collision at $r=M$. One of the particles 
that is initially ingoing, turns back at radial coordinate $0<r<M$ and appears at 
$r=M$ as an outgoing particle, while the second 
particle is ingoing. We showed that the center of mass energy of collision between 
the radially ingoing and outgoing particles
shows divergence in the near-extremal limit where Kerr spin parameter transcends the 
extremal value by an infinitesimal amount, i.e., $a=M(1+\epsilon)$ with $\epsilon \rightarrow 0^{+}$.
This process overcomes many limitations and finetunings involved in an analogous 
high-energy collision process between the 
two ingoing particles which occurs close to the event horizon of the maximally 
spinning BH \cite{BSW, Harada}.

We further showed that the particles which are produced in the ultra-high energy 
particle collisions in the 
overspinning Kerr spacetime can escape to infinity with divergent energies \cite{Patil3}.
This is a consequence 
of the collisional Penrose process which allows us to extract rotational energy from
the ergoregion of the Kerr spacetime. 
The efficiency of the  collisional Penrose shows divergence in the near-extremal 
limit for the collisions which occur at 
$r=M$, making it possible to extract large amount of energy from the overspinning Kerr geometry.
This is in the stark 
contrast with the BH case where efficiency is shown to be always finite
with an upper bound of $14$ \cite{Scht}. 
Thus near-extremal NS spacetime can possibly be the source of the 
ultra-high energy cosmic rays and neutrinos. 

Interestingly, as we showed earlier in this section, gyro precession frequency shows
a sharp increase or decline close to $r=M$ in near-extremal overspinning Kerr
spacetime as we decrease its radial coordinate along the constant value of $\theta$. 
This is precisely the location where 
ultra-high energy collisions and collisional Penrose process with divergent efficiency 
occurs. Thus a thought experiment to lower  gyro which 
we described in this paper kills two birds with the same bullet. Firstly it allows us
to identify the spacetime geometry 
which is conducive to the high-energy processes as it can tell us whether the 
geometry is overspinning and near-extremal. Secondly it also helps us to locate 
region in space which can host high-energy processes as gyro
frequency exhibits peculiar trend exactly at this location. This coincidence is quite remarkable.

\section{Frame-dragging effect in accretion disks in a Kerr Geometry \label{Lense-Thirring}}

\begin{figure}
\begin{center}
\includegraphics[width=0.5\linewidth,angle=0]{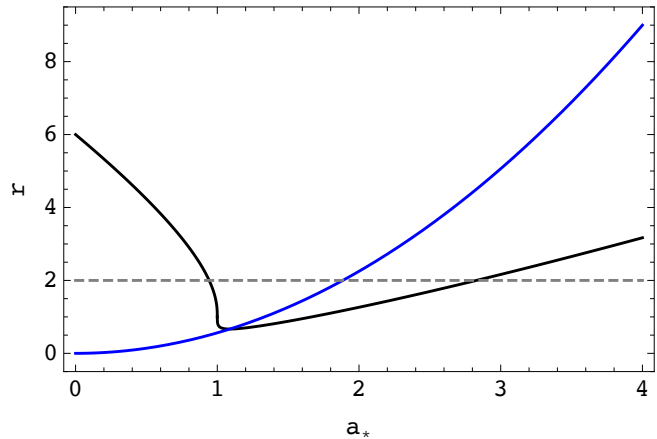}
\caption{\label{isco}Three radial quantities (in units of $M$) for prograde orbits,
namely the ISCO radius (black), the ergoradius (dashed gray), 
and the radius ($r_{0}$) at which the precession frequency ($\O_{\rm nod}$) vanishes (blue), 
plotted as functions of the dimensionless Kerr parameter $a_*=a/M$. The 
ISCO radius lies outside the ergoregion for all $a_*$ except 
$0.943 \leq a_* \leq 2.838$. $r_{0}$ meets the ISCO radius at $a_*=1.089$. 
This has the implication that for smaller values of $a_*$, the LT frequency 
is always positive and does not vanish 
for any radius. For larger values, there is a domain of $r$ for which 
this frequency becomes negative, signifying that the LT effect switches sign.
Since this feature is exhibited for $a_*=1.089 > 1$, the 
LT frequency in a BH spacetime never vanishes. We point 
out that NSs with $a_*<1.089$ also do not display vanishing
LT frequency.}
\end{center}
\end{figure}

In order to study the accretion disk around a spinning BH,
one needs to study the stable circular orbits in the Kerr space-time. 
The last or innermost stable circular orbit (ISCO) marks the inner 
boundary of this disk. The ISCO radius depends on the Kerr parameter $a_*$,
as shown in FIG.~\ref{isco}. This is a key underlying physical feature that 
can distinguish BHs from NSs, as we will see in this section.

The three fundamental frequencies for the accretion disk, 
namely the Keplerian frequency $\O_{\phi}$, vertical epicyclic frequency
$\O_{\th}$, and the radial epicyclic frequency $\O_r$ are derived for the Kerr metric
\cite{ok,ka} (in geometrized units) as, 
\begin{eqnarray}
\O_{\phi} &&=\pm \f{M^{\f{1}{2}}}{(r^{\f{3}{2}} \pm a M^{\f{1}{2}})}
 \label{fr1}
 \\
\O_r &&=\O_{\phi}\left(1-\f{6M}{r} \pm \f{8aM^{\f{1}{2}}}{r^{\f{3}{2}}}
 -\f{3a^2}{r^2} \right)^{\f{1}{2}} 
 \\
\O_{\th} &&=\O_{\phi}\left(1\mp\f{4aM^{\f{1}{2}}}{r^{\f{3}{2}}}
 +\f{3a^2}{r^2} \right)^{\f{1}{2}}
 \label{fr3}
\end{eqnarray}
where the upper sign is applicable for direct orbit and 
the lower one for retrograde orbit. These frequencies are related 
to the precession of the orbit and orbital plane. Precession of the orbit is 
measured by the periastron precession frequency $(\O_{\rm per})$, and 
orbital plane precession is measured by the nodal plane precession or Lense-Thirring 
precession frequency $(\O_{\rm nod})$ \cite{lt}. These two frequencies are defined as \cite{bs}
\begin{eqnarray} 
\O_{\rm per}= \O_{\phi}-\O_r, 
\label{per}
\\
\O_{\rm nod}= \O_{\phi}-\O_{\th}.
 \label{no} 
\end{eqnarray}

Orbital plane precession arises only due to the rotation of the spacetime. In a 
non-rotating spacetime, $\O_{\phi}$ is always equal to $\O_{\th}$, and hence the
Lense-Thirring precession is entirely
absent. However, periastron precession occurs both in rotating and 
non-rotating spacetimes. We note that the square of the radial
epicyclic frequency $\O_r^2$ vanishes at the ISCO, and is negative 
for smaller radii, which shows a radial instability for such orbits. 
Outside the ISCO, $\O_r^2$ is always positive and $\O_{\th}^2$ is 
always non-zero and positive in a rotating spacetime. The same cannot be said about 
$\O_{\rm nod}$. 
For example, the LT precession frequency 
(Eq.(\ref{no})) can be zero at $r=r_{0}$ given by,
\begin{eqnarray}
\O_{\rm nod}(r_{0}) = 0 \Rightarrow r_{0}=\f{9}{16}~a_*^2M=0.5625~a_*^2 M .
 \label{r0}
\end{eqnarray}
Since $r_{0}$ is always less than $r_{\mbox{ISCO}}$
($6M \geq r_{\mbox{ISCO}} \geq M$ \cite{bpt}) and even
inside the event horizon for a BH ($0 \leq a_* \leq 1$), 
the LT precession frequency never becomes zero
for a BH spacetime. We now discuss the location of the
ISCO in a NS spacetime and argue that its
relation with $r_{0}$ has implications for distinguishing BH and
NS spacetimes.

FIG.~\ref{isco} shows that the ISCO radius decreases with 
increasing $a_*$ for prograde orbits
for both BHs and NSs up to  $a_*=\sqrt{32/27} \approx 1.089$, 
and then increases \cite{sti,pug,ig}. Therefore, the minimum ISCO radius, having the value 
$r_{\mbox{ISCO}}=2M/3$, occurs for $a_*=1.089$. As 
seen from FIG.~\ref{isco}, the ISCO lies on or inside the ergosurface 
for $0.943 \leq a_* \leq 2.838$. 
For each $a_*$ value, there exists a radius ($r_{0}$) at which there is no frame-dragging 
effect, and hence the LT precession vanishes. This radius 
is less than the ISCO radius for $a_*<1.089$ (FIG.~\ref{isco}), but this may not be observationally 
important, as the accretion disk extends up to $r_{\mbox{ISCO}}$. FIG.~\ref{isco} also shows that
$r_{0}$ equals $r_{\mbox{ISCO}}$ for $a_* = 1.089$ \cite{ch}, and 
is greater than $r_{\mbox{ISCO}}$ for $a_*>1.089$. These make $1.089$ a special 
value of $a_*$.

\begin{figure}[!h]
\begin{center}
\subfigure[BH ISCO and NS ISCO are located at
 $2.32~M$ and $0.67~M$ respectively]{
\includegraphics[width=3in,angle=0]{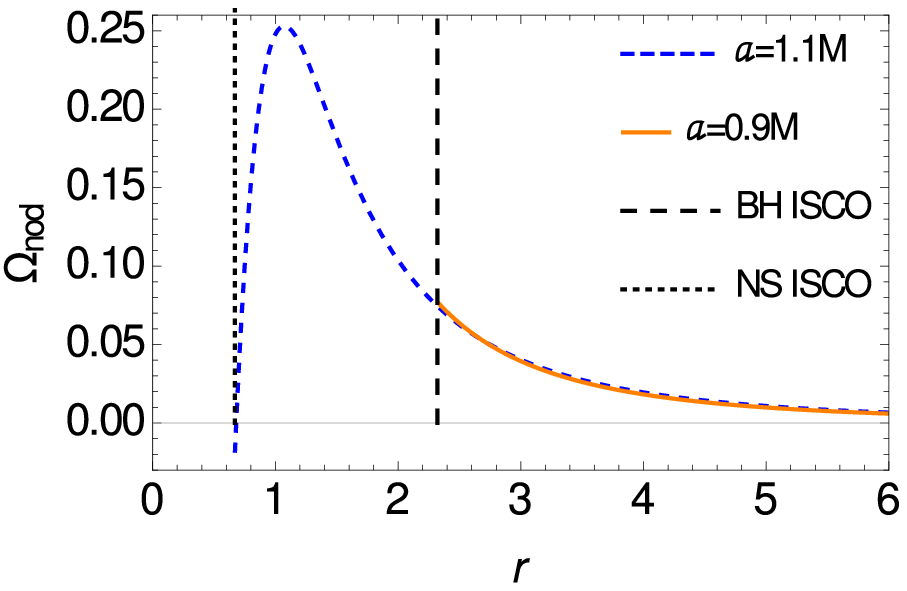}} 
\subfigure[ISCO is located at $1.26~M$]{
\includegraphics[width=3in,angle=0]{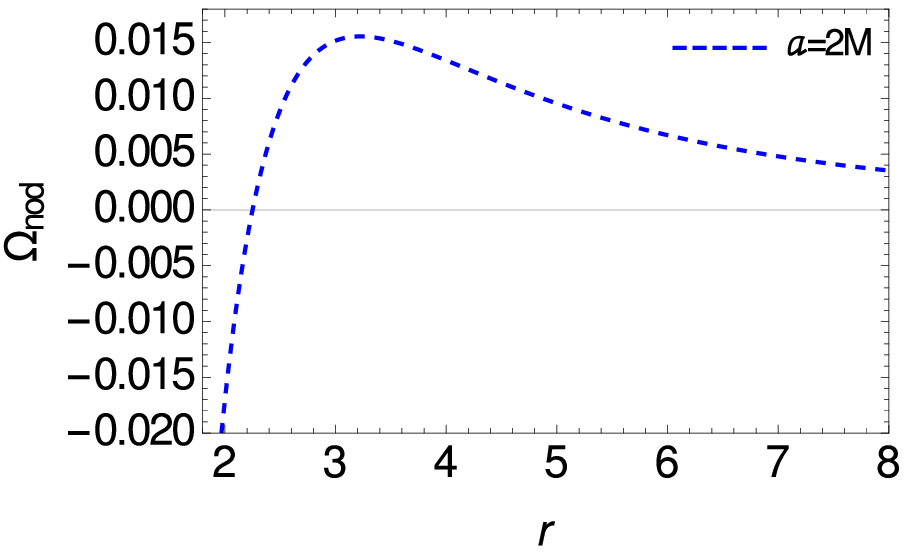}} 
\caption{\label{Nodal}Variation  of $\O_{\rm nod}$
(in units of $M^{-1}$) versus $r$ (in units of $M$). It is seen from the plots
that nodal plane precession frequency $\O_{\rm nod}$ always increases as one 
approaches to a BH but in case of a NS, we obtain a peak value of $\O_{\rm nod}$ 
for all $a_*>1$. $\O_{\rm nod}$ vanishes in a particular orbit of
radius $r_{0}$ for $a_* \geq 1.089$ and it becomes negative 
(which means that the LT precession reverses direction) in all the
orbits which are in the range $r_{0} > r \geq r_{\rm ISCO}$ for $a_* > 1.089$.}
\end{center}
\end{figure}

\begin{figure}[!h]
\begin{center}
\subfigure[~For black holes]{
\includegraphics[width=3in,angle=0]{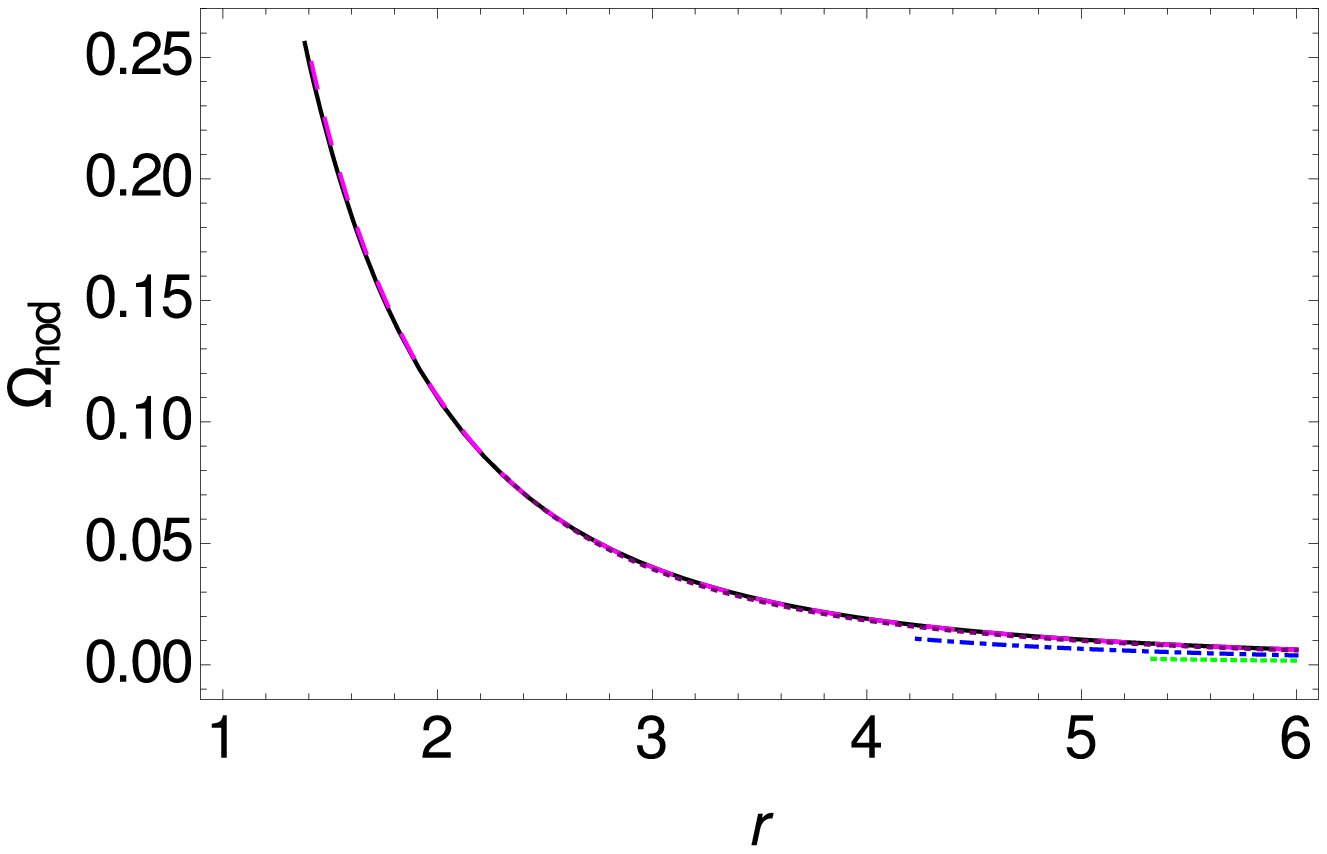} }
\subfigure[~For naked singularities]{
\includegraphics[width=3in,angle=0]{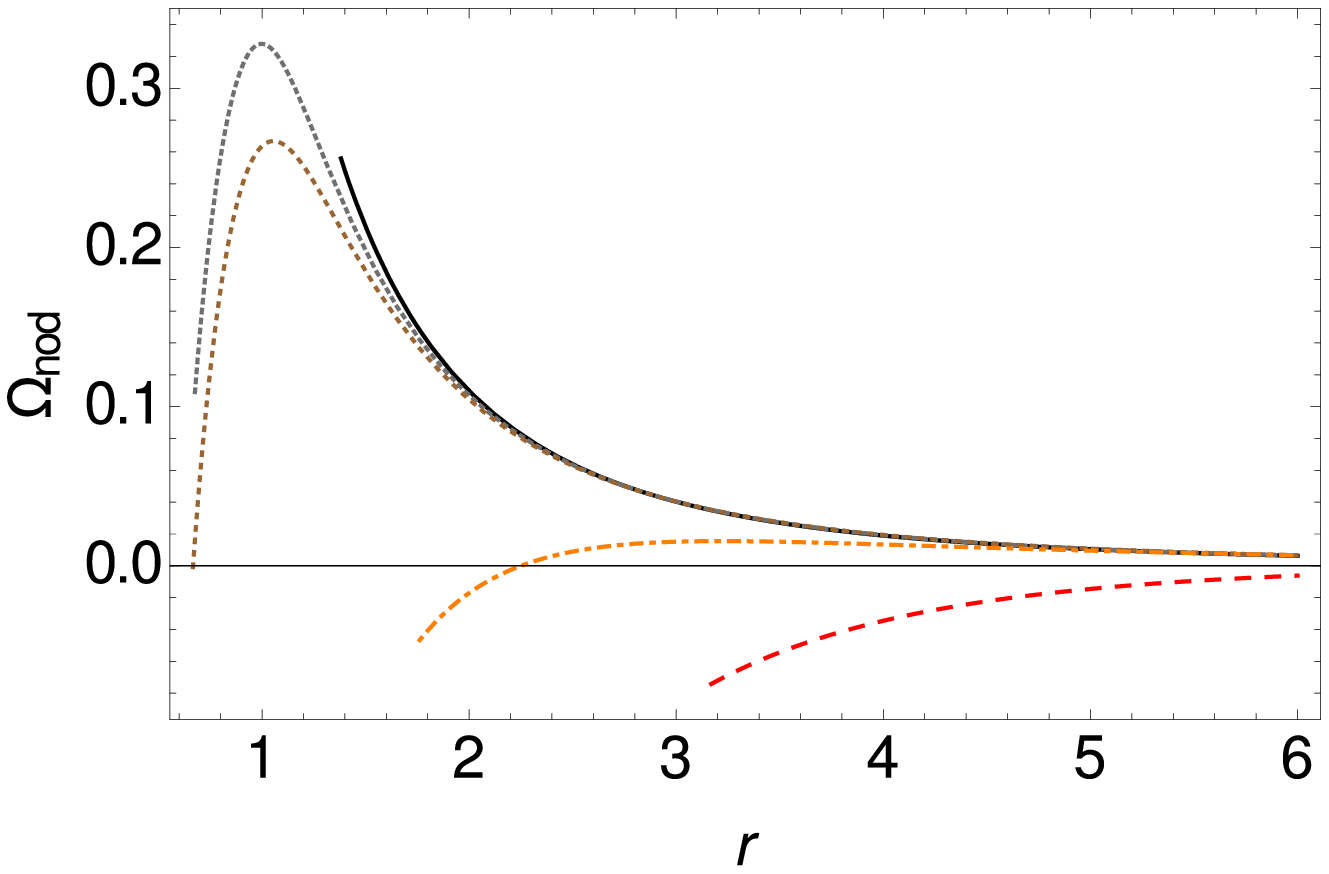}}
\caption{\label{NodalGen} We show the radial variation of the precession
frequency ($\O_{\rm nod}$) (in units of $M^{-1}$) for different 
$a_*$ BHs in the left panel and NSs on the right panel. For 
BHs, we plot $\O_{\rm nod}$ between their respective ISCO radii 
and $r=6$. We have used $a_*=$1 
(black), .9999 (large dashing, magenta), .9 (dotted, purple), 
.5 (dot-dashed, blue) and .2 (tiny dashing, green). $\O_p$ 
decreases with increasing $r$ always for BHs. For NSs, we plot 
$\O_{\rm nod}$ between their respective ISCO radii and $r=6$. We 
have included also the extremal BH case to demonstrate the clear
change in characteristic features. We have used $a_*=1$ (black), 
1.05 (tiny dashed, gray), 1.089 (dotted, brown), 2 (dot-dashed, orange),
4 (medium dashing, red). For NSs,  as we increase $r$, $\O_{\rm nod}$ always
increases initially at the ISCO radius, reaches a peak value and decreases.
Negative $\O_{\rm nod}$ implies that the sense of precession has changed. 
These are characteristic features of NSs.}
\end{center}
\end{figure}

\begin{figure}[!h]
\begin{center}
\includegraphics[width=3in,angle=0]{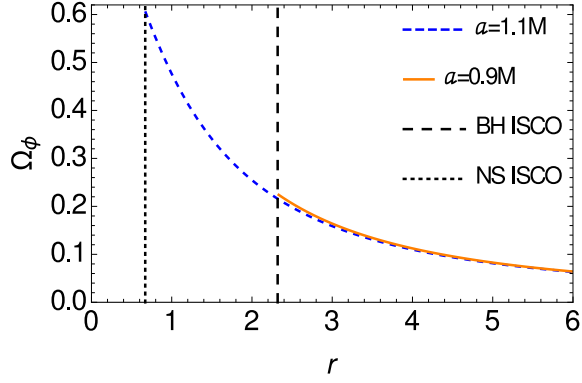} 
\caption{\label{Kepler} Variation  of $\O_{\phi}$ (in units of  $M^{-1}$) 
versus $r$ (in units of $M$). BH ISCO and NS ISCO are located at
 $2.32~M$ and $0.67~M$ respectively. The plots show that Kepler frequency
$\O_{\phi}$ for a NS is much higher than for a BH at their 
respective ISCOs for $\epsilon=\pm 0.1$, i.e., $a_*=0.9$ and $a_*=1.1$.
The difference between the values of Kepler frequencies of a NS and a 
BH decreases with decreasing the value of $\epsilon$.}
\end{center}
\end{figure}

\begin{figure}[!h]
\begin{center}
\subfigure[]{
\includegraphics[width=3in,angle=0]{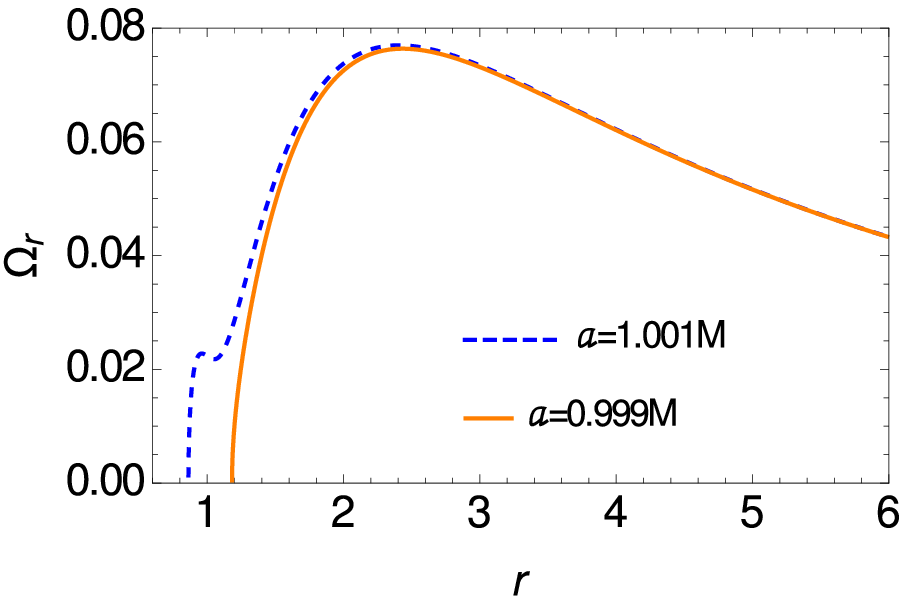}} 
\subfigure[]{
\includegraphics[width=3in,angle=0]{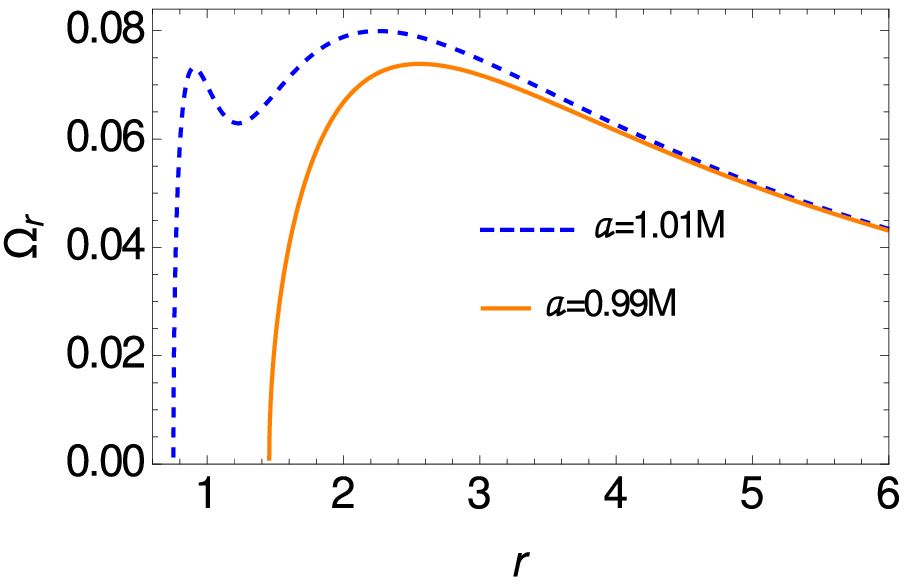}} 
\subfigure[]{
\includegraphics[width=3in,angle=0]{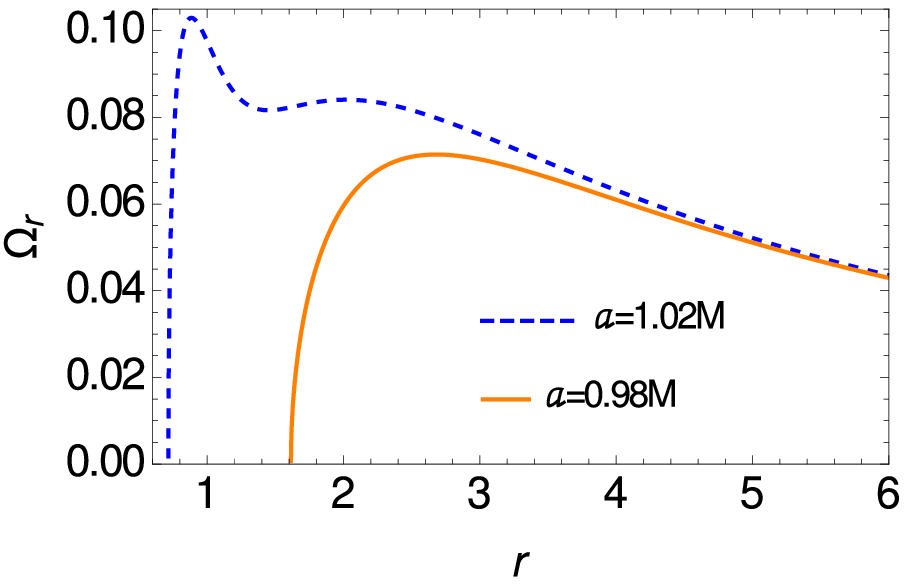}} 
\subfigure[ ]{
\includegraphics[width=3in,angle=0]{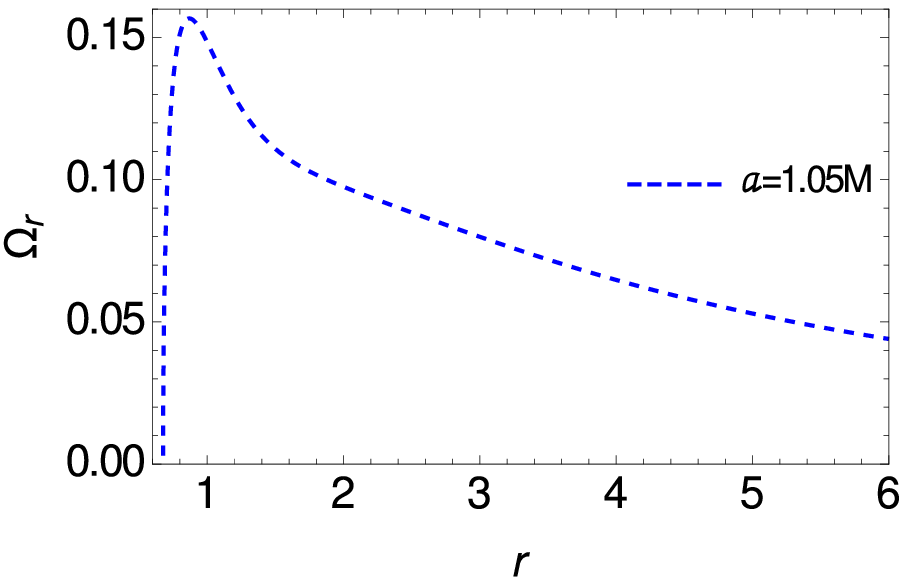}}
\caption{\label{Or} Variation  of $\O_r$ (in units of $M^{-1}$) versus 
 $r$ (in units of $M$). The plots show that $\O_r$ vanishes at their
 respective ISCOs of a BH and a NS, which is expected but it can be seen 
that a small `kink' appears in some of these near-extremal NS cases. This feature
is quite clear for $a_* \gtrsim 1.001$ and it disappears for $a_*=1.05$.}
\end{center}
\end{figure}

\begin{figure}[!h]
\begin{center}
\subfigure[BH ISCO and NS ISCO are located at
 $1.18~M$ and $0.86~M$ respectively]{
\includegraphics[width=3in,angle=0]{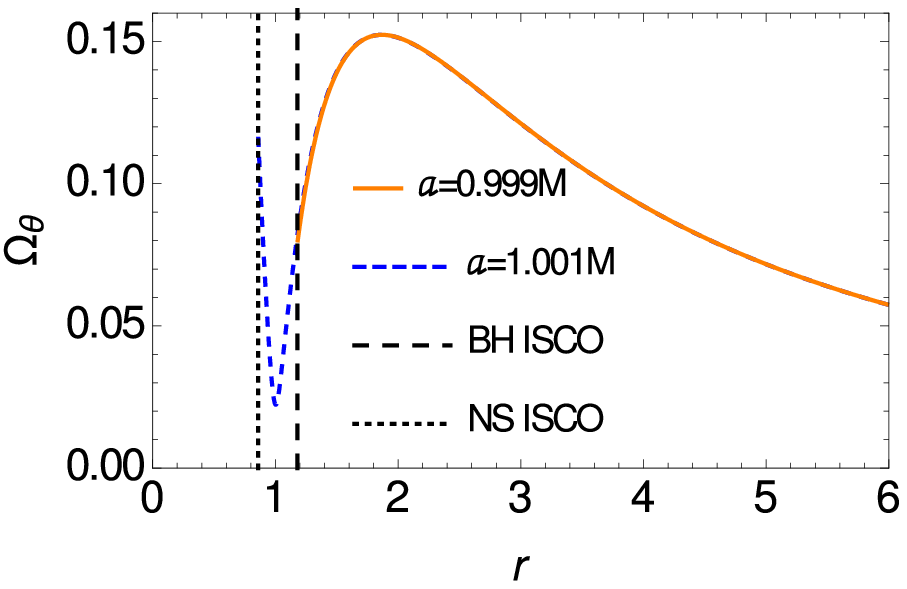}} 
\subfigure[BH ISCO and NS ISCO are located at
 $1.45~M$ and $0.75~M$ respectively]{
\includegraphics[width=3in,angle=0]{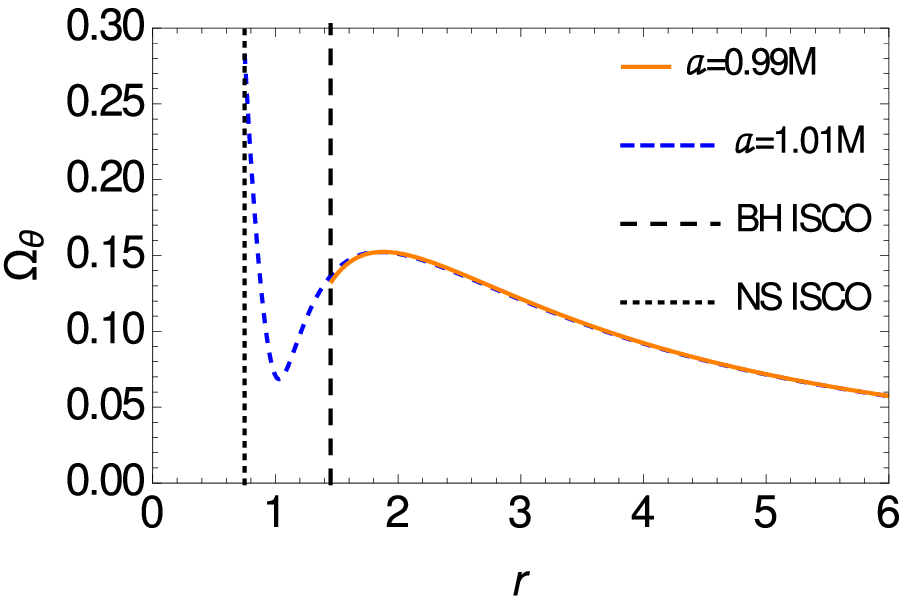}} 
\subfigure[BH ISCO and NS ISCO are located at
 $2.32~M$ and $0.67~M$ respectively]{
\includegraphics[width=3in,angle=0]{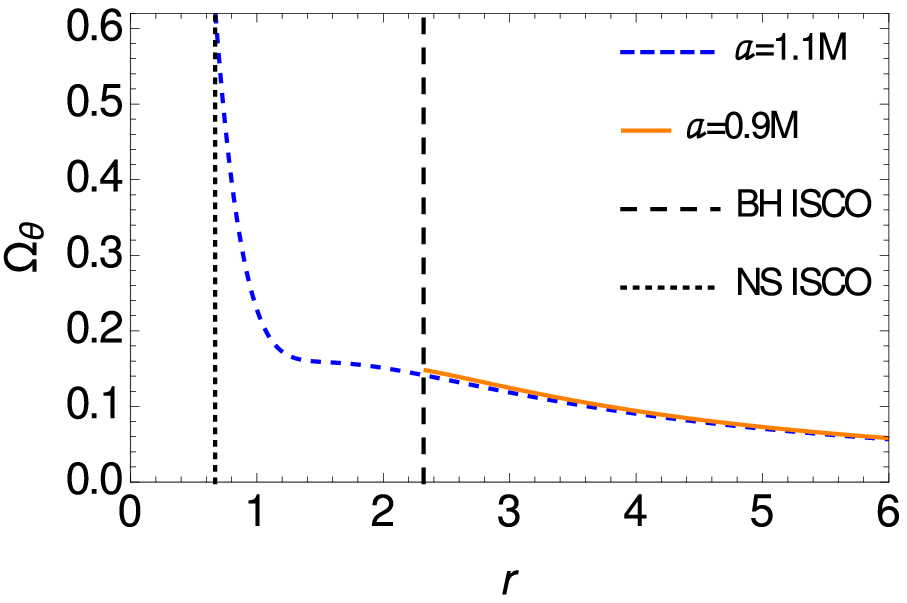}} 
\caption{\label{Oth} Variation  of $\O_{\th}$ (in units of $M^{-1}$)
versus $r$ (in units of $M$). The plots show that a local minima 
is always appeared outside of the ISCO for $1 < a_* < 1.1$ in
$\O_{\th}$ curves, in principle. This feature is completely absent
in the case of a BH.}
\end{center}
\end{figure}

\begin{figure}[!h]
\begin{center}
\includegraphics[width=3in,angle=0]{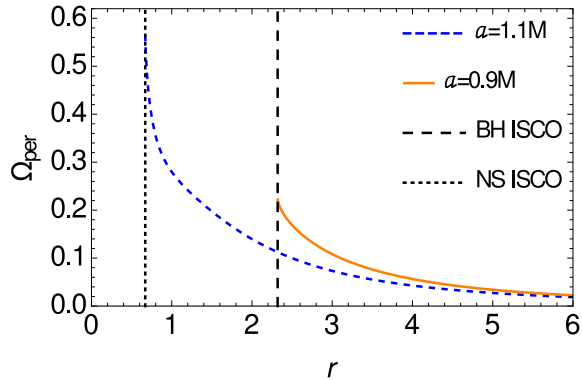} 
\caption{\label{Oper} Comparison between the periastron precession 
frequencies ($\O_{\rm per}$ in units of $M^{-1}$) of a BH and a NS
at their respective ISCOs for $\epsilon=\pm 0.1$.
BH ISCO and NS ISCO are located at $2.32~M$ and $0.67~M$ respectively.}
\end{center}
\end{figure}

\begin{table}
 \begin{tabular}{| l | c | r | l | c | l |}
  \hline                        
$a_*$  & $r_{\rm ISCO}$ & $\nu_{\phi}$ & $\nu_{\theta}$ & 
$\nu_{\rm nod}$ at $r_{\rm ISCO}$ & $\nu_{\rm nod}$ (in Hz) 
\\
 & (in $M$) & (in Hz) & (in Hz) & (in Hz) & at $r_p$ (mentioned in parentheses)
 \\ 
\hline                      
0.1 & 5.67 & 234 & 231 &  3  & \\ 
\hline                      
0.2 & 5.32 & 255 & 247 &  8  & \\ 
\hline                      
0.3 & 4.98 & 279 & 265 &  14  & \\ 
\hline                      
0.4 & 4.61 & 309 & 287 &  22  & \\ 
\hline                      
0.5 & 4.23 & 346 & 312 &  34  & \\ 
\hline                      
0.6 & 3.82 & 395 & 342 &  53  & \\ 
\hline                      
0.7 & 3.39 & 458 & 378 &  80  & \\ 
\hline                      
0.8 & 2.91 & 552 & 421 &  131  & \\ 
 \hline                      
0.9 & 2.32 & 718 & 472 &  246  & \\ 
   \hline
0.98 & 1.61 & 1053  & 462 & 590 &   \\ 
 \hline
0.99 & 1.45 & 1163  & 420 & 743 & \\ 
   \hline 
0.999 & 1.18 & 1395 & 252  & 1142 & \\ 
   \hline 
0.9999 & 1.07 & 1510 & 118 & 1392 & \\ 
   \hline 
0.99999 & 1.03 & 1556 & 54 & 1502 & \\ 
   \hline 
0.999999 & 1.016 & 1572 & 30 &  1542 & \\ 
   \hline 
 1.0  & 1 & 1591 & 0 &  1591 &  \\ 
   \hline 
1.000001 & 0.98 & 1615 &  40 &  1575 & 1586 (.998) \\ 
   \hline 
1.00001 & 0.96 & 1640 & 84 &  1556 & 1584 (.994) \\ 
   \hline 
1.0001  & 0.93 & 1677 & 157  &  1520 & 1571 (.992) \\ 
   \hline 
1.001 & 0.86 & 1769 & 370 &  1399 & 1528 (.976) \\ 
   \hline 
 1.01   & 0.75 & 1918 & 900 &  1017 & 1367 (.952) \\ 
\hline 
 1.02   & 0.71 & 1967 & 1199 &  768 & 1261 (.962) \\ 
    \hline 
     1.04   & 0.68 & 1988 & 1538 &  450 & 1104 (.976) \\ 
    \hline 
     1.06   & 0.67 & 1979 & 1745 &  234 &  1004 (.984) \\ 
    \hline 
     1.08   & 0.667 & 1959 & 1894 & 65 &  886 (1.04) \\ 
    \hline 
$\sqrt{32/27}$  & 2/3 & 1949 & 1949 &  0 & 847 (1.05) \\ 
$\approx 1.089$  &  &    & &  &  \\ 
   \hline 
1.1 & 0.67 & 1935 & 2013 &  -77 & 804 (1.07) \\ 
   \hline 
2  & 1.26 & 932 & 1588 &  -655 & 49 (3.2) \\ 
   \hline 
  4  & 3.17 & 330 & 566 &  -236 & 2 (12.3) \\ 
   \hline 
 6   & 5.38 & 172 & 288 & -116 & 0.25 (27.7) \\ 
   \hline 
\end{tabular}
\caption{An object of mass $M=10M_{\odot}(=15$ km) has been considered to calculate 
$\nu_{\phi}$ (Kepler frequency), $\nu_{\theta}$ (vertical epicyclic 
frequency) and $\nu_{\rm nod}$ (nodal plane precession frequency) using Eq.(\ref{fr1})
and Eq.(\ref{fr3}). For an example, the conversion factor between 
$\nu_{\phi}$ and $\O_{\phi}$ is as follows :
$\nu_{\phi}$ (in kHz)$=\O_{\phi}$ (in km$^{-1}$).$\f{300}{2\pi M}$ and so on. 
For other values of $M$,
the values of $\nu_{\phi}, \nu_{\theta}, \nu_{nod}$ (column no. 3, 4, 5 \& 6)
of the table have to be multiplied by $10M_{\odot}/M$. The values in the 
parentheses of Column no. 6 show the position of the peak of $\nu_{\rm nod}$.} 
  \label{table_1}
\end{table}

In case of BHs, the LT frequency increases with decreasing $r$
up to the inner edge of the accretion disk (see FIG.~\ref{Nodal}
and Panel (a) of FIG.~\ref{NodalGen}.
But for NSs, the LT frequency attains 
a maximum at $r=r_p$ which occurs always at $r (r=r_p) > r_{\mbox{ISCO}}$ 
(see TABLE \ref{table_1}), and then decreases as
$r$ decreases (FIG.~\ref{Nodal} and Panel(b) of \ref{NodalGen}).
As shown in FIGs.~\ref{isco}, \ref{Nodal} and Panel (b) of \ref{NodalGen}, the LT precession frequency 
becomes negative for $r_{\mbox{ISCO}} \leq r < r_{0}$, in case of $a_*>1.089$. This means
that the direction of LT precession is reversed.
The maximum value of $\O_{\rm nod}$ ($= 1/2M$) occurs for $a_* = 1$ at $r = 
r_{\mbox{ISCO}}$. We also note that 
\begin{eqnarray}
\frac{d\O_{\rm nod}}{dr}}|_{r_{\rm ISCO} &<& 0\ \mbox{for BH},\\
\frac{d\O_{\rm nod}}{dr}}|_{r_{\rm ISCO} &>& 0\ \mbox{for NS},
\end{eqnarray}
and hence $\O_{\rm nod}$ decreases (increases) with $r$ at
$r_{\rm ISCO}$ for BHs (NSs).

The profiles of other frequencies, i.e., $\O_{\phi}$, $\O_r$, $\O_{\th}$ 
and $\O_{\rm per}$, can also show differences between 
BHs and NSs.
FIG.~\ref{Kepler} shows that $\O_\phi$ behaves similarly for BHs 
and NSs, but much larger values are possible for the latter,
simply because the disk can extend up to much lower radii.
FIG.~\ref{Or} shows that a small additional peak appears 
in the plot of $\O_r$ profile for the near extremal value of $a_*$ in the 
case of a NS. Such a peak appears at small radius values, where 
an accretion disk cannot exist in case of a BH.
As $a_*$ increases, this peak becomes more prominent for $a_*=1.01$
and it becomes the only peak for $a_* \sim 1.05$.
Such an additional peak does not appear for the case of a BH.
Similarly, a minimum occurs in the $\O_{\th}$ profiles for
NSs with $1 < a_* < 1.1$ near the 
radius $r=M$ (FIG.~\ref{Oth}). Such a minimum does not occur
for a BH. Besides, $\O_{\th}$ for $a_*$ roughly
above 1.01 can attain a much higher value than that for BHs.
Finally, the periastron precession frequencies for NSs
can attain values much higher than those for BHs (FIG.\ref{Oper}).

\subsection*{Observational aspects}\label{Observational}

BH X-ray binary (BHXB) sources show a plethora of timing features
in X-rays \cite{bs14}. Most notable among them are 
high-frequency (HF) quasi-periodic oscillations (QPOs) and three types of
low-frequency (LF) QPOs. Sometimes two HF QPOs are seen together. Their
frequencies are observed to be in the range of several tens to several hundreds of Hz.
For example, while XTE 1650--500 has shown HF QPOs in the range of $50-270$ Hz,
4U 1630-47 has shown such QPOs in $150-450$ Hz \cite{bs12}.
The three LF QPOs are denoted with types `A', `B' and `C', and their frequencies
are typically in the ranges $6.5-8$ Hz, $0.8-6.4$ Hz and $0.01-30$ Hz respectively.
While several models exist to explain these QPOs, they are often associated
with the relativistic precession (RP) of the accretion disk, and hence with
the frequencies $\O_\phi$, $\O_r$, $\O_{\th}$, $\O_{\rm per}$
and $\O_{\rm nod}$. The RP model was originally conceived to explain
the twin kilo-Hertz (kHz) QPOs and a low-frequency QPO of neutron star
low-mass X-ray binaries \cite{sv98, sv99}. Following this idea,
frequencies of the C-type LF QPO, the lower frequency HF QPO and the higher frequency HF QPO
of BHXBs are identified with $\O_{\rm nod}$, $\O_{\rm per}$ and $\O_\phi$
respectively \cite{inm}. This can be useful to measure both
the mass ($M$) and $a_*$ of the compact object, as demonstrated by \cite{motta14}.
Table~\ref{table_1} shows that the observed LF QPOs could be identified
with $\O_{\rm nod}$ only for $a_* < 0.5$ and $a_* \sim 1.089$ (see also \cite{sp09}).
In fact, $\O_{\rm nod}$ could have a much higher value for $a_*$ closer to 1 for both
BH and NS, and it could be possible to identify an HF QPO with $\O_{\rm nod}$ (Table~\ref{table_1}).
If $a_*$ is very close to $1$, $\O_{\rm nod}$ value is quite high (Table~\ref{table_1}), and
such high frequency QPOs could be detected in future for BH and NS with $a_* \approx 1$.
While there are uncertainties in the specific identifications of observed
frequencies with the theoretical ones, the recent discovery of the C-type quasi-periodic
variation of the broad relativistic iron line energy from the BHXB H1743--322
strongly suggests that the inner accretion disk of this source is indeed
tilted and precessing \citep{ingram}. Therefore, the theoretical 
dependencies of various frequencies on $a_*$, as discussed in this section,
have potential to distinguish between a BH and a NS.

How could this be done? Here we give some examples.
Note that most of the BHXBs are transient sources, and an accretion disk is formed only during 
an outburst. Even for the persistent BHXBs, source state often changes, which implies
changes in accretion components. So it is expected that the accretion disk of
a given BHXB sometimes advances towards the central object, and sometimes recedes,
depending on the source intensity and spectral states.
If QPOs are connected to the natural frequencies mentioned above, then such a dynamics
of the disk would mean changes in QPO frequencies, as these frequencies depend on
the radial distance. And we do observe evolution of QPO frequencies. As a BH 
and a NS have significantly different theoretical radial profiles of
frequencies, it could be, in principle, possible to distinguish them by tracking the
evolution of QPO frequencies as the disk advances or recedes. For example, $\O_{\rm nod}$
for a BH will monotonically increase, and will attain the maximum value, if the
disk advances up to the ISCO radius. But $\O_{\rm nod}$ for a NS
will first increase, will attain the maximum value, and then will decrease,
as the disk advances up to $r_{\rm ISCO}$, which can be quite different from
the $r_{\rm ISCO}$ of BHs. In fact, in case of a NS,
the absolute value of $\O_{\rm nod}$ can become zero and then increase again.
Whether this will happen, and radial locations of the maximum and zero values
of $\O_{\rm nod}$ depend on $a_*$. Therefore, the Lense-Thirring precession
can provide a way to distinguish between a BH and a NS.
Similarly, the maximum possible value of $\O_\phi$ depends on $a_*$.
Finally, according to the above mentioned model, $\O_r$ is interpreted as
the separation between two HF QPO frequencies. Therefore, the qualitatively different 
$\O_r$ radial profiles for NSs with $a_* \lesssim 1.05$
can be useful to distinguish them from BHs.

\section{Conclusion \label{con}}~
The precession frequencies of the spin of test gyros attached to timelike
stationary observers in the BH case are finite both in and outside of the ergoregion
but become arbitrarily large as one considers a gyro located in an orbit close to the 
horizon,  $r \sim r_{+}$, in any direction ($0 < \th \leq \pi/2$). In contrast to this,
for NS, the precession frequencies of such gyros remain finite and regular even if one
considers those close to $r= 0$, for all $\th \nsim \pi/2$. Since the ring singularity
itself is present at $r=0, \theta=\pi/2$, the gyro frequency diverges in the limit of
approach to this region. For gyros placed increasingly closer to a BH or NS in the 
equatorial plane, their precession frequencies diverge in both the cases in different
ways, that is, it diverges close to the horizon in the case of BH whereas the divergence
occurs close to $r=0$ for NS. 

Interestingly, we have shown that the spin precession frequency stays finite in the limit of approach to the horizon for a zero angular momentum observer (ZAMO) ($q=0.5$) and diverges in all other cases.
Further, we have shown that a sharp rise/fall in the modulus of the precession frequency
is a tell-tale indication of the existence of a near extremal naked singularity and the 
location of this feature is at $r=M$. The specific maxima/minima 
structure in the radial profile of the modulus of the precession frequency $\O_p$ for
a near extremal naked singularity (for example, two local maxima or two local minima 
or two peaks or three peaks with a plateau etc) would indicate how fast the observer 
is moving with respect to the ZAMO frequency.

As we find, the nodal plane precession frequency, related to the accretion disc, has distinctive features that can be used to characterize both black holes and naked singularities, and we summarize them here. We can use these features to potentially make a statement regarding the existence of a NS:
(i) A maxima or a peak is obtained for $\O_{\rm nod}$ at some $r=r_{p}(a_*)$ for all $a_*>1$, indicating the existence of a NS. 
(ii) $\O_{\rm nod}$ vanishes at $r=r_{0}$ for a NS with $a_* \geq 1.089$ and becomes negative (which means that the LT precession reverses direction) in all orbits with $r_{0} > r \geq r_{\mbox{ISCO}}$, for a NS with $a_* > 1.089$. 
(iii) Additionally, $\O_{\rm nod}$ shows a \lq peculiar\rq\ effect: $\O_{\rm nod} \propto r^n$  (where $n \gtrsim 0$, see FIG.\ref{Nodal}) in the region $r_{0} \leq r < r_{p}$. This curve does not follow the inverse cube law of distance like other astrophysical objects. All these features are completely absent in the case of BH and this would be reflected in the observation of frequencies of QPOs. 

Finally, it can be seen from FIG.\ref{Or} and FIG.\ref{Oth} that $\O_r$ and $\O_{\th}$ 
also have characteristic differences \cite{ste} in the cases of BH and NS by which it
may be possible to detect a NS if $\O_r$ and $\O_{\th}$ are the observationally
measurable quantities \cite{sto}. It follows that if all of the above mentioned features
are never observed, then we can conclusively state that Kerr naked singularities do not
exist, or at least their abundance may be extremely small. 
\\

{\bf Acknowledgments:} 
We thank the referee for the valuable comments and suggestions. 
MP acknowledges support from the NCN grant Harmonia 6 (UMO2014/14/M/ST9/00707).


\begin{thebibliography}{99}
\bibitem{eht} http://www.eventhorizontelescope.org/
\bibitem{cm} C. Chakraborty, P. Majumdar, {\it Class. Quantum Grav.} {\bf 31}, 075006 (2014)
\bibitem{ckj} C. Chakraborty, P. Kocherlakota, P. S. Joshi, {\it Phys. Rev.} {\bf D 95}, 044006 (2017)
\bibitem{jmn} P. S. Joshi, D. Malafarina, R. Narayan {\it Class. Quantum Grav.} {\bf 31}, 015002 (2014)
\bibitem{binib} D. Bini, A. Geralico, R. T. Jantzen, {\it Phys. Rev. D} {\bf 94}, 064066 (2016) 
\bibitem{biniu} D. Bini, A. Geralico, R. T. Jantzen, arXiv:1610.06513 [gr-qc] (2016) 
\bibitem{hoj} S. A. Hojman, F. A. Asenjo, {\it Class. Quantum Grav.} {\bf 30}, 025008 (2013)
\bibitem{abk} C. Armaza, M. Banados, B. Koch, {\it Class. Quantum Grav.} {\bf 33}, 105014 (2016)
\bibitem{mtw} C. W. Misner, K. S. Thorne, J. A. Wheeler, 
\emph{Gravitation}, \emph{W. H Freeman $\&$ Company} (1973)
\bibitem{str} N. Straumann, {\it General Relativity with applications 
to Astrophysics}, Springer, Berlin (2009)
\bibitem{jan} R. T. Jantzen, P. Carini, D. Bini, {\it Annals Phys.} {\bf 215}, 1 (1992)
\bibitem{cmb} C. Chakraborty, K. P. Modak, D. Bandyopadhyay, {\it ApJ} {\bf 790}, 2 (2014)
\bibitem{ccb} D. Chatterjee, C. Chakraborty, D. Bandyopadhyay, {\it JCAP} {\bf 01} (2017) 062
\bibitem{bd}  J. M. Bardeen, \emph{Astrophys. J} {\bf 162}, 71 (1970)
\bibitem{bpt} J. M. Bardeen, W. H. Press, S. A. Teukolsky, ApJ {\bf 178}, 347 (1972)
\bibitem{horava} E. G. Gimon, P. Horava, {\it Phys. Lett.} {\bf B 672}, 299 (2009)
\bibitem{jh} J. B. Hartle, {\it Gravity:An introduction to 
Einstein's General relativity}, Pearson (2009)
\bibitem{chiba} K. Sakina, J. Chiba, {\it Phys. Rev. D} {\bf 19}, 2280 (1979)
\bibitem{JoshiMalafarina} D. Malafarina, P. S. Joshi, arXiv:1603.02848v1 (2016)
\bibitem{lb} Z. Li, C. Bambi, {\it JCAP} {\bf 03} (2013) 031
\bibitem{Patil1} M. Patil, P.S. Joshi, { \it Class. Quantum Grav.} {\bf 28}, 235012 (2011) 
\bibitem{Patil2} M. Patil, P.S. Joshi, {\it Phys. Rev. D} {\bf 84}, 104001 (2011)
\bibitem{BSW} M. Banados, J. Silk, S. West,  { \it Phys.Rev.Lett.} {\bf 103}, 111102 (2009)
\bibitem{Harada} T. Harada and M. Kimura {\it Class. Quantum Grav.} {\bf 31} 243001 (2014)
\bibitem{Patil3} M. Patil, T. Harada, K. Nakao, P.S. Joshi, M. Kimura {\it Phys. Rev.} {\bf D 93}, 104015 (2016) 
\bibitem{Scht} J. Schnittman, {\it Phys. Rev. Lett.} {\bf 113}  261102, (2014)
\bibitem{ok} A. T. Okazaki, S. Kato, J. Fukue, {\it PASJ} {\bf 39}, 457 (1987)
\bibitem{ka} S. Kato, {\it PASJ} {\bf 42}, 99 (1990)
\bibitem{lt} J. Lense,  H. Thirring, {\it Phys. Z.} {\bf 19}, 156-163, (1918)
\bibitem{bs} T. M. Belloni, L. Stella, {\it Space Sci Rev} {\bf 183}, 43, (2014)
\bibitem{sti} Z. Stuchlik, {\it Astronomical Institutes of Czechoslovakia,
Bulletin} {\bf 31}, 129 (1980)
\bibitem{pug} D. Pugliese, H. Quevedo, R. Ruffini, {\it Phys. Rev. D} {\bf 84}, 044030 (2011)
\bibitem{ig} I. F. Ranea-Sandoval, H. Vucetich, {\it Relativity and Gravitation}
{\bf 157}, 435 (2014)
\bibitem{ch} S. Chandrasekhar, {\it The Mathematical 
Theory of Black Holes}, Oxford (1992)
\bibitem{bs14}  T. M. Belloni, L. Stella, {\it Space Science Reviews} {\bf 183}, 43 (2014)
\bibitem{bs12} T. Belloni, A. Sanna, M. M\'endez, {\it MNRAS} {\bf 426}, 1701 (2012)
\bibitem{sv98} L. Stella, M. Vietri, {\it ApJ} {\bf 492}, L59 (1998)
\bibitem{sv99} L. Stella, M. Vietri, {Phys. Rev. Lett.} {\bf 82}, 17 (1999)
\bibitem{inm} A. Ingram, S. Motta, {\it MNRAS} {\bf 444}, 2065 (2014)
\bibitem{motta14} S. Motta et al., {\it MNRAS} {\bf 437}, 2554 (2014) ;
S. Motta et al., {\it MNRAS} {\bf 439}, L65 (2014)
\bibitem{sp09} L. Stella, A. Possenti, {\it Space Sci. Rev.} {\bf 148}, 105 (2009)
\bibitem{ingram} A. Ingram et. al, {\it MNRAS} {\bf 461}, 1967 (2016)
\bibitem{ste} G. T\"or\"ok, Z. Stuchlik, {\it Astronomy and Astrophysics}
{\bf 437}, 775 (2005)
\bibitem{sto} Z. Stuchlik, J. Schee, {\it Class. Quantum Grav.} {\bf 29}, 
065002 (2012)
\end{thebibliography}
\end{document}